\documentclass[doublecol]{epl2} 

\usepackage{graphicx,color}
\graphicspath{{./}{./figures/}}
\usepackage{dcolumn}
\usepackage{bm}
\usepackage{amsmath}
\usepackage{amssymb}
\usepackage{hyperref}
\usepackage{commath}
\usepackage{makecell}
\usepackage{amsthm}
\usepackage{subfig}

\usepackage{color}

\hypersetup{
    colorlinks=true,       
    linkcolor=blue,          
    citecolor=magenta,        
    filecolor=red,      
    urlcolor=cyan           
}
 
\definecolor{labelkey}{cmyk}{.4,.2,0,0}

\newcommand{\be}{\begin{equation}}
\newcommand{\ee}{\end{equation}}
\newcommand{\bea}{\begin{eqnarray}}
\newcommand{\eea}{\end{eqnarray}}

 \def\XXint#1#2#3{{ \setbox0=\hbox{$#1{#2#3}{\int}$} \vcenter{\hbox{$#2#3$}}\kern-.5\wd0}}

\shorttitle{} 
\title{Nonequilibrium dynamics of noninteracting fermions in a trap}

\author{David S. Dean \inst{1} \and Pierre Le Doussal \inst{2} \and Satya N. Majumdar \inst{3} \and Gr\'egory Schehr \inst{3}  }
\shortauthor{}

\institute{        
\inst{1} Univ. Bordeaux, CNRS, Laboratoire Ondes et Mati\`ere  d'Aquitaine
(LOMA), UMR 5798, F-33400 Talence, France\\
\inst{2} Laboratoire de Physique de l'Ecole Normale Sup\'erieure,
PSL University, CNRS, Sorbonne Universit\'es, 
24 rue Lhomond, 75231 Paris, France.\\
\inst{3} LPTMS, CNRS, Univ. Paris-Sud, Universit\'e Paris-Saclay, 91405 Orsay, France\\
}
 

\abstract{We consider the real time dynamics of $N$ noninteracting fermions in $d=1$.
They evolve in a trapping potential $V(x)$, starting from the equilibrium state in a potential $V_0(x)$. 
We study the time evolution of the Wigner function $W(x,p,t)$ in the phase space $(x,p)$, and the associated kernel which encodes all correlation functions. At $t=0$ the Wigner function for large $N$ is 
uniform in phase space inside the Fermi volume, and vanishes at the Fermi surf over a
scale $e_N$ being described by a universal scaling function related to the Airy function.
We obtain exact solutions for the Wigner function, the 
density, and the correlations in the case of harmonic and inverse square potentials, for several
$V_0(x)$. In the large $N$ limit, near the edges where the
density vanishes, we obtain limiting kernels (of the Airy or Bessel types) that retain the form found
in equilibrium, up to a time dependent rescaling. 
For non-harmonic traps the evolution
of the Fermi volume is more complex. Nevertheless we show that, for
intermediate times, the Fermi surf is still described by the same
equilibrium scaling function, with a non-trivial time and space dependent width
which we compute analytically.  We discuss the multi-time correlations and obtain
their explicit scaling forms valid near the edge for the harmonic oscillator. Finally,
we address the large time limit where relaxation to the Generalized Gibbs Ensemble (GGE) was found to occur in the
"classical" regime $\hbar \sim 1/N$. Using the diagonal ensemble we compute the
Wigner function in the quantum case (large $N$, fixed $\hbar$) and show that it agrees with the GGE. We also obtain the higher order (non-local) correlations in the diagonal ensemble.
}

\begin{document}

\maketitle

There was recent progress in the theoretical study of noninteracting fermions in a
confining trap at equilibrium \cite{GPS08,castin,Kohn,Eis2013,marino_prl,us_finiteT,DPMS:2015,fermions_review,Dubail,DubailStephan2017}, motivated in part by progress
in manipulating and imaging cold atoms \cite{BDZ08,Cheuk:2015,Haller:2015,Parsons:2015,Omran2015}.
 Although they are
noninteracting, because of Pauli exclusion they exhibit non trivial quantum
correlations. In particular at zero temperature 
and in one dimension $d=1$, 
it was shown that the correlations are universal and related to random matrix theory (RMT) \cite{mehta,forrester}.
For instance, for $N$ fermions in a harmonic trap $V(x) = \frac{1}{2} m \omega^2 x^2$,
the positions $x_i$ are in one-to-one correspondence with the eigenvalues of an $N \times N$
random matrix drawn from the Gaussian unitary ensemble. This implies that
the average density (with unit normalization) is given for large $N$ by the Wigner semi-circle $\rho_N \simeq \rho_{sc}$, 
\be \label{sc} 
\rho_{sc}(x) = \frac{2}{\pi x_e^2} \sqrt{x_e^2-x^2}
\ee 
which has edges at $x=\pm x_e$, $x_e=\sqrt{\frac{\hbar}{m \omega}} \sqrt{2 N}$. In the bulk, i.e. away from the edges, the density correlations
are described by the sine-kernel which can be obtained through the local density approximation
(LDA) \cite{castin} or in a more controlled way using the connection to RMT \cite{fermions_review}.
Near the edge, within a scale $w_N \sim N^{-1/6}$, the fluctuations are enhanced and are described by the 
so-called Airy kernel \cite{Eis2013,marino_prl,us_finiteT}. This result extends to a large class of smooth potentials \cite{fermions_review}.
In presence of hard edges, similar results were shown in terms of the Bessel kernel
\cite{FFGW03,UsHardBox,Cunden1D}.
Both kernels are well known to describe soft and hard edges in RMT. Extensions
to finite temperature $T>0$, $d>1$ and interactions were studied in~\cite{us_finiteT,LiechtyT,DPMS:2015,UsHardBoxLong,torquato,stephan}. 

It is natural to ask how these properties behave under real time quantum dynamics.
For equilibrium dynamics this was studied in \cite{UsPeriodic}, and related to
the extended Sine and Airy processes in imaginary time. Here we consider 
non-equilibrium quantum quenches, first from zero temperature $T=0$, and later from
finite temperature) and study the time evolution of the
density and correlations. In the $T=0$ quench, the many body system is prepared in the ground state of a given
Hamiltonian ${\cal H}_0$ and evolves under the Hamiltonian ${\cal H} \neq {\cal H}_0$.
Such quantum quenches in traps have been studied in a number of works,
however most of them focus on interacting bosons, in particular in $d=1$ in the Tonks-Girardeau limit of infinite repulsion \cite{Minguzzi2005,QuinnHaque2014,ColluraCalabreseTonksTrap,ColluraKormos2018} and were studied for finite $N$.
Although in this limit
there is a formal relation to noninteracting fermions
\cite{FFGW03,Atas}, the correlations, beyond the density, are
different. For instance, the possible edge universality of the time dependent fermion problem 
at large $N$ has not been addressed. Very recently the evolution
in the bulk was investigated in the limit $\hbar \sim 1/N$, 
$N \to \infty$ (which we call here "classical"), and
at large time thermalization was found to occur for a one-particle
observable \cite{Kulkarni2018}. 

We consider noninteracting fermions in the quantum regime for fixed $\hbar$.
The initial and evolution many body Hamiltonians 
thus take the forms ${\cal H}_0= \sum_{i=1}^N H_0(x_i,p_i)$ and ${\cal H}= \sum_{i=1}^N H(x_i,p_i)$, 
in terms of the single particle Hamiltonians (below we work in units where $m=1$)
\be \label{HH0} 
H_0(x,p) = \frac{p^2}{2} + V_0(x) 
\quad , \quad H(x,p) = \frac{p^2}{2} + V(x) 
\ee
A convenient way to study noninteracting fermions \cite{Wiegmann2011,BettelheimGlazman} is via the Wigner function $W(x,p,t)$
(see definition below) which plays the role of a (quasi-)probability density in the phase space $(x,p)$
\cite{Case2008}.
In particular $\int dp ~ W(x,p,t) = \tilde \rho(x,t)= N \rho_N(x,t)$, the time dependent number density. 
We choose the initial condition $W(x,p,t=0)$ as the
Wigner function associated to the ground state of ${\cal H}_0$, 
studied in \cite{Balatz1,Balatz2,UsWigner} for various
potentials $V_0$. 
It was shown that in the large $N$ limit 
it takes the form in the bulk
\bea \label{Wbulk} 
W(x,p,t=0) \simeq \frac{1}{2 \pi \hbar} \theta( \mu - H_0(x,p) ) 
\eea 
with $\theta(x)$ the Heaviside step function.
It depends on $N$ only through $\mu$, the Fermi energy, i.e. the energy of the highest occupied single particle state,
which is an increasing function of $N$.
The curve $(x_e,p_e)$ where $W$ in \eqref{Wbulk} vanishes, i.e. $\mu=H_0(x_e,p_e)$, defines an edge 
in phase space (the Fermi surf $S_0$, enclosing the Fermi volume $\Omega_0$). At large $N$, i.e. for large $\mu$ the step has a finite width $e_N=e(x_e,p_e) \ll \mu$, i.e.
near the edge and for smooth potentials $V_0(x)$, the Wigner 
function takes the universal scaling form 
\be \label{Wedge} 
\! \! \! W(x,p,t=0) \simeq \frac{1}{2 \pi \hbar} {\cal W}(a) ~ , ~ {\cal W}(a) = \int_{2^{2/3} a}^{+\infty} {\rm Ai}(u) du
\ee
in terms of the scaled distance $a$ to the Fermi surf
\be \label{a}
\! a=\frac{H_0(x,p)-\mu}{e(x,p)} ~ , ~  e(x,p) = \frac{\hbar^{\frac{2}{3}}}{2^{\frac{1}{3}}}
[p^2 V_0''(x) + V_0'(x)^2 ]^{\frac{1}{3}}
\ee

\begin{figure}[h]
\centering
\includegraphics[width =0.9 \linewidth]{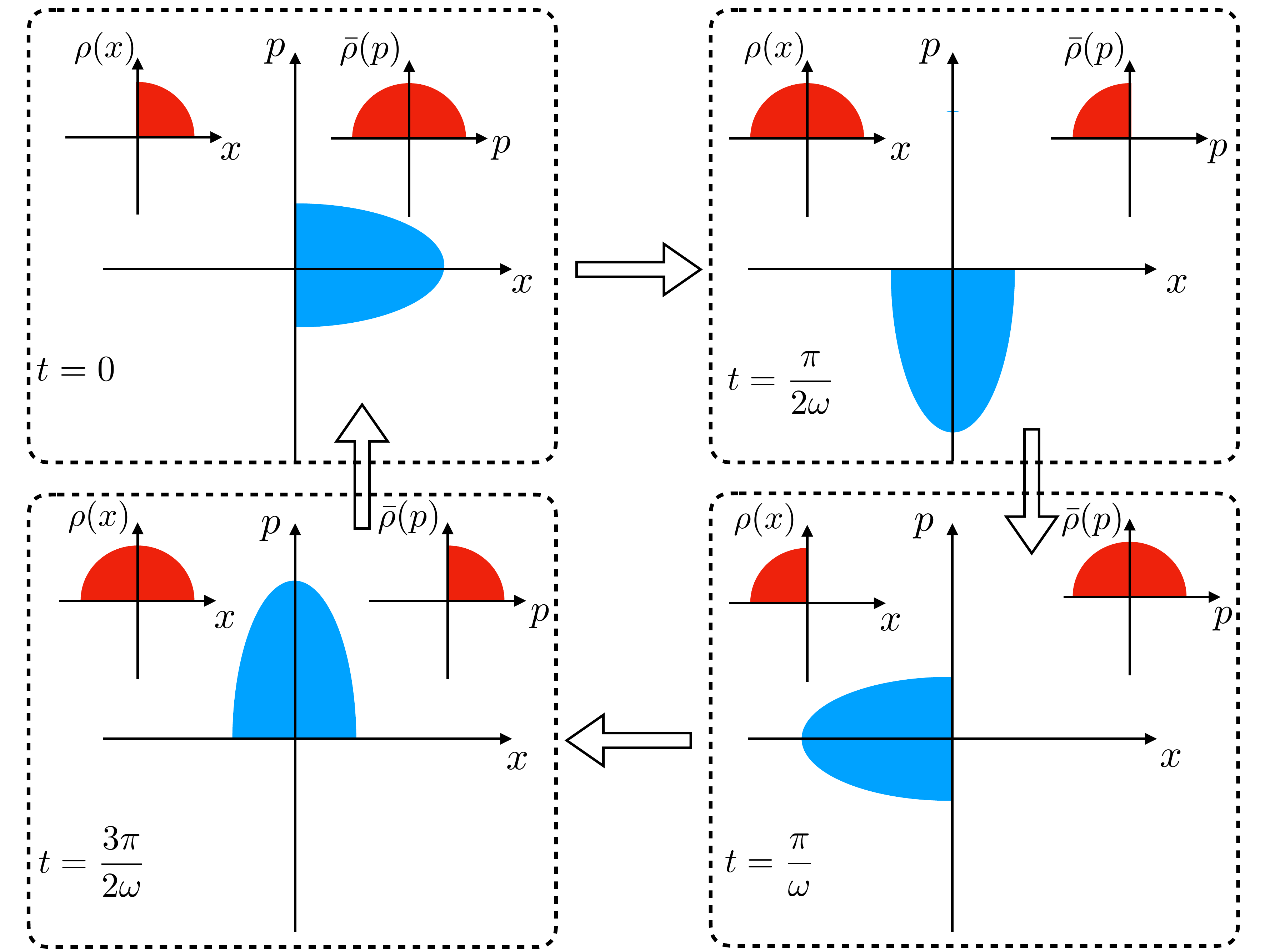}
\caption{Sketch of the ($\frac{2 \pi}{\omega}$ periodic) time evolution of the support of the Wigner function (in blue) for
the quench from a half-oscillator to an oscillator. Red: densities in $x$ ($\rho$) and in $p$
($\bar \rho$).}
\label{fig1}
\end{figure}

Our goal in this paper is to study the time dependent Wigner function starting from this initial condition,
e.g. \eqref{Wbulk}, \eqref{Wedge},
and to understand how its bulk and edge properties evolve 
with time, in particular whether their universality is 
preserved under nonequilibrium dynamics.
We
obtain exact solutions for time dependent harmonic and inverse square potentials, and calculate the Wigner function and the density for various $V_0(x)$. For large $N$, near the edges 
where the density vanishes, 
we obtain limiting kernels (of the Airy or Bessel types) that retain the form found
in the statics, up to a time dependent rescaling. It implies the occurrence of the
Tracy-Widom distribution \cite{TracyWidomAiry} 
for the position of the rightmost fermion \cite{us_finiteT}.
For generic non-harmonic traps,
at large $N$ and intermediate times, the Wigner function is still uniform on the Fermi volume,
whose evolution can be more complex. We show that at generic points the
Fermi surf is still described by the same universal function as in
equilibrium \eqref{Wedge}, \eqref{W_b}, with a space and time dependent
width that we calculate. We express multi-point and multi-time correlations
in terms of the Wigner function and its associated kernel. In the infinite time limit, we obtain exact 
formula for the Wigner function under a thermalization hypothesis, which we show coincide
with the prediction of the Generalized Gibbs Ensemble (GGE) \cite{EsslerReview,GGECalabrese}.

The time evolved $N$ body wave function keeps the form of a Slater determinant
\be
\Psi(x_1,\cdots x_N;t) = \frac{1}{\sqrt{N!}} \det_{1 \leq k,j \leq N} \psi_{k}(x_j,t)   \label{prop2}
\ee 
where the $\psi_k(x,t)$ are the solutions of $i \hbar \partial_t \psi= H \psi$ with initial condition $\psi_k(x,t=0)=\phi^0_k(x)$, where $\phi^0_k(x)$, $k \geq 1$, are the single particles eigenstates of $H_0$ in increasing order 
of energies $\epsilon^0_k$.
We are interested in the quantum probability 
$|\Psi(x_1,\cdots x_N;t)|^2 = \frac{1}{N!} \det_{1 \leq i,j \leq N} K_\mu(x_i,x_j,t)$,
described by the time dependent kernel
\be \label{kernel} 
K_\mu(x,x',t) = \sum_{k=1}^N  \psi_k^*(x,t) \psi_k(x',t) 
\ee
where the $\psi_k(x,t)$'s form an orthonormal basis for all $t$. 
All $m$-point correlations $R_m(x_1,\dots x_m;t):=\frac{N!}{(N-m)!} \prod_{j=m+1}^N \int dx_j
|\Psi(x_1,\cdots x_N;t)|^2$ are determinants, $R_m(x_1,\dots x_m;t) = \det_{1\leq i,j \leq m} K_\mu(x_i,x_j,t)$,
i.e. the $x_i$'s form a determinantal point process \cite{johansson,borodin_determinantal}. 
One defines the $N$-body Wigner function
\bea \label{Wdef1} 
&& W(x,p,t) = \frac{N}{2 \pi \hbar} \int_{-\infty}^{+\infty} dx_2 .. dx_N dy ~ e^{ \frac{i p y}{\hbar}} \\
&& \times 
\Psi^*(x+ \frac{y}{2},x_2,\cdots x_N;t)  \Psi(x- \frac{y}{2}, x_2,\cdots x_N;t)  \nonumber 
\eea 
which is related to the kernel via \cite{UsWigner}
\be \label{Wrel} 
W(x,p,t) = \frac{1}{2 \pi \hbar} \int_{-\infty}^{+\infty} dy ~ e^{ \frac{i p y}{\hbar}} 
K_\mu(x+ \frac{y}{2},x- \frac{y}{2},t)
\ee 
Using the Schr\"odinger equation for the $\psi_k(x',t)$ in \eqref{kernel} we
see that the kernel $K_\mu(x,x',t)$ satisfies 
\be \label{eqK}
\hbar \partial_t K_\mu= - \frac{i \hbar^2}{2} (\partial_x^2 - \partial_{x'}^2) K_\mu 
+ i (V(x) - V(x')) K_\mu
\ee
which, via Fourier transformation, leads to the evolution equation
for $W(x,p,t)$, i.e. the Wigner equation (WE)
\be \label{exactW} 
\partial_t W = \left( - p \partial_x  + \frac{i}{\hbar} V(x - \frac{i \hbar}{2} \partial_p) - \frac{i}{\hbar}  V(x + \frac{i \hbar}{2} \partial_p) \right) 
W
\ee
Although derived here for any $N$, this equation does not explicitly depend on $N$, hence
it is identical to the 
equation for $N=1$ \cite{Case2008}. It was also obtained
in second quantized form in \cite{WadiaFermi,Kulkarni2018}. The crucial difference
thus lies in the initial condition, discussed above in the large $N$ limit, see
\eqref{Wbulk}-\eqref{Wedge}-\eqref{a} at $T=0$. 
In the classical limit, i.e. setting $\hbar=0$ in the WE, one obtains the Liouville equation (LE)
\be
\label{Liouville} 
\partial_t W = \left( - p \partial_x  + V'(x) \partial_p \right) W
\ee 
It verifies $dW(x(t),p(t),t)/dt=0$ along the classical trajectories 
$\dot x= p$ and $\dot p=-V'(x)$. Denoting $x_0(x,p,t)$, $p_0(x,p,t)$ the
initial conditions (at $t=0$) as functions of the final ones (at $t$), 
the general solution of the LE can thus be written as
$W(x,p,t) = W_0(x_0(x,p,t), p_0(x,p,t))$ where $W_0(x,p)=W(x,p,0)$ 
is the initial Wigner function. 

Let us start with the simplest solvable case of the harmonic oscillator, $V(x)= \frac{1}{2}  \omega^2 x^2$.
Since $V'''(x)=0$, the WE  \eqref{exactW}, reduces to the LE \eqref{Liouville}. Hence the solution is
\bea \label{WHO1} 
&& \!\!\!  \!\!\!  \!\!\!  W(x,p,t) \\
&&  \!\!\!  \!\!\!  \!\!\!  = W_0(x\cos(\omega t) - \frac{p}{\omega} \sin(\omega t) , p \cos( \omega t) 
+ \omega x \sin(\omega t)) \nonumber
\eea
which is a simple rotation in phase space of the initial Wigner function and is valid for any $N$. 
Let us specify the
initial Hamiltonian $H_0$ to be a harmonic oscillator, i.e. $V_0(x)=\frac{1}{2}  \omega_0^2 x^2$
in \eqref{HH0}.
Hence the initial density $\rho_N(x,t=0)$ is the semi-circle \eqref{sc} for 
large $N$ with $x_e=\sqrt{2 \mu}/\omega_0= \sqrt{2N \hbar/\omega_0}$.
Substituting the initial condition \eqref{Wbulk} into \eqref{WHO1} one immediately
finds that for large $N$ the density remains a semicircle with $x_e \to x_e(t)$
\be \label{xet}
x_e(t)= \frac{\sqrt{\mu}}{\omega_0 \omega} 
\sqrt{\omega^2+ \omega_0^2 + (\omega^2-\omega_0^2) \cos(2 \omega t)}
\ee 
and $\mu = N \hbar \omega_0$. In fact a solution of the general time
dependent harmonic oscillator, $H(t)=\frac{p^2}{2} + \frac{1}{2} \omega(t)^2 x^2$,
can be obtained for the same initial condition 
using the `rescaling' method 
\cite{Popov1969,Gritsev2010}. The wave functions are given by
\bea
\psi_k(x,t) = \frac{e^{i \frac{L'(t)}{2 L(t) \hbar} x^2 - i \frac{\epsilon_k^0 \tau(t)}{\hbar}} }{\sqrt{L(t)}} \phi^0_k( \frac{x}{L(t)} ) \;,
\eea 
where $\tau(t) = \int_0^t dt'/L^2(t')$ and 
$L(t)$ satisfies Ermakov's equation 
\be \label{erm} 
\partial^2_t L(t) + \omega(t)^2 L(t) = \frac{ \omega_0^2}{L(t)^3}  
\ee
with $L(0)=1$, $L'(0)=0$. The kernel and the density are
thus given in terms of their initial values, for any $N$ by 
\bea \label{solu1} 
K_\mu(x,x',t) = \frac{e^{i \frac{L'(t)}{2 L(t) \hbar} ({x'}^2-x^2)} }{L(t)} K_\mu(\frac{x}{L(t)}, \frac{x'}{L(t)},0) 
\eea 
and $\rho_N(x,t) = K_\mu(x,x,t)/N = \frac{1}{L(t)} \rho_N(\frac{x}{L(t)},0)$.
In addition we obtain \cite{SM} the exact formula for 
\be \label{solu2}
W(x,p,t)= W_0(\frac{x}{L(t)}, p L(t) - x L'(t))
\ee 
which reduces to \eqref{WHO1} for $\omega(t)=\omega$. 
Hence we see that for large $N$ the density remains a semi-circle
with an edge at $x_e(t)=L(t) x_e$, formula \eqref{xet}
being recovered in the case $\omega(t)=\omega$. For the kernel,
the scaling forms obtained at equilibrium, namely the sine-kernel (in the bulk) and the
Airy kernel (at the edge) \cite{fermions_review}, are preserved, up to a
change in the scale. In particular the density at the
edge reads
\be \label{rhoedge} 
\rho(x,t)=\frac{1}{w_N(t)} F_1(\frac{x-x_e(t)}{w_N(t)})
\ee 
with $F_1(y)={\rm Ai}'(y)^2 - y {\rm Ai}(y)^2$ and a width
$w_N(t)=w_N L(t)$ and $w_N=\sqrt{\frac{\hbar}{2 \omega_0}} N^{-1/6} \sim \mu^{-1/6}$.

The Wigner function method allows to study a variety of initial conditions.
One example is $H_0$ given by the $\omega_0$ harmonic oscillator 
restricted to $x>0$ 
with an impenetrable wall at $x=0$. The initial kernel, $K^+_\mu$,
is obtained
via the image method $K^+_\mu(x,x')=(K_\mu(x,x')-K_\mu(x,-x'))\theta(x) \theta(x')$.
The associated Wigner function in the bulk is unity on the half-ellipse 
$H_0(x,p)<\mu$ with $x>0$. Under evolution with $\omega(t)=\omega$, this
half ellipse rotates clockwise (see Fig. \ref{fig1}). Consider for
simplicity $\omega=\omega_0$. Defining $x=x_e \sin \theta$, $\theta \in [-\frac{\pi}{2},\frac{\pi}{2}]$,
where $x_e=\sqrt{2 \mu}/\omega_0$,
the density reads for $0<\omega t< \pi/2$
\be \label{soluhalf} 
\rho_N(x,t) =  \frac{2}{\pi x_e} \begin{cases} 
& 0 \quad , \quad \sin \theta < - \sin \omega t \\
& \frac{\sin(\omega t + \theta)}{\sin \omega t} \quad , \quad |\sin \theta| < \sin \omega t \\
& 2 \cos \theta  \quad , \quad \sin \theta > \sin \omega t
\end{cases} 
\ee 
and in addition satisfies $\rho_N(x,t)=\rho_N(-x,\frac{\pi}{\omega} - t)$
and is time periodic of period $2\pi/\omega$. 
Hence it is not a semi-circle, and there is now a moving front
(e.g. at $x=-x_e \sin \omega t$ for $\omega t<\frac{\pi}{2}$)
where the density vanishes linearly (see \cite{SM} for details).
A formula for $W$ can also be obtained for any 
$\omega(t)$ and arbitrary initial condition (see \cite{SM}).


It is possible to extend the exact solution 
to the case of an additional $1/x^2$ wall at $x=0$, following \cite{Kanasugi1995}. Consider
\be  \label{defValpha}
H(x,p,t) = \frac{p^2}{2} + \frac{1}{2} \omega(t)^2 x^2 +  \frac{\alpha(\alpha-1) \hbar^2}{2 x^2} \quad , \quad x>0
\ee
and denote $\omega_0=\omega(0)$. The 
initial kernel has an explicit expression \cite{SM} in terms of
Laguerre polynomials for any $N$. 
%
At large $N$, the density in the bulk is a time-dependent half semi-circle 
\be
\!\!\!\! \rho_N(x,t) \simeq \frac{1}{L(t)} \tilde \rho(\frac{x}{L(t)},0) ~,~ 
\tilde \rho(x)= \frac{4  \theta(x)}{\pi x_e^2} \sqrt{x_e^2-x^2}
\ee
where $L(t)$ is the same 
solution of \eqref{erm}, e.g. $L(t)=x_e(t)/x_e$ for $\omega(t>0)=\omega$.
Near the wall the kernel takes the form
\be
K_\mu(x,y,t) \simeq
\frac{2 k_F^2 \sqrt{x x'}}{L(t)^2} 
K^{\rm Be}_\nu( \frac{k_F^2 x^2}{L(t)^2},\frac{k_F^2 y^2}{L(t)^2}) 
\ee
up to a phase factor, where 
$\mu= \frac{\hbar^2}{2} k_F^2$, $\nu=\alpha + \frac{1}{2}$, in terms of the Bessel kernel 
\be \label{K_bessel}
\!\!\!\!\!  K^{\rm Be}_\nu(u,v)= \frac{ \sqrt{v} J'_{\nu}(\sqrt{v}) J_{\nu}(\sqrt{u}) 
- \sqrt{u} J'_{\nu}(\sqrt{u}) J_{\nu}(\sqrt{v}) }{2 (u-v)} 
\ee
characteristic of the "hard-edge" universality class of RMT \cite{UsHardBoxLong}, thus
preserved by the dynamics in this case.

Consider now generic smooth confining potentials $V(x)$, $V_0(x)$, with $V'''(x) \neq 0$.
Our strategy is as follows.
We first argue that, for large $N$ (or equivalently large $\mu$), $W$ obeys Liouville equation \eqref{Liouville}
(and at $T=0$ is a theta function, i.e.
takes value in $\{0,1\}$). We then study the solution to the LE, leading to Burgers equation
and fermion hydrodynamics. Finally, we obtain the leading (quantum) corrections to that picture, of foremost
importance near the Fermi surf. 

Let us expand \eqref{exactW} to the first "quantum" order as
\be \label{cubic}
\big(\partial_t + p ~ \partial_x - V'(x) \partial_p + \frac{\hbar^2}{24} V'''(x) \partial_p^3 \big) W =0
\ee 
To check that the quantum terms are subdominant, we can
evaluate each term at $t=0$ using the initial condition \eqref{Wedge}-\eqref{a}. The derivatives
are non-zero only near the Fermi-surf: consider a generic point there, $(x_e,p_e)$,
such that $p_e^2 \sim V_0(x_e) \sim \mu$, leading to $e_N \sim (\hbar \mu/x_e)^{2/3}$. As $t$ increases, 
we assume that $V(x)$ and its derivatives still scale the same
way as $V_0(x)$, i.e. $V^{(n)}(x_e) \sim \mu/x_e^n$. Each order in $\hbar$ brings $\partial_x$ acting on
$V$ and $\partial_p$ acting on $W$, hence, using the scaling form \eqref{Wedge}-\eqref{a},
a factor either $\sim \hbar/(x_e p_e)$ or
$\sim \hbar p_e/(x_e e_N)$. The latter dominates, but is $\sim \hbar/(x_e^{1/3} \mu^{1/6}) \ll 1$
at large $\mu$. 

Under the Liouville equation the Fermi volume $\Omega_t$ and surf $S_t$ are simply
transported by the (phase-space area preserving)
classical equation of motion, i.e. 
\be
W(x,p,t)= \frac{1}{2 \pi \hbar} \theta\big(\mu- \frac{p_0(x,p,t)^2}{2} - V_0(x_0(x,p,t))\big) 
\ee
There are several parametrizations to describe the
transported (and possibly wildly deformed) Fermi volume $\Omega_t$ \cite{SM}. 
In the simplest case one can write
\be \label{WS} 
W(x,p,t) = \frac{1}{2 \pi \hbar} \theta(p - p_-(x,t))  \theta(p_+(x,t)-p) 
\ee
for $x \in {\cal I} = [x^-_e(t),x^+_e(t)]$ the (single) support of the 
number density $\tilde \rho(x,t)=\frac{1}{2 \pi \hbar} (p_+(x,t) - p_-(x,t))$.
For $x \in {\cal I}$, each $p_\pm$ must then satisfy the Burgers equation
\cite{footnote3} 
\be \label{burgers} 
\partial_t p(x,t) + p(x,t) \partial_x p(x,t) + V'(x)=0
\ee
It recovers the standard free fermion hydrodynamics with a velocity field 
$v = \frac{1}{2} (p_+ + p_-)$ with
$\tilde \rho v= \int dp ~ p W$, and
the local kinetic energy density
$\int dp ~ \frac{p^2}{2} W = \frac{1}{2} \tilde \rho v^2 + \frac{\hbar^2}{6} \pi^2 \tilde \rho^3$
\cite{SM}.
The equilibrium is recovered for $V(x)=V_0(x)$ with
\bea
p_\pm(x,t)= \pm p_\mu^0(x) \quad , \quad p_\mu^0(x) = \sqrt{2 (\mu- V_0(x))_+} 
\eea 
and $W$ is time independent. The form \eqref{WS} using \eqref{burgers}
automatically satisfies Liouville equation. 

We now study the behavior of the Wigner function near the transported 
Fermi-surf $S_t$. We know that at $t=0$ it is given by \eqref{Wedge}-\eqref{a}
and has a thickness $e_N=e(x_e,p_e)$. We thus look for a solution of \eqref{cubic} for $p \approx p_+(x,t)$ of the form 
\bea \label{Wscal} 
W(x,p,t) \simeq F(\tilde a) \quad , \quad \tilde a = \frac{p - p_+(x,t)}{D(x,t)} 
\eea
with a time-dependent thickness $D(x,t)$, $F(+\infty)=0$ and $F(-\infty)=1$.
Remarkably, plugging this form
into Eq.\eqref{cubic} allows to determine uniquely \cite{SM}
\be \label{D}
D(x,t) = - (\frac{\hbar^2}{2} \partial^2_{x} p_+(x,t))^{1/3}
\ee
and $F'''(a) = 4 a F'(a)$, which is obeyed by $F(a) = {\cal W}(a)={\rm Ai}_1(2^{2/3} a)$. It also provides the correct matching with the initial condition,
since $- (\frac{\hbar^2}{2} \partial^2_{x} p_\mu^0(x))^{1/3} =\frac{e(x,p)}{p}|_{p=p_\mu^0(x)}$
on the Fermi surf. The solution corresponding to $\theta(p-p_-)$ is obtained similarly
with $\tilde a\to -\tilde a$. Furthermore
we show \cite{SM} that for smooth potentials, adding the missing higher order terms
in $\hbar$ of the WE 
in Eq.\eqref{cubic} does not change the result. 
Thus derivatives up to and including $V'''$ determine the 
dynamical width, while only $V''$ and $V'$ enter the static width $e_N$.
Hence we have found the universal scaling form near a generic 
point of the transported Fermi-surf $S_t$.
The width \eqref{D} can be calculated for any $V(x)$ solving the Burgers equation \eqref{burgers}
with initial condition $p_+(x,t=0)=p_\mu^0(x)$. Note that related 
results were obtained
\cite{Berry1977,Berry1979,Filippas2003,Filippas2006} for a single particle in the
semi-classical limit, and for fermions at equilibrium 
\cite{Balatz1,Balatz2}. Other expressions can be
obtained \cite{SM} at non-generic points, e.g. when $\partial^2_{x} p(x,t)=0$ 
\cite{Filippas2003,UsMulticritical,UsWigner}.
We assumed here that
the Fermi surf maintained its integrity, which in some cases
requires finite time. More generally, the section $\Omega_t(x)$ 
of $\Omega_t$ (such that $p \in \Omega_t(x)$ iff $(x,p) \in \Omega_t$)
can contain multiple intervals $\Omega_t(x) = \cup_{j=1}^{n(x)} [p^j_-(x,t),p^j_+(x,t)]$
which merge or disappear as $x$ varies. Locally, however,
as in the statics as long as one can find a smooth local parametrization
of the Fermi surf, similar scaling form as \eqref{Wscal} should hold. This dynamical
scenario remains valid over a time scale when the Fermi surf does not change drastically from the initial
condition. Beyond this time scale, the dynamics
becomes quite different, as discussed later. However, it is hard to estimate precisely this time scale.

The above results can be extended to a quench from an initial system
prepared at finite temperature $T=1/\beta$, in the grand canonical (GC)
ensemble with chemical potential $\tilde \mu$. The equal time correlations
are again determinantal with the GC kernel (overbars denote GC averages)
\be \label{kernelGC} 
K_{\tilde \mu}(x,x',t) = \sum_{k=1}^{+\infty} \bar n^0_k  \psi_k^*(x,t) \psi_k(x',t) 
\ee
with $\bar n^0_k=1/(1+e^{\beta (\epsilon^0_k-\tilde \mu)})$ the mean occupation number of
the energy level $\epsilon^0_k$ of $H_0$. The GC Wigner function $W_{\tilde \mu}$
(defined in \cite{UsWigner}) is again the Fourier transform of the GC kernel,
as in \eqref{Wrel}. It is clear from \eqref{kernelGC} that it obeys \eqref{eqK}
hence $W$ still the solution of \eqref{exactW}, albeit with a different initial
condition. For large $N$, in the bulk it reads
\be \label{WT} 
W_{\tilde \mu}(x,p,t=0) \simeq [2 \pi \hbar (1+ e^{\beta(H_0(x_0(x,p,t),p_0(x,p,t))-\tilde \mu)})]^{-1} \;.
\nonumber
\ee 
Introducing the important dimensionless parameter $b= e_N/T$,
the (same) Fermi surf is still well defined at low $T = e_N/b \ll \mu$ (with 
$\tilde \mu \simeq \mu$) and 
Eq. \eqref{Wedge} still applies with
\be \label{W_b} 
{\cal W}(a) \to {\cal W}_b(a) =\int_{-\infty}^{+\infty} dy {\rm Ai}(y)/(1+ e^{b(a - y 2^{-2/3})})
\ee
a universal function depending only on 
the dimensionless parameter $b$ (with ${\cal W}_{+\infty}(a)={\cal W}(a)$ in the $T=0$ limit).
Formula \eqref{WHO1}, \eqref{solu1}, \eqref{solu2},
readily apply with, for large $N,\tilde \mu$, these new initial conditions.
Eqs. \eqref{soluhalf}, \eqref{K_bessel}, \eqref{rhoedge}
admit $T>0$ extensions \cite{SM}.

From the linearity of the W-equation \eqref{exactW} one can
express its solution at $T>0$ in terms of the $T=0$ solution~as
\be
W_{\tilde \mu}(x,p,t)=\int d\mu' \frac{\partial_{\mu'} W(x,p,t)|_{\mu=\mu'}}{1+ e^{\beta(\mu'-\tilde \mu)}} 
\ee 
since it holds at $t=0$ (equilibrium) as shown in \cite{UsWigner}.
At low $T$ near the Fermi surf, it allows to generalize Eq.~\eqref{Wscal} 
\be
W_{\tilde \mu}(x,p,t) \simeq  {\cal W}_{b(x,t)}(\tilde a) , ~
b(x,t) = \beta \frac{- (\frac{\hbar^2}{2} \partial_x^2 p_+(x,t,\mu))^{1/3}}{\partial_\mu p_+(x,t,\mu)}
\ee
where the dependence in $\mu$ of $p_+(x,t,\mu)$ is indicated explicitly,
and the parameter $b$ acquires a dependence in $x,t$, which
however disappears for the harmonic 
oscillator for which $b(x,t)=b=\beta (\hbar \omega_0)^{2/3} \mu^{1/3}$,
as for equilibrium. 

We now study the multi-time quantum correlations.
As for equilibrium dynamics \cite{UsPeriodic}
one shows from the Eynard-Mehta theorem \cite{borodin_determinantal} that
they are determinants based on the non-equilibrium space-time
extended kernel~\cite{footnote4} 
\be \label{extended} 
K_{\tilde \mu}(x,t;x',t') =  \sum_k (\bar n_k   -  \theta_+(t'-t)) \psi^*_k(x,t) 
\psi_k(x',t') 
\ee
with $\theta_+(x)=1$ for $x>0$ and $\theta_+(x)=0$ for $x \leq 0$.
For instance the density-density correlation for $t_1 < t_2$ reads
$\langle \hat \rho(x_1,t_1) \hat \rho(x_2,t_2) \rangle =
\det_{1 \leq i,j \leq 2} K_{\tilde \mu}(x_i,t_i;x_j t_j)$.
%
Consider now $H$ to be the $\omega(t)$-harmonic oscillator
and use the rescaling method. If $H_0$ is the
$\omega_0$-harmonic oscillator, one easily obtains, up to a phase factor \cite{SM}
\be 
K_{\tilde \mu}(x,t;x',t') = 
\frac{1}{\sqrt{L(t) L(t')}} 
K_{\tilde \mu}^{\rm eq}(\frac{x}{L(t)},\tau(t);\frac{x'}{L(t')},\tau(t')) 
\ee
where $L(t)$ obeys \eqref{erm}, and $\tau(t)=\int_0^t dt'/L(t')^2$.
For $\omega(t)=\omega$, then $L(t)=x_e(t)/x_e$ as given in \eqref{xet}
and $\tan \omega_0 \tau(t)= \frac{\omega_0}{\omega} \tan \omega t$.
Here $K_{\tilde \mu}^{\rm eq}(x,t;x',t')$ is the {\it equilibrium} two time kernel of the 
$\omega_0$-harmonic oscillator studied in \cite{UsPeriodic},
given by \eqref{extended} with $\psi_k^*(x,t) \psi_k(x',t') \to
{\phi^0_k}^*(x) \phi^0_k(x')
e^{- i \omega_0 k (t'-t)}$. This relation between $K_{\tilde \mu}$ and $K_{\tilde \mu}^{\rm eq}$ is
special to the harmonic oscillator. 
$K_{\tilde \mu}^{\rm eq}$ is periodic in both times of 
period $2 \pi/\omega_0$, and $K_{\tilde \mu}$ of period $2 \pi/\omega$. 
Within a period $K_{\tilde \mu}^{\rm eq}$ decays very fast 
in $t-t'$ with a time scale $\sim 1/N$ in the bulk and $\sim 1/N^{1/3}$ at the edge
\cite{UsPeriodic}. One can thus expand near $t \simeq t'$ and one obtains
the following scaling forms for $K_{\tilde \mu}$. In the bulk
\be
\!\!\! K_{\tilde \mu}(x,t;x',t') \simeq \frac{1}{\xi_{x,t}} {\cal K}^{\rm bulk,r}(\frac{x-x'-v_{x,t}(t-t')}{\xi_{x,t}},\frac{\hbar(t-t')}{\xi_{x,t}^2})
\ee
where ${\cal K}^{\rm bulk,r}$ is the real time continuation of the equilibrium extended sine kernel
given at $T=0$ e.g. in Eq. (70-71) of \cite{UsPeriodic} and \cite{SM}.
The width $\xi_{x,t}=2 \hbar L(t)^2/(\omega_0 \sqrt{x_e(t)^2-x^2}) \sim N^{-1/2}$
and the velocity $v_{xt}= x L'(t)/L(t) \sim N^{1/2}$ are new features of the
non-equilibrium dynamics ($v_{xt}=0$ at equilibrium). Near the edge
it takes the scaling form 
\bea
&& \!\!\! \!\!\! \!\!\! K_{\tilde \mu}(x,t;x',t') \simeq \\
&& \!\! \!\!\! \!\!\! \!\!\! \! \! \! \frac{1}{w_N(t)} {\cal K}_b^{\rm edge,r}(\frac{x-x_e(t)}{w_N(t)},
\frac{x'-x_e(t) + v^e_t (t-t') }{w_N(t)},\frac{2 \hbar (t-t')}{w_N(t)^2}) \nonumber
\eea
where $v_e(t)=x_e(t) L'(t)$ is the velocity of the edge and
${\cal K}_b^{\rm edge,r}$ is the equilibrium extended Airy kernel
(continued to real time)
given e.g. in Eq. (156) of \cite{UsPeriodic} and \cite{SM}.

Finally we discuss the large time behavior. In the "classical" limit
$\hbar \sim 1/N$, $N \to \infty$, it was shown recently \cite{Kulkarni2018} that
for generic confining potentials, e.g. different from the HO,
the Wigner function has a large time stationary limit. It was found 
to coincide with the prediction of the GGE. Here we investigate this
question for fixed $\hbar$. Starting from the time-dependent kernel
\eqref{kernelGC}, and expanding on the basis of the eigenfunctions
$\phi_\ell(x)=\langle x|\phi_\ell\rangle$ of $H$ (of energies $\epsilon_\ell$)
we obtain
\be
\! \! K_{\tilde \mu}(x,y,t) \! =\sum_{k,\ell,\ell'} e^{- \frac{i}{\hbar} (\epsilon_\ell - \epsilon_{\ell'}) t} 
\langle \phi_\ell | \phi_k^0 \rangle \bar n^0_k \langle \phi_k^0 | \phi_{\ell'} \rangle 
\phi_\ell^*(x) \phi_{\ell'}(y) \nonumber 
\ee 
For a confining potential the spectrum of $H$ is non-degenerate,
hence the only non-oscillating terms are $\ell=\ell'$. Keeping only these
terms one obtains
\be \label{nuell} 
\! \! K^{\rm di}_{\tilde \mu}(x,y)  = \sum_{\ell=1}^{\infty} \nu_\ell \phi_\ell^*(x) \phi_{\ell}(y) \\
~,~ \nu_\ell= \langle \phi_\ell | \frac{1}{1 + e^{\beta(H_0- \tilde \mu)}}|\phi_\ell \rangle
\ee
where the $0 \leq \nu_\ell \leq 1$ are "effective mean occupation numbers",
with $\sum_\ell \nu_\ell= \sum_k \bar n_k^0 = \bar N$ (i.e. $N$ at $T=0$),
as expected from particle number conservation. This is 
the so-called diagonal approximation (DA) \cite{EsslerReview} 
which, in the present case, can
also be obtained as a time average, i.e. 
$K^{\rm di}_{\tilde \mu}(x,y) = \lim_{\tau\to \infty} \frac{1}{\tau}\int_0^\tau dt K_{\tilde \mu}(x,y,t)$.
Under certain conditions, including absence of time periodicity, the large time limit exists 
$K^{\infty}_{\tilde \mu}(x,y) = \lim_{t \to \infty} K_{\tilde \mu}(x,y,t)$ and then 
$K_{\tilde \mu}^{\infty}=K_{\tilde \mu}^{\rm di}$. Let us define a fonction
$\nu_{\tilde \mu}(\epsilon)$ such that $\nu_{\tilde \mu}(\epsilon_\ell)=\nu_\ell$
for all $\ell$. Then one can obtain the diagonal Wigner function
\be \label{Wdiag}
W_{\tilde \mu}^{\rm di}(x,p)=\int d\mu' \nu_{\tilde \mu}(\mu') \partial_{\mu'} W^H_{\mu'}(x,p)
\ee 
in terms of the Wigner function, $W^H_{\mu}(x,p)$, associated to the ground state
of ${\cal H}$ i.e to the kernel $K^H_\mu(x,y)=\sum_\ell \theta(\mu-\epsilon_\ell) \phi_\ell^*(x) \phi_{\ell}(y)$
(one first shows that $K_{\tilde \mu}^{\rm di}$ and $K_\mu^H$ are related via \eqref{Wdiag}, which implies
\eqref{Wdiag} for their Wigner functions). Note that if one defines
$\hat \nu_{\tilde \mu}(\epsilon)=\sum_\ell \nu_\ell \delta(\epsilon-\epsilon_\ell)$,
then $\hat \nu_{\tilde \mu}(\epsilon) = \nu_{\tilde \mu}(\epsilon) \rho(\epsilon)$,
where $\rho(\epsilon)=\sum_\ell \delta(\epsilon-\epsilon_\ell)$ is the density of states.
The above holds for arbitrary $N,\tilde \mu$. 

For large $N$, we insert in
\eqref{Wdiag} the known form of the Wigner function 
from \eqref{Wedge}, \eqref{a}, \eqref{W_b}, i.e.
$W^H(x,p) \simeq \frac{1}{2 \pi \hbar} {\cal W}_b(a)$
and obtain one of our main result 
\be
W_{\tilde \mu}^{\rm di}(x,p) \simeq \frac{-1}{2 \pi \hbar e(x,p)} 
\int d\mu' \nu_{\tilde \mu}(\mu') 
{\cal W}'_b(\frac{H(x,p)-\mu'}{e(x,p)})
\ee 
where at $T=0$, ${\cal W}'_{+\infty}(a)=- 2^{1/3} {\rm Ai}(2^{2/3} a)$.
If one neglects the quantum width of the edge, $\delta \mu' \sim e_N \sim \hbar^{2/3}$ for
fixed $\mu'$, it is equivalent to insert instead 
$W^H(x,p) \simeq \frac{1}{2 \pi \hbar} \theta(\mu-H(x,p))$ in \eqref{Wdiag}. This leads to
\be \label{Wdiagbulk}
W_{\tilde \mu}^{\rm di}(x,p) \simeq \frac{1}{2 \pi \hbar} \nu_{\tilde \mu}(H(x,p))
\ee
an approximation valid when $e_N$
is much smaller than the width $\delta \mu'$ over which the
effective occupation number 
$\nu_{\tilde \mu}(\mu')$ drops from $1$ (for $\mu' \ll \tilde \mu$)
to $0$ (for $\mu' \gg \tilde \mu$). This is the case in the
classical limit $\hbar \sim 1/N$ which we now consider
\cite{footnote7}. We note that $\frac{1}{N} \hat \nu_{\tilde \mu}(\epsilon)$ is the probability density of the energy $H$
of a single particle in the initial state. In the classical limit it is evaluated through
a phase space average over the initial Wigner function $W(x,p,0)$, which together with 
\eqref{Wdiagbulk} leads to 
\bea \label{nuclass} 
&& W_{\tilde \mu}^{\rm di}(x,p) \simeq W_{\tilde \mu}^{\rm di,sc}(x,p) =
\frac{\hat \nu^{\rm sc}_{\tilde \mu}(\epsilon)}{2 \pi \hbar \rho^{\rm sc}(\epsilon)}|_{\epsilon=H(x,p)} \\
&& \simeq \int \frac{dx_0 dp_0 \delta(H(x,p)-H(x_0,p_0))}{2 \pi \hbar \rho^{\rm sc}(H(x,p))}  W(x_0,p_0,0)
 \nonumber 
\eea
where $\rho^{sc}(\epsilon)$ is the semi-classical density of states. 
This is our result in the classical limit $\hbar \sim 1/N$. In the simplest
case considered in \cite{Kulkarni2018} (i.e. $T=0$ and a single orbit at each energy $\epsilon$) 
one has $\rho^{sc}(\epsilon)=\frac{T(\epsilon)}{2 \pi \hbar}$ where $T(\epsilon)$
is the period of the classical orbit at energy $\epsilon$. Inserting
$W(x_0,p_0,0) = \frac{1}{2 \pi \hbar} \theta(\mu- H_0(x_0,p_0))$, integrating over $p_0$ in
\eqref{nuclass}, and putting all together \cite{SM}, one recovers
the large time limit result of \cite{Kulkarni2018} (obtained by a quite
different method)
valid in the classical case. Our formula \eqref{Wdiag}
is thus a good candidate for the large time limit of the Wigner function
in the quantum regime. 

The GGE thermalization hypothesis \cite{EsslerReview,GGECalabrese,Murthy2018} states that for local observables $O$,
$\lim_{t \to +\infty} \langle O(t) \rangle = {\rm Tr} {\cal D}_{\rm GGE} O$, 
where the density matrix ${\cal D}_{\rm GGE}$ involves an extensive
number of conserved quantities. For non-interacting fermions, as discussed in \cite{EsslerReview,Kulkarni2018},
these are the occupation numbers $c^\dagger_\ell c_\ell$
of single particle energy levels $\epsilon_\ell$ of $H$. The natural candidate is thus
${\cal D}_{\rm GGE}  = \frac{1}{Z_{\rm GGE}}  e^{\sum_\ell f_\ell c^\dagger_\ell c_\ell}$
where the $f_\ell$ are determined from the initial condition. One has
$\langle c^\dagger_\ell c_\ell \rangle_{t=0} = \nu_\ell$, where the $\nu_\ell$ are exactly the
ones obtained in \eqref{nuell} \cite{SM}.
Hence one finds $\langle c^\dagger_\ell c_\ell \rangle_{\rm GGE} = \frac{1}{1 + e^{- f_\ell}}= \nu_\ell$.
Thus we find that at the level of the single particle observables (the kernel and
Wigner function) the result of the GGE is identical to the diagonal
approximation.

What about the multi-point time dependent correlations ?
Since for any initial state equal to an eigenstate of ${\cal H}_0$, labeled by a set
${\bf n}^0 = \{ n^0_k\}$ of occupation numbers, $n^0_k=0,1$, the time evolved state is determinantal, 
correlations in this state have the form
\be
\!\!\!\! R_{m,{\bf n}^0}(x_1,\cdots,x_m,t) = \det_{1 \leq i,j \leq m} \sum_{k=1}^{+\infty} n^0_k \psi^*_k(x_i,t) \psi_k(x_j,t) \nonumber
\ee
Expanding on the eigenstates of $H$ now involves a sum over two sets of $m$-fermion eigenstates
of ${\cal H}_m$.
The DA restricts that sum to identical eigenstates in both sets, 
and one obtains \cite{SM} (after a grand canonical average)
 \be \label{resultR} 
 R_m^{\rm di}(x_1,\cdots,x_m)  = \sum_{1 \leq \ell_1<\dots<\ell_p} |\det \phi_{\ell_i}(x_j)|^2 
\det_{p \times p} \nu_{\ell_i,\ell_j} 
\ee
where 
$\nu_{\ell,\ell'} = \langle \phi_{\ell} |(1 + e^{\beta(H_0-\mu)})^{-1}  | \phi_{\ell'}  \rangle
= \langle c_\ell^\dagger c_{\ell'} \rangle_{t=0}$. 
This DA for $p \geq 2$ differs from the prediction of the
simplest GGE ansatz discussed above. Indeed, the latter predicts
that the $m$-point correlation $R_m$ is equal to the $m \times m$ determinant
built from the kernel $K^{\rm GGE}= K^{\rm di}_{\tilde \mu}$ given in \eqref{nuell}.
Agreement $R_m^{\rm di} = R_m^{GGE}$ would hold only if
one could neglect the off-diagonal terms in the determinant
in \eqref{resultR}, i.e. replace $\det_{p \times p} \nu_{\ell_i,\ell_j}  \to \nu_{\ell_1} \dots \nu_{\ell_p}$,
as implied by the Cauchy-Binet identity. This occurs in some models, see a bosonic example in
\cite{GGECalabrese} which has translational invariance and the GGE was shown to hold.
Here, connecting $R_m^{\rm di}$ to the GGE (possibly in an inhomogeneous version
involving local Lagrange multipliers
$f_\ell(x)$) remains open.


In conclusion we studied the evolution of the Wigner function
for noninteracting fermions in $d=1$ after a quantum quench.
At times such that the integrity of the Fermi surf is retained, we found that 
it is described by the same universal function as 
at equilibrium, although with a space time dependent width that we calculated. 
We obtained an exact candidate formula for the large time limit of the Wigner function
for generic (e.g. non-harmonic) confining potentials, and showed
that it agrees with the GGE. We calculated the diagonal approximation
to the many point correlations.
It would be interesting to
explore $d>1$ and detect these predictions in cold atom experiments.

{\it Note added}: at completion of this work we became aware
of the recent work in \cite{DubailTrapDyn} which studies a related problem.

\acknowledgments
We thank P. Calabrese for remarks and references, A.~ Borodin, 
B.~Doyon, A. de Luca and M.~Fagotti for discussions,
and A. Krajenbrink for useful interactions at the initial stage of this work. 
We acknowledge support from ANR grant ANR-17-CE30-0027-01 RaMaTraF.


\begin{thebibliography}{99}
\bibitem{footnote3}
It is also the derivative of the eikonal equation for 
the semi-classical action (with $p=\partial_x S$),
see \cite{SM}.
\bibitem{footnote7}
At $T=0$ the relation between $N$ and $\mu$ is through
$2 \pi \hbar N=\int dx dp ~ \theta(\mu- \frac{p^2}{2} - V_0(x))$, leading to $\mu \sim (N \hbar)^{\frac{2 p}{p+2}}$ for a potential $V_0(x) \sim x^p$. 
More generally $\hbar \to 0$ with fixed $N \hbar$ corresponds to fixed $\mu$,
in which case $e_N \sim \hbar^{2/3}$.
\end{thebibliography}
\end{document}







\begin{center}

\begin{large}

{\bf Supplemental material for} \\

{\bf  "Nonequilibrium dynamics of noninteracting fermions in traps"}

\end{large}

\end{center}

\date{\today}

\author{David S. \surname{Dean}}
\affiliation{Univ. Bordeaux and CNRS, Laboratoire Ondes et Mati\`ere  d'Aquitaine
(LOMA), UMR 5798, F-33400 Talence, France}
\author{Pierre Le Doussal}
\affiliation{Laboratoire de Physique de l'Ecole Normale Sup\'erieure,
PSL University, CNRS, Sorbonne Universit\'es, 
24 rue Lhomond, 75231 Paris, France}
\author{Satya N. \surname{Majumdar}}
\affiliation{LPTMS, CNRS, Univ. Paris-Sud, Universit\'e Paris-Saclay, 91405 Orsay, France}
\author{Gr\'egory \surname{Schehr}}
\affiliation{LPTMS, CNRS, Univ. Paris-Sud,  Universit\'e Paris-Saclay, 91405 Orsay, France}

\begin{abstract} 

\bigskip

\bigskip
\tableofcontents
\end{abstract}

\maketitle

\section{Model and dynamical determinantal structure}

\subsection{Quench from zero temperature} 

We consider the real time quantum dynamics of $N$ noninteracting identical fermions evolving in $d=1$ dimensions  with the many body Hamiltonian 
${\cal H}= \sum_{i=1}^N H(x_i,p_i)$. The system is initially prepared at $t=0$ in the ground state of the Hamiltonian
${\cal H}_0= \sum_{i=1}^N H_0(x_i,p_i)$. We restrict our study to the following choice for the
single particle Hamiltonians
\be \label{HH0} 
H_0(x,p) = \frac{p^2}{2 m} + V_0(x) 
\quad , \quad H(x,p) = \frac{p^2}{2 m} + V(x)  \;.
\ee
For simplicity, we work here in units where the mass $m=1$. More generally, we will also consider
the evolution under a time dependent $H(t)$, i.e. \eqref{HH0} with $V(x) \to V(x,t)$.
We will work in term of  the eigenstates and eigenenergies of $H$ and $H_0$ which we denote by
\be
H_0 |\phi_k^0 \rangle = \epsilon_k^0 |\phi_k^0 \rangle  \quad , \quad \phi_k^0(x)= \langle x   |\phi_k^0 \rangle 
\quad , \quad
H |\phi_\ell \rangle = \epsilon_\ell |\phi_\ell \rangle \quad , \quad \phi_\ell(x)= \langle x   |\phi_\ell \rangle  \;.
\ee 
The potentials $V(x)$ and $V_0(x)$ 
are generally assumed to be confining. In this case the spectra of both $H_0$ and $H$ are discrete and the
eigenstates are non-degenerate. We label them using integers $k$ and $\ell$ in increasing order $\epsilon_1^0 < \epsilon_2^0 < \dots< \epsilon_k^0 < \ldots$ and 
$\epsilon_1 < \epsilon_2 < \dots< \epsilon_l < \ldots$. The initial state, i.e. the ground state of ${\cal H}_0$, take the form of a Slater determinant
\be
\Psi(x_1,\cdots, x_N;t=0) = \frac{1}{\sqrt{N!}} \det_{1 \leq i,j \leq N} \phi^0_{i}(x_j) \;.
\ee 
We denote by $\mu$ the Fermi energy, i.e. the energy $\epsilon^0_N$ of the highest occupied 
single particle state.
The Slater determinant property (sometimes called ``Gaussian state'' in the quantum quench literature) is preserved under time evolution. We define by $\psi_k(x,t)$  the solutions of $i \hbar \partial_t \psi= H \psi$ with initial condition $\psi_k(x,t=0)=\phi^0_k(x)$. The time evolved wave function is thus
\be
\Psi(x_1,\cdots, x_N;t) = \frac{1}{\sqrt{N!}} \det_{1 \leq i,j \leq N} \psi_{i}(x_j,t)   \label{prop2} \;.
\ee 
We are interested in the quantum probability density $|\Psi|^2$ which can be written in terms of the time dependent kernel 
$K_\mu(x,x',t)$ via
\be \label{def_K_supp}
|\Psi(x_1,\cdots, x_N;t)|^2 = \frac{1}{N!} \det_{1 \leq i,j \leq N} K_\mu(x_i,x_j,t) \quad , \quad K_\mu(x,x',t) = \sum_{k=1}^N  \psi_k^*(x,t) \psi_k(x',t) \;.
\ee
Note that the $\psi_k(x,t)$ form an orthonormal basis for all $t$ since the
Schr\"odinger unitary evolution conserves all scalar products. Consequently, the kernel is  reproducing, i.e.
$\int dx' K_\mu(x,x',t) K_\mu(x',x'',t) = K_\mu(x,x'',t)$ and satisfies ${\rm Tr}K_\mu = \int dx K_\mu(x,x)=N$. 
Hence $|\Psi(x_1,\cdots x_N;t)|^2$ is the joint probability distribution function (JPDF)
of a determinantal point process \cite{johansson,borodin_determinantal}. As a consequence
one can show that the $m$-point correlation functions can be written as determinants
\be
R_m(x_1,\dots,x_m,t)= \frac{N!}{(N-m)!} \int dx_{m+1} \dots dx_N |\Psi(x_1,\cdots x_N;t)|^2
= \det_{1 \leq i,j \leq m} K_\mu(x_i,x_j,t)
\ee
The $m$-point time-dependent density-density correlation functions can be obtained easily from these
$R_m$ functions \cite{fermions_review}. For instance $R_1(x,t)=N \rho_N(x,t)$ where $\rho_N(x,t)$ is the (time-dependent)
average density.\\

Taking time derivative of $K_\mu(x,x',t)$ defined in Eq. (\ref{def_K_supp}) and using the time-dependent Schr\"odinger equation for $\psi_k(x,t)$, it is easy to obtain the equation (10) of the text
\begin{equation}
\partial_t K_\mu= -\frac{i\hbar}{2}\,(\partial_x^2-\partial_{x'}^2)\,K_\mu +\frac{i}{\hbar}\,
\left(V(x)-V(x')\right)\, K_\mu\, .
\label{kmu1.A1}
\end{equation}
In addition, one can also express formally the time-dependent kernel in terms of the quantum Euclidean propagator, defined through
\be \label{defG} 
G(x,x',\tau) = \langle x | e^{- \tau H/\hbar }  |x' \rangle \quad , \quad \hbar \partial_\tau G(x,x',\tau) = - H_x G(x,x',\tau) \quad , \quad G(x,x',0)= \delta(x-x') \;,
\ee
where $H_x=H(x,\frac{\hbar}{i} \partial_x)$.
Using $\psi_k(x,t) = \int dx' \langle x | e^{- i H t/\hbar} |x' \rangle \phi^0_k(x')$,
we can write
\bea \label{GKG} 
&& K_\mu(x,x',t) = \int dy dy' G(x,y, - i t) K_\mu(y,y',0) G(x',y', i t)  \quad , \quad K_\mu(x,x',0) = \int_C \frac{d\tau}{2 i \pi \tau} e^{\mu \tau/\hbar} G(x,x',\tau)
\eea
where, {in the second equality we have expressed the equilibrium kernel using the Euclidean propagator as
shown in~\cite{fermions_review}}. Taking time derivatives we see that the kernel obeys the same evolution equation
as in (\ref{kmu1.A1}).


The $N$ body zero temperature Wigner function, defined in Eq. (8) in the main text, can be written as a 
Fourier transform
of the Kernel $K_\mu$ as in Eq. (9) of the main text, this was shown in the Appendix A of Ref. \cite{UsWigner}.
It is convenient to re-express Eq. (9) of the main text, upon replacing $x$ by $X$, as
\begin{equation}
W(X,p,t)= \frac{1}{2\pi \hbar}\, \int_{-\infty}^{\infty} 
dY\, e^{i\,p\, Y/\hbar}\, K_\mu(X+Y/2, X-Y/2,t)\, ,
\label{wigner1.A1}
\end{equation}
where $K_\mu(x,x',t)$ evolves via Eq. (\ref{kmu1.A1}). 
We make the change of variables, $x= X+Y/2$ and $x'= X-Y/2$, and consequently
$\partial_x^2-\partial_{x'}^2= 2\, \partial_X\partial_Y$. Taking the time-derivative of Eq. (\ref{wigner1.A1}), and
using Eq. (\ref{kmu1.A1}), one obtains
\begin{equation}
\partial_t W(X,p,t)= \frac{1}{2\pi\hbar} 
\int_{-\infty}^{\infty} dY\, e^{i\,p\,Y/\hbar}\,
\left[
-i\hbar\, \partial_X\,\partial_Y\, K_\mu+ 
\frac{i}{\hbar}\, \left(V(X+Y/2)-V(X-Y/2)\right)\, K_\mu \right]\, ,
\label{wigner2.A1}
\end{equation}
where $K_\mu \equiv K_\mu(X+Y/2, X-Y/2,t)$. The first term on the right hand side (rhs) of Eq. (\ref{wigner2.A1})
can be evaluated using integration by parts. In the second term (involving $V$'s), we first formally
expand $V(X+Y/2)= \sum_{n=0}^{\infty} V^{(n)}(X) (Y/2)^n/n!$ with $V^{(n)}(x)= d^n V(x)/dx^n$ being 
the $n$-th derivative. Each term can then
be re-expressed as derivatives with respect to $p$ and the series can then be resummed. After straightforward algebra,
this then leads to the evolution equation for $W(X,p,t)$
\bea
\partial_t W(X,p,t) = - p \partial_X W(X,p,t) +
\frac{i}{\hbar}  \left(
V(X - \frac{i \hbar}{2} \partial_p) - V(X + \frac{i \hbar}{2} \partial_p) \right)\, W(X,p,t) \, .
\label{wigner3.A1}
\eea
This is precisely Eq. (11) of the main text (with $X$ replacing $x$).
This evolution equation for the Wigner function,
which we call the $W$-equation, is valid for arbitrary $N$.

\subsection{Quench from finite temperature}
\label{sec:temperature} 

To consider a finite temperature initial state, i.e. described by an initial density matrix, we need to consider the
evolution of an arbitrary eigenstate of ${\cal H}_0$. These states are denoted by $| {\bf n}^0 \rangle$ and
are indexed by a set of occupation numbers,
introduced in the text,
${\bf n}_0= \{ n_k^0 \}_{k \geq 1}$ with $n^0_k \in \{0,1\}$, with $\sum_{k=1}^\infty n_k^0=N$. In fact $|{\bf n}_0 \rangle$ is an eigenstate
of ${\cal H}_0$ with eigenvalue $\sum_{k \geq 1} n_k^{0} \epsilon_k^0$, i.e.
\bea\label{eigen_n0}
{\cal H}_0 |{\bf n}_0 \rangle = \left( \sum_{k \geq 1} n_k^{0} \epsilon_k^0\right)  \, |{\bf n}_0 \rangle \;.
\eea

Under time
evolution they become $| {\bf n}_0 , t \rangle = e^{- i {\cal H} t/\hbar} | {\bf n}_0 \rangle$ and read
\be
\langle x_1,\cdots x_N|{\bf n}_0 , t\rangle = \Psi_{{\bf n}_0}(x_1,\cdots x_N;t) = \frac{1}{\sqrt{N!}} \det_{1 \leq i,j \leq N} \psi_{k_i}(x_j,t)   \label{prop3}
\ee 
with $ |{\bf n}_0 , t=0\rangle=|{\bf n}_0 \rangle$. Here
$1 \leq k_1<k_2<\dots k_N$ denote the occupied states, i.e. $n^0_k = \sum_{i=1}^N \delta_{k,k_i}$.
Hence for each of these states the quantum probability corresponds to a distinct determinantal process 
based on a kernel $K(x,x',t;{\bf n}^0)$
\be
|\Psi_{{\bf n}_0}(x_1,\cdots x_N;t)|^2 = \frac{1}{N!} \det_{1 \leq i,j \leq N} K(x_i,x_j,t;{\bf n}_0) 
\quad , \quad K(x,x',t;{\bf n}_0) = \sum_{k=1}^{+\infty} n^0_k \psi_k^*(x,t) \psi_k(x',t) 
\ee
Since each kernel $K(x,x',t;{\bf n}_0)$ is reproducing for any set ${\bf n}_0$ (since $n_k^2=n_k$)
with trace equal to $N$, it implies that the correlations in that state are determinantal
\be
R_{m,{\bf n}_0}(x_1,\dots,x_m,t) = \det_{1 \leq i,j \leq m} K(x_i,x_j,t;{\bf n}_0), 
\ee
and which is zero if $m> \sum_{k \geq 1} n_k^0$.

As discussed in \cite{fermions_review} for the equilibrium problem,
i.e. for $t=0$ in our case, to study finite temperature $T=1/\beta>0$, it is more convenient
to work in the grand canonical (GC) ensemble, at chemical potential $\tilde \mu$ (equal to the
Fermi energy $\mu$ at $T=0$), where the total number of fermions can fluctuate.
This is because in the canonical ensemble (fixed $N$) the correlations are {\it not} determinantal.
By contrast these correlations (defined below) are determinantal in the GC ensemble
\cite{Joh07}. 
In the large $N$ limit we expect that for local observables the two ensembles are equivalent
(see also the discussion in \cite{fermions_review,Texier2018}).\\

The grand canonical measure amounts to attributing a weight to each eigenstate of ${\cal H}_0$,
i.e. to each $|{\bf n}_0 \rangle$. This weight has a simple product form and the 
corresponding equilibrium partition sum and initial many body density matrix ${\cal D}_{\beta,\tilde \mu}(t=0)$
are defined as
\be
Z_{\beta,\tilde \mu} = \sum_N e^{\beta \tilde \mu N} Z_{\beta,N}
\quad , \quad
{\cal D}_{\beta,\tilde \mu}(t=0)= \frac{1}{Z_{\beta,\tilde \mu}} \sum_{{\bf n}_0} 
e^{- \beta( \sum_k \epsilon^0_k - \tilde \mu) n^0_k} |{\bf n}_0\rangle \langle {\bf n}_0|
\ee 
where $Z_{\beta,N}$ is the canonical partition sum
$Z_{\beta,N} = \sum_{{\bf n}_0} \delta_{\sum_k n^0_k,n}
e^{- \beta \sum_k \epsilon^0_k n^0_k}$. Hence the grand canonical measure amounts to attributing a weight $p_N = e^{\beta \tilde \mu N} Z_{\beta,N}/Z_{\beta,\tilde \mu}$ for having $N$ particles in the system at canonical equilibrium.
Under time evolution, the GC density matrix becomes ${\cal D}_{\beta,\tilde \mu}(t)$ obtained
from ${\cal D}_{\beta,\tilde \mu}(t=0)$ by replacing $ |{\bf n}_0\rangle \langle {\bf n}_0| \to 
 |{\bf n}_0,t\rangle \langle {\bf n}_0,t|$. \\

The natural definition of the correlations in the GC ensemble is obtained by averaging the correlation in
each state $|{\bf n}_0,t\rangle $ over the GC measure (denoted here and what follows by an overbar)
\be
 R^{}_m(x_1,\dots,x_m,t) = \overline{R_{m,{\bf n}_0}(x_1,\dots,x_m,t) }  := \frac{1}{Z_{\beta,\tilde \mu}} \sum_{{\bf n}_0} 
e^{- \beta( \sum_k \epsilon^0_k - \tilde \mu) n^0_k} R_{m,{\bf n}_0}(x_1,\dots,x_m,t) 
\ee
Using the property that the $n^0_k$ are independent Bernoulli random variables in the GC measure, we obtain 
\bea
R^{}_m(x_1,\dots,x_m,t) = \overline{\det_{1\leq i,j\leq m} \sum_k n^0_k \phi^{0*}_k(x_i) \phi^0_k(x_j)}
= \det_{1\leq i,j \leq m} K_{\tilde \mu}(x_i,x_j,t) 
\eea
where the time-dependent GC kernel, and its equilibrium initial conditions are given by
\bea
K_{\tilde \mu}(x,x',t)  = \sum_{k=1}^{+\infty} \bar n^0_k  \psi_k^*(x,t) \psi_k(x',t)  \quad , \quad 
K_{\tilde \mu}(x,x',t=0)=
\sum_k \bar n^0_k \phi^{0*}_k(x_i) \phi^0_k(x_j)
\eea
Here the $\bar n^0_k \equiv \overline{n^0_k}$ are the usual GC averaged occupation number (i.e $n^0_k$ averaged over the
Bernoulli distribution) 
\bea\label{Nbar}
\bar n^0_k = \frac{1}{1+e^{\beta (\epsilon^0_k-\tilde \mu)} } \quad , \quad \sum_k \bar n^0_k = \bar N,
\eea 
where the second relation gives $\bar N$, the mean number of fermions, as a function
of the chemical potential $\tilde \mu$.

It is important to note that $K_{\tilde \mu}(x,x',t)$ satisfies exactly the same evolution
equation (10) in the text as $K_\mu(x,x',t)$, the zero temperature kernel, the only difference being
the initial condition. Similarly, one defines the finite temperature Wigner function in the GC 
ensemble, see Section 3 in \cite{UsWigner}
\bea \label{defWGC} 
W_{\tilde \mu}(x,p,t) = \frac{1}{Z_{\beta,\tilde \mu}} \sum_N  \frac{N}{2 \pi \hbar} e^{\beta \tilde \mu N}
\int_{-\infty}^{+\infty} dy dx_2 \dots dx_N e^{i p y/\hbar} 
\langle x+\frac{y}{2}, x_2, \dots, x_N| {\cal D}^c_{\beta,N}(t) |  x-\frac{y}{2}, x_2, \dots, x_N \rangle
\eea
where the canonical density matrix and partition function are defined as
\bea
{\cal D}^c_{\beta,N}(t)= \frac{1}{Z^c_{\beta,N}} \sum_{{\bf n}_0} \delta_{N,\sum_k n^0_k}
e^{- \beta \sum_k \epsilon^0_k n^0_k} |{\bf n}_0,t\rangle \langle {\bf n}_0,t|
\quad , \quad Z^c_{\beta,N} = \sum_{{\bf n}_0} \delta_{N,\sum_k n^0_k}
e^{- \beta \sum_k \epsilon^0_k n^0_k} 
\eea
As shown in \cite{UsWigner} (see Appendix A) it is related to 
\bea \label{Wrel} 
W_{\tilde \mu}(x,p,t) = \frac{1}{2 \pi \hbar} \int_{-\infty}^{+\infty} dy ~ e^{ \frac{i p y}{\hbar}} 
K_{\tilde \mu}(x+ \frac{y}{2},x- \frac{y}{2},t)
\eea 
and thus satisfies exactly the same evolution equation (11) in the text, although
with a different initial condition as $T$ and $\tilde \mu$ are varied.

\section{Dynamics for the harmonic oscillator}

In this section we study the dynamics of a single particle in a time dependent harmonic potential, with Hamiltonian  
\be \label{HOt} 
H(t) = \frac{p^2}{2} + \frac{1}{2} \omega(t)^2 x^2
\ee
for various initial conditions. 

\subsection{The rescaling method}

The rescaling method \cite{Popov1969} compares
the evolution with two different single particle Hamiltonians: (i) the time dependent oscillator $H(t)$ defined in \eqref{HOt} 
and (ii) the fixed frequency oscillator of Hamiltonian
\be
H_1 = \frac{p^2}{2} + \frac{1}{2}  \omega_1^2 x^2
\ee
Let us denote by $\varphi(x,t)$ the solution of $i \hbar \partial_t \varphi=H(t) \varphi$ with an arbitrary initial condition
$\varphi(x,t=0)= \varphi_0(x)$, and 
$\varphi_1(x,t)$ the solution of $i \hbar \partial_t \varphi_1=H_1 \varphi_1$ with the same initial condition
$\varphi_1(x,t=0)= \varphi_0(x)$. Then one has the relation valid at all times $t \geq 0$
\be \label{rel1} 
\varphi(x,t) = e^{i \frac{L'(t)}{2 L(t) \hbar} x^2} \frac{1}{\sqrt{L(t)}} \varphi_1( \frac{x}{L(t)}  ,\tau(t)) 
\ee
where $\tau'(t) = 1/L(t)^2$ and $L(t)$ satisfies Ermakov's equation \cite{Ermakov} 
\be \label{erm} 
\partial^2_t L(t) + \omega(t)^2 L(t) = \frac{ \omega_1^2}{L(t)^3}  
\ee
with $L(0)=1$, $L'(0)=0$. Let us recall that the general solution of the Ermakov equation has the form
\cite{PainleveErmakov}
\bea
L(t) = \pm \sqrt{ A x_1(t)^2 + 2 B x_1(t) x_2(t) + C x_2(t)^2} \quad , \quad A C - B^2 = \omega_1^2 
\eea 
where $x_1(t)$ and $x_2(t)$ are two independent solutions of the time dependent harmonic oscillator
$\partial^2_t x_i(t) + \omega(t)^2 x_i(t)=0$. Let us recall that if $x_1(t)$ is a solution of the time
dependent harmonic oscillator, then $x_2(t)=x_1(t) \int^t dt'/x_1(t')^2$ is also a solution of the time dependent 
harmonic oscillator.

There are two interesting choices for $\omega_1$ to apply this rescaling method. The
first one is to choose $\omega_1=\omega_0$ which directly leads to the solution given in the text
for an initial state at equilibrium described by the Hamiltonian $H_0=\frac{p^2}{2} + \frac{1}{2}  \omega_0^2 x^2$
as we now explain. The second choice, $\omega_1=0$, corresponding to free evolution, allows the treatment of more general initial conditions
and is discussed later below.

\subsubsection{Harmonic oscillator initial condition} 

Let us choose first $\omega_1=\omega_0$ and focus on the evolution of the $k$-th eigenstate $\phi_k(x,t)$, starting from the initial condition $\phi_k(x,t=0) = \phi_k^0(x)$. In this special case, $\varphi_1(x,t)$ satisfies the Schr\"odinger equation $i \hbar \partial_t \varphi_1 = H_0 \varphi_1$ whose solution is
\bea\label{sol_H1H0}
\varphi_1(x,t) = \phi_k^0(x) e^{-i \frac{\epsilon^0_k\,t}{\hbar}}
\eea
 is actually independent of time as this is an eigenstate of $H_1=H_0$. Plugging this solution (\ref{sol_H1H0}) in the rhs of Eq. (\ref{rel1}) and replacing $t$ by $\tau(t)$ leads to Eq. (15) of the main text.
The rescaling method extends also to many body Hamiltonian,
as discussed in \cite{Gritsev2010}. To see it here for free fermion, we simply insert Eq. (15) of the text
in the expression of the kernel valid for arbitrary $N$
\be \label{Kresc2} 
K_{\tilde \mu}(x,x',t) = \sum_k \bar n^0_k \psi_k^*(x,t) \psi_k^*(x',t) 
=  \frac{e^{i \frac{L'(t)}{2 L(t) \hbar} ({x'}^2-x^2)} }{L(t)}
 \sum_k \bar n^0_k \phi^0_k(\frac{x}{L(t)} ) \phi^0_k(\frac{x}{L(t)} ) 
 =  \frac{e^{i \frac{L'(t)}{2 L(t) \hbar} ({x'}^2 - x^2)} }{L(t)} K_{\tilde \mu}(\frac{x}{L(t)}, \frac{x'}{L(t)},0) 
\ee 
which holds for any initial state at equilibrium at any arbitrary temperature $T$, with at $T=0$, $\tilde \mu=\mu$, and
$\bar n^0_k = \theta(\mu - \epsilon^0_k)$. This leads to Eq. (17) in the text. A similar result was obtained
in \cite{QuinnHaque2014} together with the same scaling formula for hard core bosons in the Tonks Girardeau limit. There it applies to the one body density matrix \cite{QuinnHaque2014,Minguzzi2005}, 
which identifies with our kernel only at coinciding points. The density is the same in both models.\\ 

It is important to note that the evolution equations (10) and (11) in the text, for the kernel and the Wigner
function respectively remain valid for a time dependent potential with $V(x) \to V(x,t)$. We can thus 
check explicitly that the kernel given by Eq. (17) in the text obeys the evolution equation (10) in the text
with $V(x) \to V(x,t)$. To perform this check,
one must use the fact that the equilibrium kernel $K_{\tilde \mu}(x,x',t=0)$ 
satisfies the same equation (10) in the text
with the term $\partial_t K_{\tilde \mu}$ set to zero, and $V(x) \to V_0(x)= \frac{1}{2} \omega_0^2 x^2$.\\

Since here the initial condition is a harmonic oscillator, we know that the initial density 
$\rho_N(x,t=0)$ for large $N$ is given at $T=0$ by the semi-circle in Eq. (1) of the text,
with $x_e= \sqrt{2 \mu}/\omega_0= \sqrt{\frac{\hbar}{\omega_0}} \sqrt{2 N}$. The time
evolution of the density is thus quite simple and given at all times by a semi-circle
as in Eq. (1) of the text with a moving edge $x_e(t) = L(t) x_e$. In the bulk the
time-dependent kernel at the scale of the inter-particle distance and at $T=0$ retains the form of the 
the sine kernel which holds at equilibrium 
\be \label{Ksine} 
K_{\tilde \mu}(x,x',t) \simeq \frac{1}{\xi_{x,t}} {\cal K}^{\rm bulk}(\frac{x-x'}{\xi_{x,t}}) \quad , \quad 
{\cal K}^{\rm bulk}(z)=\frac{\sin 2 z}{\pi z} 
\ee
with the width $\xi_{x,t}=2 \hbar L(t)^2/(\omega_0 \sqrt{x_e(t)^2-x^2}) \sim N^{-1/2}$. The same formula
holds for a finite temperature quench, in the regime $T \sim N$, with ${\cal K}^{\rm bulk}(z) \to {\cal K}_y^{\rm bulk}(z)$ given by Eq. (118) in 
\cite{fermions_review} with $y=\mu/T$. Near the moving edge for $x \simeq x_e(t)$ the kernel takes the scaling form
given by the Airy kernel 
\be \label{scedge} 
 K_{\tilde \mu}(x,t;x',t') \simeq  \frac{1}{w_N(t)} {\cal K}^{\rm edge}(\frac{x-x_e(t)}{w_N(t)},
\frac{x'-x_e(t)}{w_N(t)}) \quad , \quad 
{\cal K}^{\rm edge}(z,z')= \int_0^{+\infty} du {\rm Ai}(z+u) {\rm Ai}(z'+u) 
\ee
with a time dependent width 
$w_N(t) = w_N L(t)$ and $w_N=\sqrt{\frac{\hbar}{2 \omega_0}} N^{-1/6} \sim \mu^{-1/6}$.
Taken at coinciding points, it yields the formula (19) in the text for the form of the density around the edge.
The same scaling form \eqref{scedge} holds at finite temperature in the regime 
$T \sim N^{1/3}$ with
${\cal K}^{\rm edge} \to {\cal K}^{\rm edge}_b$ given in Eq. (132) in \cite{fermions_review} 
with $b= \frac{\hbar \omega_0}{T} N^{1/3}$. \\

Once the kernel is known, one can compute the Wigner function from Eq. (9) in the
text. For any $N$ one has
\bea
&& W(x,p,t) = \frac{1}{2 \pi \hbar}  \int dy e^{i p y/\hbar} e^{- i \frac{L'(t)}{L(t) \hbar}  x y} 
\frac{1}{L(t)} K(\frac{x}{L(t)} + \frac{y}{2 L(t)}, \frac{x}{L(t)} - \frac{y}{2 L(t)},0) \nonumber \\
&& = \frac{1}{2 \pi \hbar}  \int dY e^{i (p L(t) - i  L'(t) x)  Y/\hbar} 
 K(\frac{x}{L(t)} + \frac{Y}{2}, \frac{x}{L(t)} - \frac{Y}{2},0) 
 = W(\frac{x}{L(t)} , p L(t) - L'(t) x, 0) \label{W1} 
\eea 
where $Y=y/L(t)$. This leads to the formula (18) in the text. The initial condition $W(x,p,0)$
is known exactly as a sum over Laguerre polynomials $L_k$ \cite{UsWigner} 
\be \label{WN}
W(x,p,0)=\frac{1}{\pi} \sum_k \bar n^0_k 
 (-1)^k L_k(2 r^2) e^{-r^2} \quad , \quad r^2 = \frac{\omega_0^2 x^2 + p^2}{\hbar \omega_0} \quad , \quad 
 \bar n^0_k=\frac{1}{1+ e^{\beta (\hbar \omega (k+\frac{1}{2})-\tilde \mu)}}.
\ee 
At $T=0$ the sum is from $k=1$ to $k=N$ with $\bar n^0_k=1$. From this formula
one shows that at large $N$ the Wigner function takes the form given in Eqs. (3) and (4) in the text
at $T=0$ (bulk and edge) and in Eq. (33) and (34) at finite $T$ (bulk and edge).

If one writes the last term in \eqref{W1} using the formula \eqref{WN},
the variable $r^2$ becomes $
r^2 = p^2 B(t)^2 
+ A(t) (x-x_1)^2$, 
with $A=L'(t)^2+\frac{\omega
   _0^2}{L(t)^2}$ and $x_1=\frac{p L(t)^3 L'(t)}{L(t)^2 L'(t)^2+\omega
   _0^2}$. Integrating over $x$ one thus finds that the density in momentum space, $\bar \rho_N(p,t)$, is also given by a change of scale of its initial form, although the scale factor $B(t)$ is more involved
\be
\bar \rho_N(p,t) = \int dx W(x,p,t) = B(t) \bar \rho_N(B(t) p,t=0) \quad , \quad B(t)= \frac{L(t)}{\sqrt{1 + \frac{L(t)^2 L'(t)^2}{\omega_0^2}} }
\ee
where $\bar \rho_N(p,t=0)$ is a semi-circle {QuinnHaque2014}. Hence the momentum density retains a a semi-circle form.
\\

The simplest case is when $H$ is a time independent oscillator, $\omega(t)=\omega$. The solution of the
Ermakov equation is then $L(t)=x_e(t)/x_e$ where $x_e(t)$ is given formula (14) in the text
and $x_e=\sqrt{2 \mu}/\omega_0$. One can check 
in this case that the formula \eqref{W1}, i.e. (18) in the text, for the Wigner function, coincides with Eq. (13) 
in the text obtained by transporting the Wigner function along the classical trajectories. 
With this is mind one should note, that for the harmonic oscillator $H_0$, the Wigner function 
from \eqref{WN} depends only on the combination $r^2$ (total energy),
and one can verifiy explicitly that 
$(\frac{x}{L(t)})^2 \omega_0^2 +  (p L(t) - L'(t) x)^2$
is indeed exactly equal to 
$(p \cos( \omega t) 
+ \omega X \sin(\omega t) )^2 + \omega_0^2 (x \cos(\omega t) - \frac{p}{\omega} \sin(\omega t))^2$
using $L(t)=x_e(t)/x_e$ given in Eq. (14) of the text.\\

Note that the results of this section are valid for any choice of the $\bar n_k^0$, i.e.
they extend to the finite temperature quench. They are also valid if the initial state
is any $N$ body eigenstate $|{\bf n_0}\rangle$ of the HO.\\

{Finally we note that one can also treat, in a similar way, the case of an applied constant linear potential $E(t)x$ , i.e. the
evolution Hamiltonian
\be \label{HOt2} 
H(t) = \frac{p^2}{2} + \frac{1}{2} \omega(t)^2 x^2 + E(t)x 
\ee}
The equation \eqref{rel1} is replaced by
\be \label{rel12} 
\varphi(x,t) = e^{i \frac{L'(t)}{2 L(t) \hbar} x^2 + \frac{i}{\hbar} x (a'(t) - \frac{a(t) L'(t)}{L(t)})
+ \frac{i}{\hbar} F(t)} \frac{1}{\sqrt{L(t)}} \varphi_1( \frac{x- a(t)}{L(t)}  ,\tau(t)) 
\ee
where $L(t)$ and $\tau(t)$ are given by the same equations as for $E(t)=0$.
There is a shift of the center   $a(t)$, which satisfies the classical
equation of motion
\be \label{eqa} 
a''(t) + \omega(t)^2 a(t) + E(t) = 0 
\ee 
with $a(t)=0$ and $a'(t)=0$. {The additional phase factor $F(t)$ is determined
from 
$F'(t)=\frac{a(t) a'(t) L'(t)}{L(t)}-\frac{1}{2} a'(t)^2-\frac{a(t)^2
   L'(t)^2}{2 L(t)^2}+\frac{\omega _1^2 a(t)^2}{2 L(t)^4}$
   with $F(0)=0$. 

The above results \eqref{Kresc2}, \eqref{Ksine} and \eqref{scedge}
immediately generalize to this case upon the shift $x \to x - a(t)$ 
in all the spatial arguments and a modified phase factor. One can
also generalize the result Eq. (18) of the text for the Wigner function. The solution
now reads
\be \label{WE1} 
W(x,p,t) = W_0\left( \frac{x - a(t)}{L(t)} , (p-a'(t)) L(t) - (x-a(t)) L'(t) \right) 
\ee 
This is obtained either from the kernel, or alternatively (and more simply) by 
substituting \eqref{WE1} into the (time dependent version of the) 
Wigner equation (11) in the text,
which reads in this case
\be \label{WEE} 
\partial_t W = - p \partial_x W + (\omega(t)^2 x + E(t)) \partial_p W. 
\ee
Using that the initial condition $W_0$ obeys the Wigner equation
for the $\omega_0$ harmonic oscillator, i.e. $0=- p \partial_x W_0 + \omega_0^2 x \partial_p W$,
as well as Ermakov's equation for $L(t)$, i.e. Eq. (16) in the text, and Eq. \eqref{eqa} for $a(t)$,
one check can directly that indeed \eqref{WE1} is the correct solution.}

\subsubsection{General initial condition} 

A second way to apply the rescaling method is to choose $\omega_1=0$.
In this case the evolution under $H_1$ is simply a free evolution and is straightfoward  to solve. 
One obtains now 
\bea \label{gritsev2} 
K_{\tilde \mu}(x,x',t) = e^{ i \frac{L'(t)}{2 L(t) \hbar} ((x')^2-x^2)} \frac{1}{L(t)} K^{\rm free}_{\tilde \mu}(\frac{x}{L(t)}, \frac{x'}{L(t)},\tau(t)) 
\eea 
where $K^{\rm free}_{\tilde \mu}(z,z',\tau)$ is the time-dependent kernel associated to the free evolution
(under $H_1=\frac{p^2}{2}$) with initial condition $K^{\rm free}_{\tilde \mu}(z,z',0) =
K_{\tilde \mu}(z,z',0)$ which here can be chosen arbitrarily. Note that the equation 
which determines $L(t)$ is now the linear oscillator equation
\be \label{E0} 
\partial^2_t L(t) + \omega(t)^2 L(t)=0 \quad , \quad \tau'(t)=1/L(t)^2 
\ee
To obtain $K_{\tilde \mu}^{\rm free}$ it is easier to consider its associated Wigner function $W^{\rm free}$
(by Fourier transform according to equation (9) in the text). The latter evolves
according to Eq. (12) with $V(x)=0$, i.e. $\partial_t W^{\rm free} = - p \partial_x W^{\rm free}$ 
which leads to
\be
W^{\rm free}(x,p,\tau) = W(x - p \tau, p, 0)   
\ee
Using the inverse Fourier relation to equation (9) in the text, we obtain
\be
K_{\tilde \mu}(x,x',t)= \int dp ~ e^{- i p (x-x')/\hbar} ~ W(\frac{x+x'}{2},p,t) 
\ee
Applying this equation to obtain $K^{\rm free}$ from $W^{\rm free}$ and inserting into
\eqref{gritsev2}, we obtain the kernel at time $t$ in terms of the initial Wigner function as
\bea \label{Karbitrary1}
K_{\tilde \mu}(x,x',t) = e^{ i \frac{L'(t)}{2 L(t) \hbar} ((x')^2-x^2)} \frac{1}{L(t)}
\int_{-\infty}^{+\infty} dp e^{- i p \frac{x-x'}{\hbar L(t)} } ~
W( \frac{x+x'}{2 L(t)} - p \tau(t) , p,0), 
\eea 
which is valid for an arbitrary initial condition.

It is interesting to note that the formula for the corresponding Wigner function is
simpler. Indeed, Fourier transforming $K_{\tilde \mu}(x,x',t)$ according to Eq. (9)
in the text, and integrating over $y$ and $p$,
we obtain the solution for the evolution under the $\omega(t)$-oscillator starting from an arbitrary
initial condition as
\be \label{formula2} 
W(x,p,t) = W\left(x ( \frac{1}{L(t)} + \tau(t) L'(t) ) - \tau(t) L(t) p, L(t) p - L'(t) x,0 \right), 
\ee
where we recall that $L(t)$ and $\tau(t)$ are solutions of \eqref{E0}
with $\tau(0)=0$, $L'(0)=0$, $L(0)=1$.\\

Consider as an example the case $\omega(t)=\omega$. One now obtains $L(t) = \cos(\omega t)$ and
$\tau(t)=\frac{1}{\omega} \tan(\omega t)$ recovering the equation (13) in the text. 
Note that in this method $L(t)$ does not remain positive
hence there are cancellations which eventually lead to the final physical result. 

We can now check that \eqref{formula2} is compatible with the transport of the Wigner
function along the classical trajectories 
\be
W(x,p,t)= W(x_0(x,p,t), p_0(x,p,t),0) \quad \Leftrightarrow \quad W(x(x_0,p_0,t),p(x_0,p_0,t),t)= W(x_0, p_0,0)
\ee
where $x(x_0,p_0,t)$, $p(x_0,p_0,t)$ denote the classical trajectories in the $\omega(t)$-oscillator starting from
$(x_0,p_0)$ at $t=0$. Similarly, $x_0(x,p,t)$ and $p_0(x,p,t)$ denote the reciprocal function (initial condition
as a function of the final condition). To this aim we solve  Hamilton's equations
\bea
\dot x = p  \quad  \quad \dot p = - \omega(t)^2 x 
\eea 
by writing $x(t) = A(t) x_0 + B(t) p_0$ and $p(t) = C(t) x_0 + D(t) p_0$. We obtain the conditions
equations 
\bea \label{eq2} 
\dot A = C \quad , \quad \dot B = D \quad , \quad \dot C= - \omega^2 A \quad , \quad \dot D = - \omega^2 B
\eea 
with initial conditions $A(0)=D(0)=1$ and $B(0)=C(0)=0$. One can check that for all
times $AD-BC=1$. The inversion is thus
\bea
x_0 = x_0(x,p,t) = D(t) x - B(t) p \quad , \quad 
p_0 = p_0(x,p,t) = - C(t) x + A(t) p
\eea 
It is now easy to check that $W(x_0(x,p,t), p_0(x,p,t))$ indeed satisfies the Wigner equation
(12) of the text with $V'(x) \to V'(x,t)=\omega(t)^2 x$. We now identify
\bea
D(t) = \frac{1}{L(t)} + \tau(t) L'(t) \quad , \quad B(t) = \tau(t) L(t) \quad , \quad A(t) = L(t) \quad , \quad C(t)=L'(t)
\eea 
and it is easy to check that these functions satisfy the equations in \eqref{eq2}.\\

{In the presence of an additional external field $E(t)$ the solution of the Wigner equation 
with an arbitrary initial condition is simply given by \eqref{formula2} where 
$x \to x-a(t)$ and $p \to p - a'(t)$ in the r.h.s. of the equation. The functions
$L(t)$ and $\tau(t)$ are again solutions of \eqref{E0}
with $\tau(0)=0$, $L'(0)=0$, $L(0)=1$, and $a(t)$ is solution of \eqref{eqa} 
with $a(t)=0$ and $a'(t)=0$. This is easily checked,
for an arbitrary initial condition $W(x,p,0)$, by inserting that solution
into the Wigner equation \eqref{WEE}. The reason why the generalization to a time dependent external field is so simple for the
Wigner function 
is because for the oscillator the (time dependent) WE identifies with the Liouville equation. One can check 
that if $\tilde W$ is solution of the LE with 
$\partial_x \tilde V(x,t)$, then $W(x,p,t)=\tilde W(x-a(t),p-a'(t),t)$ is solution of the
LE with $\partial_x V(x,t)=\partial_x \tilde V(x-a(t),t) - a''(t)$. Thus, if $\tilde W$ is the
solution for the $\omega(t)$ harmonic oscillator with $E(t)=0$, i.e.
$\partial_x \tilde V(x,t)= \omega(t)^2 x$, then $W$ is the solution 
for the $\omega(t)$ harmonic oscillator in presence of the field $E(t)= - \omega(t)^2 a(t) - a''(t)$
which is exactly Eq. \eqref{eqa}.} \\

An interesting question is about what happens for more general initial conditions, specifically
where $V_0(x)$ is different from the HO on the full line. For instance we know that the sine-kernel describes correlations  in
the bulk, for the initial state, for quite general $V_0(x)$. Near the edge, the Airy kernel holds for a smooth
$V_0(x)$, while the Bessel kernel holds for hard wall potentials \cite{UsShortReview}. It is interesting to ask if any aspects of this universality in the initial state survive a subsequent time evolution under a new Hamiltonian, in what follows we address this question.

\subsection{Half-oscillator initial condition} 

One interesting example of non-harmonic initial condition is the half-harmonic oscillator 
\be
V_0(x) = \begin{cases} & \frac{1}{2} \omega_0^2 x^2  \quad , \quad  x>0  \\
&+\infty ~ \quad , \quad ~~ x<0
\end{cases}
\ee
i.e. with an impenetrable wall at $x=0$. The wall is removed at $t=0$ and the
evolution the occurs with  $V(x)=\frac{1}{2} \omega^2  x^2$ on the full line. 
For simplicity we restrict ourselves to the case  $\omega_0=\omega$ and to $T=0$. 

Let us denote by $\chi_n(x)$, $n=0,1,\dots$ the real eigenfunctions of the $\omega$-harmonic oscillator on the full line.
The eigenfunctions of $H_0$ are then $\sqrt{2} \chi_n(x)$ with $n$ odd (so they are normalized on the half-axis),
and which are odd functions of $x$. 
The initial kernel associated with the ground state of the oscillator on the half line $K_\mu^+$ can be obtained by the image method, in terms of the kernel $K_\mu$ 
on the full line
\bea
&& K^+_\mu(x,x',0) = 2 \sum_{k \geq 1} \theta(\mu-\epsilon^0_{2 k-1})  \chi_{2k-1}(x) \chi_{2k-1}(x') \theta(x) \theta(x')
= (K_\mu(x,x',0) - K_\mu(x,-x',0) ) \theta(x) \theta(x')  \\
&& K_\mu(x,x',0) = \sum_{n \geq 0} \theta(\mu-\epsilon^0_{n})  \chi_{n}(x) \chi_{n}(x')
= \sum_{k \geq 1} \theta(\mu-\epsilon^0_{2 k-1})  \chi_{2k-1}(x) \chi_{2k-1}(x') +
\sum_{k \geq 0} \theta(\mu-\epsilon^0_{2 k})  \chi_{2k}(x) \chi_{2k}(x') \nonumber 
\eea
where the even eigenfunctions cancel in the difference. Note that the relation
between the Fermi energy $\mu$ and $N$ is now 
\be
\mu = \hbar \omega (2 N+\frac{1}{2}) 
\ee
instead of $\mu=\hbar \omega (N+\frac{1}{2})$ for the oscillator on the full line. The initial state 
corresponds to the model with a hard wall and a smooth potential studied
in \cite{UsHardBox,UsHardBoxLong}. One knows that for large $N$, the image
part of the kernel is important only in a width $1/k_F$ near the wall where
$k_F = \sqrt{2 \mu}/\hbar$. Outside this region, for $x>0$, i.e. in the bulk,
one can neglect the image part of the kernel, hence 
$K^+_\mu(x,x',0) \simeq K_\mu(x,x',0)$ for $x,x'>0$. The density in the bulk 
is thus a half semi-circle, with $x_e=\sqrt{2 \mu}/\omega$
\be \label{hsc} 
\rho_N(x,t=0) \simeq \frac{4  \theta(x)}{\pi x_e^2} \sqrt{x_e^2-x^2}
\ee 
normalized to unity on the positive axis. The Wigner function in the bulk is given
by equation (3) in the text, i.e.
\be
W(x,p,t=0) \simeq \frac{1}{2 \pi \hbar} \theta( \mu - \frac{p^2}{2} - \frac{1}{2} \omega^2 x^2  ) \theta(x), 
\ee
such that $\int dp W(x,p,0)=N \rho_N(x)$. It is thus uniform on the initial Fermi volume $\Omega_0$ which is 
a half ellipse in the $(x,p)$ plane as
represented in Fig. 1 of the text. Near $x=0$, the kernel takes the scaling form 
\cite{UsHardBox,UsHardBoxLong}
\bea \label{hw} 
K^+_\mu(x,x',0) = k_F {\cal K}^{\rm e}(k_F x, k_F x') \quad , \quad 
{\cal K}^{\rm e}(z, z') = [ \frac{\sin(z-z')}{\pi(z-z')} - \frac{\sin(z+z')}{\pi(z+z')} ] \theta(z) \theta(z')
\eea 
Note that the first term corresponds to the usual sine kernel, and is consistent
(in the absence of a wall) with $k_F =2/\xi_{x=0,t=0}$ where $\xi_{x,t}$ is defined in
the text below Eq. (38). \\

We now study the dynamics starting from this initial condition, for large $N$, and focus first on the bulk.
From the equation (13) in the text we see that the Wigner function is now 
uniform on the Fermi volume $\Omega_t$, which is the half ellipse $\Omega_0$ 
rotated counter-clockwise by an angle $\omega t$ in phase space. We now calculate the
time dependent density using the relation
\bea
\tilde \rho_N(x,t)= \frac{p_+(x,t)-p_-(x,t)}{2 \pi \hbar}, 
\eea 
and we denote by $[x_e^-(t),x_e^+(t)]$ the support of the density (which here is a single interval).
It is convenient
to use the variables $\tilde p=p/\sqrt{2 \mu}$ and $\tilde x=x/x_e$. In these
variable $\Omega_t$ is just a half circle of unit radius rotated counter-clockwise by $\omega t$. 

\begin{figure}
\includegraphics[width = 0.9\linewidth]{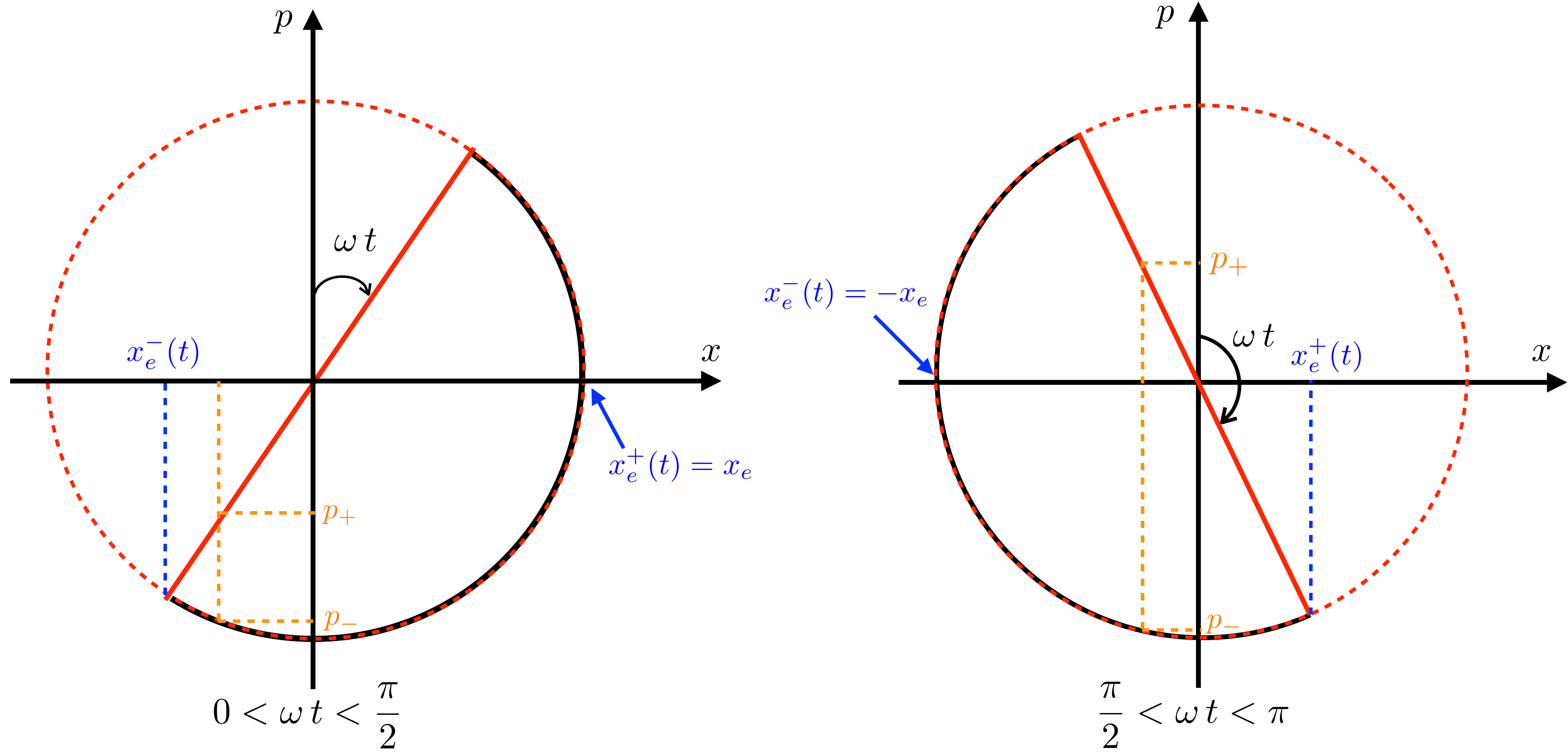}
\caption{Phase space picture in rescaled coordinates ($p$ in units $\sqrt{2 \mu}$ and 
$x$ in units of $x_e=\sqrt{2 \mu}/\omega_0$, denoted by $\tilde p$ and $\tilde x$ in the text).
The Fermi volume is delimited by the red straight line and the solid half-circle, which both rotate
with time with angle $\omega t$. The edges of the time dependent spatial density 
are indicated in blue. The Fermi momenta $p_+$ and $p_-$ given in \eqref{50}
are indicated in orange.}
\label{Fig1} 
\end{figure}

\begin{figure}
\includegraphics[width = 0.8\linewidth]{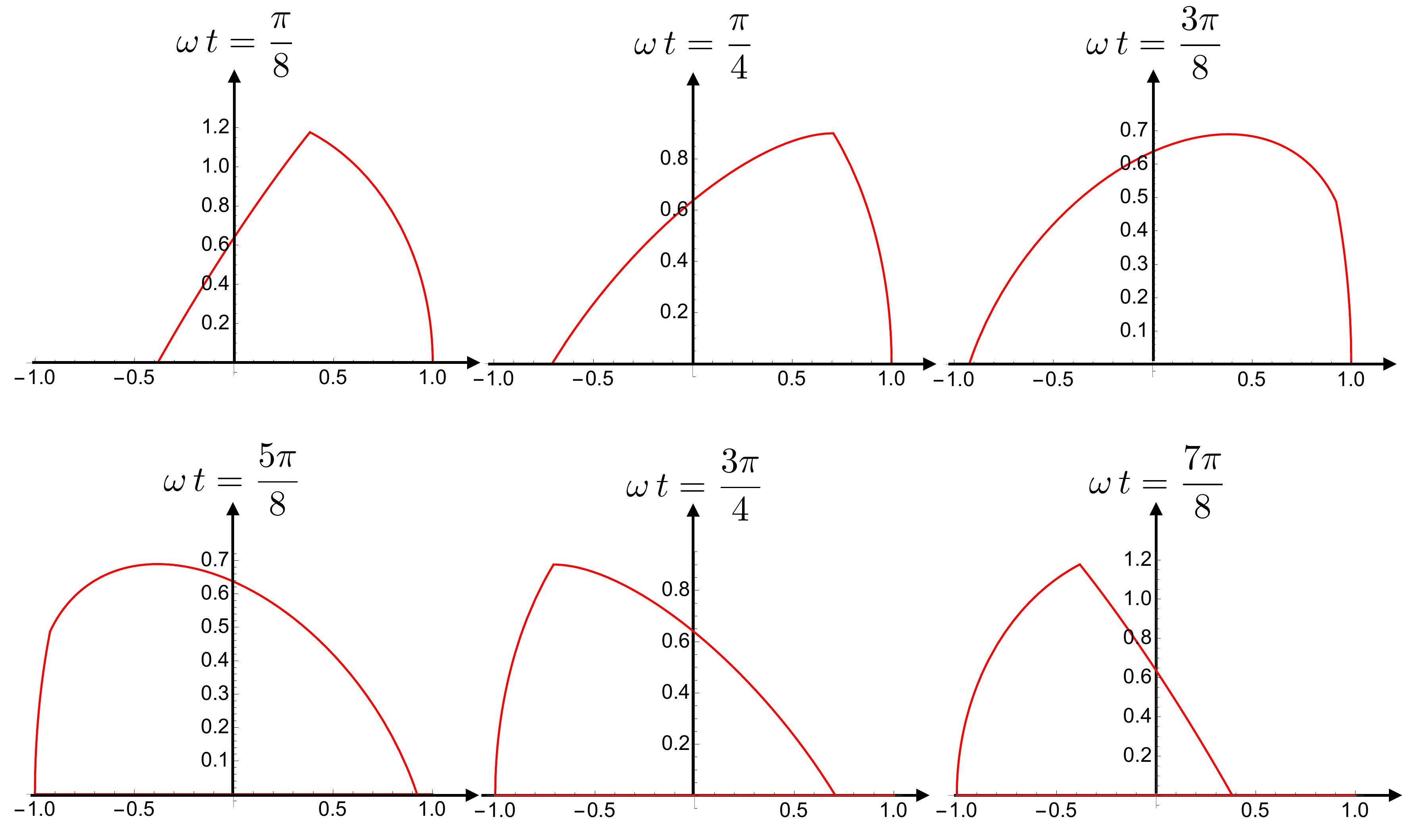}
\caption{Plots of the dimensionless number density $x_e \tilde \rho_N(x,t)$ 
as a function of $x/x_e$ for various times. The top figures are in the time
interval $[0,\pi/(2 \omega)]$ and correspond to the formula 
\eqref{densityvst}.} 
\label{Fig2} 
\end{figure}

Let us start by discussing the time interval
$0< \omega t < \pi/2$. In this case, we find that the density $\rho_N(x,t)$ is given exactly by Eq. (20) of the main text which reads
\be \label{soluhalf} 
\rho_N(x,t) =  \frac{2}{\pi x_e} \begin{cases} 
& 0 \quad , \quad \sin \theta < - \sin \omega t \\
& \frac{\sin(\omega t + \theta)}{\sin \omega t} \quad , \quad |\sin \theta| < \sin \omega t \\
& 2 \cos \theta  \quad , \quad \sin \theta > \sin \omega t
\end{cases} 
\ee 
where $x = x_e \sin \theta$ and $x_e = \sqrt{2\mu}/\omega_0$. The density $\rho_N(x,t)$ vs $x$ for different values of $t$ is shown in Fig. \ref{Fig2}. Indeed, as one can
see on the Fig. \ref{Fig1}, left panel,
there are three regions

(i) for $\tilde x<- \sin \omega t$ the density vanishes $\rho_N(x,t)=0$, as given in the first line of Eq. (20) of the main text.

(ii) for $- \sin \omega t < \tilde x<\sin \omega t$ one has
\be \label{50} 
\tilde p_+ = \tilde x \cot \omega t \quad , \quad \tilde p_- = - \sqrt{1- \tilde x^2}
\ee 
Hence the number density is given by
\be \label{densityvst} 
\tilde \rho_N(x,t)= \frac{\sqrt{2 \mu}}{2 \pi \hbar}  \left(\frac{x}{x_e} \cot \omega t + \sqrt{1- \frac{x^2}{x_e^2}} \right)
\ee
Defining $x= x_e \sin \theta$, with $\theta \in [-\frac{\pi}{2},\frac{\pi}{2}]$, the formula for the density $\rho_N(x,t)= \tilde \rho(x,t)/N$ can be simplified to yield the formula in the second line of (20) in the text. 
We have used  $\sqrt{1- \frac{x^2}{x_e^2}} = \cos \theta$ and that at large $N$, 
$N \simeq \frac{\mu}{2 \hbar \omega}$ which leads to $\frac{\sqrt{2 \mu}}{2 \pi \hbar N}= \frac{2}{\pi x_e}$.

%

(iii) for $\tilde x> \sin \omega t$, the density has not changed since $t=0$ and is still exactly equal to the original semi-circle given by formula \eqref{hsc},
which can be rewritten as $\rho_N(x,t)=  \frac{4}{\pi x_e} \cos \theta$ as given in the third line of (20) in the text.

The evolution of $\rho_N(x,t)$ in the first quarter period $0 < \omega t < \frac{\pi}{2}$ is
shown in Fig. \ref{Fig2}, top panel. 
In this time interval, the density vanishes 
linearly at the " moving lower edge" $x_e^-(t)=- x_e \sin \omega t$. Writing $\delta \tilde x= 
\frac{x -  x_e^-(t)}{x_e}$
we have $\sin \theta \simeq - \sin \omega t + \delta \tilde x$ and
$\cos \theta \simeq \cos \omega t + \delta \tilde x \tan \omega t$ leading to
\be
\rho_N(x,t) = \frac{4}{\pi x_e^2 \sin 2 \omega t} (x -  x_e^-(t)) + O((x -  x_e^-(t))^2)
\ee 
Note that the amplitude diverges at $t\to 0^+$ and $t \to \pi/2^-$, as the density
becomes of semi-circular shape near $x_e^-(t)$ in these limits. 
The fact that the density vanishes linearly suggests that the limiting kernel
which describes the edge is not the Airy kernel. 

At the mirror image point $x=x_e \sin \omega t$, i.e. $\theta=\omega t$, the density is continuous
(between regimes (ii) and (iii)) 
with $\rho_N(x=x_e \sin \omega t,t) = \frac{4}{\pi x_e} \cos \omega t$, but with a discontinuity in
the first derivative. Beyond this point, for $x_e \sin \omega t < x <x_e$, 
the initial semi-circle density has survived. It vanishes at the upper edge $x_e^+(t)=x_e$,
which is thus time independent in the time interval $0<\omega t < \pi/2$. 

As discussed in the text, once the density is known for $0< \omega t<\frac{\pi}{2}$
it can be obtained at all $t$ by noting the symmetry (see the Fig. 1 in the text)
$\rho_N(x,t)=\rho_N(-x,\frac{\pi}{\omega} - t)$, and the fact that it is time periodic with 
period $2\pi/\omega$. This implies, for instance, that in the time interval $\frac{\pi}{2} < \omega t <\pi$
the upper edge $x_e^+(t)$ recedes from $x=x_e$ to $x=0$,
and the lower edge is fixed at $x_e^-(t)=-x_e$. This can be seen from the right panel in
Fig. \ref{Fig1}, and from the bottom panel of Fig. \ref{Fig2} (see also
Figure 1 in the text).\\

To obtain the evolution near the edge, we use Eq. \eqref{Karbitrary1}, which is valid for
an arbitrary initial condition. In this equation we replace the initial Wigner function by its
expression in terms of the initial kernel (from (9) in the text). Performing
the change of variable $p= \frac{x+x'}{2 L(t) \tau(t)} - \frac{q}{\tau(t)}$ it reads
\bea 
K_{\tilde \mu}(x,x',t) = \frac{e^{ i \frac{L'(t)}{2 L(t) \hbar} ((x')^2-x^2)}}{2 \pi \hbar L(t) \tau(t)}
\int_{-\infty}^{+\infty} dq  dy ~ e^{\frac{i}{\hbar}  (\frac{x+x'}{2 L(t) \tau(t)} - \frac{q}{\tau(t)}) (y - \frac{x-x'}{L(t)}) } ~
K_\mu(q  + \frac{y}{2}, q- \frac{y}{2},0)
\eea
Up to this point the result is exact (and valid for any initial condition $H_0$). If we take the initial
kernel to be the sine kernel, i.e. replace $K_\mu(q  + \frac{y}{2}, q- \frac{y}{2},0) \to \frac{\sin k_F y}{\pi y} $,
we obtain the same result as the rescaling of the initial sine-kernel as in Eq. 
\eqref{Kresc2}. This is consistent with our previous results for $H_0$ being the harmonic oscillator,
but it holds in fact in the bulk for essentially any initial condition. Let us also note that if we
replace the initial kernel instead with the image kernel,
$K_\mu(q  + \frac{y}{2}, q- \frac{y}{2},0) \to \frac{\sin 2 k_F q}{2 \pi q} $, we find again
that it evolves by simple rescaling as in Eq. 
\eqref{Kresc2}. 

For the half-harmonic 
oscillator we must replace $K_\mu \to K_\mu^+$ and the integration region is limited to that where
both arguments of the initial kernel are positive, hence $
\int_{-\infty}^{+\infty} dq  dy \to \int_{0}^{+\infty} dq  \int_{-2q}^{2 q} dy$.
%
One can replace in the r.h.s. the kernel by its edge form \eqref{hw} whenever
the integral is dominated by the region $y \simeq q \simeq 0$. This gives
\bea \label{Karbitrary3}
K_{\tilde \mu}(x,x',t) = \frac{e^{ i \frac{L'(t)}{2 L(t) \hbar} ((x')^2-x^2)}}{2 \pi \hbar L(t) \tau(t)}
\int_{0}^{+\infty} dq  \int_{-2q}^{2 q} dy 
 ~ e^{\frac{i}{\hbar}  (\frac{x+x'}{2 L(t) \tau(t)} - \frac{q}{\tau(t)}) (y - \frac{x-x'}{L(t)}) } ~
\left( \frac{\sin k_F y}{\pi y}  - \frac{\sin 2 k_F q }{2 \pi q} \right)
\eea
\\

It would be interesting to generalize the present study to the case of an initial condition given by a $\omega_0$ harmonic oscillator for $x >0$ in the presence of a $1/x^2$ wall, which is removed at $t=0$. This initial condition is described near the wall by a Bessel kernel, as we now discuss. 

\section{Harmonic oscillator in presence of $1/x^2$ potential}

As claimed in \cite{Kanasugi1995} (which we checked explicitly) the rescaling method 
extends to the following choice of $H(t)$ and $H_1$
\be \label{HtH0} 
H(t)= \frac{p^2}{2} + \frac{1}{2} \omega(t)^2 x^2 +  \frac{\alpha(\alpha-1) \hbar^2}{2 x^2} \quad , \quad H_1= \frac{p^2}{2} + \frac{1}{2} \omega_1^2 x^2 +  \frac{\alpha(\alpha-1) \hbar^2}{2 x^2} 
\ee
This system is defined on $x>0$ in presence of an impenetrable repulsive barrier with $\alpha > 1$,
alternatively one can choose $V(x)$ and $V_0(x)$ to be positive infinite for $x \leq 0$. 
Let us study the case $H_0=H_1$ with $\omega_1=\omega_0$. The initial condition is now a 
harmonic oscillator on the half line $x>0$, plus a $1/x^2$ wall at $x=0$. Note that for $\alpha=1$ it is the half oscillator studied
above, but with a different evolution Hamiltonian $H(t)$.
Let us calculate the corresponding kernel.
The eigenfunctions of $H_0$ are expressed using Laguerre polynomials and labeled by the integer $n=0,1..$
\bea
\phi_n(x) = A_n e^{- \omega_0 x^2/2 \hbar} x^\alpha L_n^{\alpha-1/2}(\omega_0 x^2/\hbar) \quad , \quad 
\epsilon_n= \hbar \omega_0 (2 n+ \alpha + \frac{1}{2})
\quad , \quad |A_n|^2 = \frac{2 (\omega_0/\hbar)^{\alpha+1/2} n!}{\Gamma(n+\alpha + 1/2)}
\eea
The relation between the Fermi energy $\mu$ and $N$ is now $\mu =\hbar \omega_0 (2 n+ \alpha + \frac{1}{2})$. 
The initial condition of the dynamics at $t=0$ is thus the $T=0$ kernel corresponding to the ground state of ${\cal H}_0$ 
\be
K_\mu(x,y,0)=\sum_{n=0}^{N-1} \phi_n^*(x) \phi_n(y)  = 2 (\frac{\omega_0}{\hbar})^{ \alpha +1/2} (x y)^{\alpha} e^{- \frac{ \omega_0 (x^2+y^2)}{2 \hbar}} 
\sum_{k=0}^{N-1} \frac{k!}{\Gamma(k+\alpha + 1/2)} 
 L_k^{\alpha-1/2}(\frac{\omega_0 x^2}{\hbar})  L_k^{\alpha-1/2}(\frac{\omega_0 y^2}{\hbar}) 
\ee 
We use the identity \cite{Grad} (see formula 8.974 p. 1002), for $n=0,1,..$ 
\bea
(x-y) {n+\alpha-1/2 \choose n} \sum_{k=0}^n \frac{L_k^{\alpha-1/2}(x) L_k^{\alpha-1/2}(y) }{{k+\alpha-1/2 \choose k} }
= (n+1) \big[ L^{\alpha-1/2}_{n}(x) L^{\alpha-1/2}_{n+1}(y) - L^{\alpha-1/2}_{n+1}(x)  
L^{\alpha-1/2}_{n}(y) \big]
\eea 
Hence
\be \label{Kalpha}
K_\mu(x,y,0)= e^{- \frac{ \omega_0 (x^2+y^2)}{2 \hbar}} 
\frac{2 N! (\frac{\omega_0}{\hbar})^{ \alpha -1/2} (x y)^{\alpha} }{\Gamma(N+ \alpha-1/2)}
~ \frac{L^{\alpha-1/2}_{N-1}(\omega_0 x^2/\hbar) L^{\alpha-1/2}_{N}(\omega_0 y^2/\hbar) 
- L^{\alpha-1/2}_{N}(\omega_0 x^2/\hbar)  L^{\alpha-1/2}_{N-1}(\omega_0 y^2/\hbar)}{x^2-y^2} 
\ee
a formula valid for any $N$. By construction this kernel is reproducible.
The average initial density is thus given, for any $N$, by
\bea
\rho_N(x,t=0)=e^{- \frac{ \omega_0 x^2}{ \hbar}} 
\frac{2 N! (\frac{\omega_0}{\hbar})^{ \alpha +1/2} x^{2 \alpha} }{\Gamma(N+ \alpha-1/2)}
\left(L_{N-1}^{\alpha -\frac{1}{2}}\left(\frac{\omega_0 x^2}{\hbar}\right) L_{N-1}^{\alpha
   +\frac{1}{2}}\left(\frac{\omega_0 x^2}{\hbar} \right)-L_{N-2}^{\alpha +\frac{1}{2}}\left(\frac{\omega_0 x^2}{\hbar}\right) L_N^{\alpha -\frac{1}{2}}\left(\frac{\omega_0 x^2}{\hbar} \right)\right)
   \eea
This family of kernels indexed by $\alpha$ are well known in RMT. Indeed, there is a one to one
correspondence between the $T=0$ quantum JPDF of the positions of the fermions $x_i$ in this potential
and the JPDF of the eigenvalues $\lambda_i=\omega_0 x_i^2/\hbar$ of the Wishart-Laguerre ensemble of
random matrices with $\beta=2$ and
$M-N=\nu=\alpha - \frac{1}{2}$ \cite{Nadal_Majumdar}
\be
P_{\rm joint}(\lambda_1,\cdots,\lambda_N)=\frac{1}{Z_N}\prod_{i<j}|\lambda_i-\lambda_j|^{2}\prod_{k=1}^{N}\lambda_k^{\nu} e^{-N \lambda_k}\;.
\ee
In the large $N$ limit, the mean density of eigenvalues converges  to the Mar\v cenko-Pastur distribution \cite{forrester}, which upon the change of variables leads to the half semi-circle given in Eq. (22) of the text.

From the rescaling method we know that for any $N$ the time dependent kernel $K_\mu(x,y,t)$, evolving
with $H(t)$ given in \eqref{HtH0} keeps the scaling form 
given by Eq. (17) in the text. 
where $L(t)$ is the same solution of the Ermakov's equation (16) in the text as for
the harmonic oscillator. For the simplest case $\omega(t>0)=\omega$ one 
has $L(t)=x_e(t)/x_e$, where $x_e(t)$ is given by Eq. (14) in the text and $x_e=\sqrt{2 \mu}/\omega_0$. In the limit of large $N$ there are two regions:

\begin{enumerate} 

\item[$\bullet$] The bulk, for $x >0$ and $k_F x \gg 1$
where $k_F = \sqrt{2 \mu}/\hbar$. In this region one can in fact neglect the 
$1/x^2$ potential to leading order at large $\mu$. Hence to leading order the right edge
of the initial Fermi gas is at $x \simeq x_e$. The density in the bulk is thus
the (time-dependent) half-semi circle given in (22) in the text.


\item[$\bullet$] The second region is the edge $k_F x \sim O(1)$. In that region at large $N$ one shows that the
Laguerre kernel becomes the Bessel kernel which describes the "hard edge" universality class
of RMT \cite{forrester}. This is easy to see by neglecting the $x^2$ term
in the potential in \eqref{HtH0}. In that case the Bessel kernel 
is the exact kernel (for any $N$) for $H_1=H_0$ setting $\omega_1=\omega_0=0$ (see for
instance the detailed calculation in \cite{UsHardBoxLong} Section 6.2).
For the time evolution, it leads to the formula (23) in the text for the time-dependent kernel near the edge. 
Up to a time dependent scale, it remains the Bessel kernel which thus describes the edge
at all times for this choice of $H(t)$ and $H_0$. 

\end{enumerate}

\section{Generic smooth confining potentials} 

\subsection{Classical evolution: Liouville equation}

As discussed in the text, for generic potential $V(x)$ with $V'''(x) \neq 0$ the WE
differs from the LE by "quantum terms". In the "classical" limit, $\hbar \sim 1/N$ and $N$ large
considered in \cite{Kulkarni2018} the Wigner function satisfies two exact properties
at $T=0$

(i) $(2 \pi \hbar W)^2=2 \pi \hbar W$, hence at any point $(x,p)$ in phase space, either $W(x,p,t)=0$ 
or $2 \pi \hbar W(x,p,t)=1$. Thus $2 \pi \hbar W$ is the characteristic function
of $\Omega_t$ the Fermi volume.

(ii) The Liouville equation is obeyed exactly, for any $V(x)$. Hence $\Omega_t$ is
simply transported from the initial Fermi volume by the classical equations of motion.

It is absolutely not obvious that these two properties necessarily apply at large 
$N$, i.e. large $\mu$, but for fixed $\hbar$. We have argued in the text that, when $V(x)$ and 
$V_0(x)$ are both confining and smooth, and of "similar" scales, the quantum terms in the
WE are subdominant at large $\mu$ at $t=0$. This should ensure that it remains
true in some time window for $t>0$.

We thus start by analyzing the Liouville equation
\be \label{LE2} 
\partial_t W(x,p,t) = - p \partial_x W(x,p,t) + \partial_x V(x) \partial_p W(x,p,t) 
\ee 
and later will consider the "quantum corrections". 
Consider the classical trajectories in phase space, $\dot x(t) = p(t)$ and $\dot p(t)= - V'(x(t))$.
The Wigner function is constant along such trajectories
$\frac{d}{dt} W(x(t),p(t),t) = \partial_t W(x(t),p(t),t)  + \dot x(t) \partial_x W(x(t),p(t),t) + \dot p(t) 
\partial_p W(x(t),p(t),t) = 0$ if one uses both the LE and the classical equations of motion. 
Let us denote $(x(t)=x_c(x_0,p_0,t),p(t)=p_c(x_0,p_0,t))$ the classical trajectory 
as a function of the initial point $(x_0,p_0)$. One can thus write the solution of
the WE as 
\begin{equation}
W(x,p,t) = \int dx_0 dp_0 W(x_0,p_0,0) \delta(x- x_c(x_0,p_0,t))\delta(p-p_c(x_0,p_0,t))
\end{equation}
Since the phase space volume element $dx dp$ is preserved by the classical evolution and because of
the unicity of the solutions, one can also write the general solution of LE as in the text as
$W(x,p,t) = W(x_0(x,p,t),p_0(x,p,t),0)$, where
$x_0(x,p,t)$, $p_0(x,p,t)$ denote the initial conditions (at $t=0$) as functions of the final ones (at $t$).
Note that all the above extends to the case of a time dependent potential $V(x,t)$. 

Let us now consider an initial condition prepared at $T=0$. For large $N$ the initial condition is a theta function
$W(x,p,0)= \frac{1}{2 \pi} \theta(\mu-\frac{p^2}{2} - V_0(x))$, hence under the LE the Wigner function
\bea
W(x,p,t) = \frac{1}{2 \pi} \theta(\mu-\frac{p_0(x,p,t)^2}{2} - V_0(x_0(x,p,t))) 
\eea
remains a theta function at all times, which is unity inside the transported Fermi volume $\Omega_t$ 
and zero outside. 

There are various parameterizations for the boundary of $\Omega_t$, i.e. the transported Fermi surf $S_t$. The first one is in terms of the (multiple) roots $p(x,t)$ of the equation
\bea \label{eqmu} 
\mu = \frac{p_0(x,p(x,t),t)^2}{2} + V_0(x_0(x,p(x,t),t))
\eea 
Consider for simplicity $V(x)$ to be time independent.
One can show that any such root satisfies the Burgers equation Eq. (28) in the text.
Indeed, from energy conservation 
\be
\frac{p_0(x,p,t)^2}{2} + V(x_0(x,p,t)) = \frac{p^2}{2} + V(x)
\ee 
one has the relations
\bea
p_0 \partial_p p_0 + V'(x_0) \partial_p x_0 = p \quad , \quad  p_0 \partial_x p_0 + V'(x_0) \partial_x x_0 = V'(x)
\eea
On the other hand, taking derivatives of \eqref{eqmu} w.r.t. $t$ and w.r.t. $x$
and using the (time reversed) equation of motions 
$\partial_t p_0=V'(x_0)$ and $\partial_t x_0 = - p_0$ 
gives 
\bea
&& \partial_t p = \frac{p_0 (V'_0(x_0) - V'(x_0))}{p_0 \partial_p p_0 + V_0'(x_0) \partial_p x_0} \quad , \quad  \partial_x p = - \frac{V'_0(x_0) \partial_x x_0 + p_0 \partial_x p_0}{p_0 \partial_p p_0 
+ V_0'(x_0) \partial_p x_0} .
\eea
Using the above relations we can show that 
\bea\label{defGt}
\partial_t p + p \partial_x p + V'(x) = (1 + \partial_x p_0 \partial_p x_0 - \partial_x x_0 \partial_p p_0) 
\frac{p_0 (V_0'(x_0)-V'(x_0))}{p_0 \partial_p p_0 + V_0'(x_0) \partial_p x_0} \;.
\eea 
Consider now the first factor $J(t) =  1 + \partial_x p_0 \partial_p x_0 - \partial_x x_0 \partial_p p_0$ appearing on the rhs of Eq. (\ref{defGt}).
One can evaluate the time evolution $dJ(t)/dt$ using the (time reversed) equation of motions 
$\partial_t p_0=V'(x_0)$ and $\partial_t x_0 = - p_0$. After a straightforward algebra, we find $dJ(t)/dt = 0$. This reflects the conservation of 
the volume in phase space. In addition, from the initial condition where
$x=x_0$ and $p=p_0$, we see that $J(0)=0$. Hence $J(t)=0$ for all time $t \geq 0$. Consequently Eq. (\ref{defGt}) reduces to the Burger's equation
\beq \label{burgers_supp}
\partial_t p + p \partial_x p + V'(x) = 0 \;.
\eeq
Hence any root of \eqref{eqmu}, $p(x,t)$, 
satisfies the Burgers equation (BE)
\be \label{burgers} 
\partial_t p(x,t) + p(x,t) \partial_x p(x,t) + V'(x)=0
\ee

The transported Fermi volume $\Omega_t$ can be parameterized by its "section" in phase space at coordinate $x$, $\Omega_t(x)$, given as a function of $x$. It is defined such that $p \in \Omega_t(x)$ iff $(x,p) \in \Omega_t$. 
In general $\Omega_t(x)$ is made of multiple intervals (we can call $n(x,t)$ its number),
i.e. $\Omega_t(x) = \cup_{j=1}^{n(x,t)} [p^j_-(x,t),p^j_+(x,t)]$, where the edges 
merge and disappear, or appear in pairs, as $x$ spans $]-\infty,+\infty[$. Each of these
roots $p^j(x,t)$ satisfies the BE inside its interval of existence. 

Let us consider the simplest case, where one can write
\be \label{WS} 
W(x,p,t) = \frac{1}{2 \pi \hbar} \theta(p - p_-(x,t))  \theta(p_+(x,t)-p) 
\ee
Since the number density is $\tilde \rho(x,t)= \int dp W(x,p,t)=
\frac{1}{2 \pi \hbar} (p_+(x,t) - p_-(x,t))$, the form \eqref{WS} is valid for 
$x \in {\cal I} = [x^-_e(t)^-,x^+_e(t)]$, the support of the density, 
which is assumed here to be a single interval. At $x=x_e^\pm(t)$, in
generic situations, the density vanishes smoothly and $p_+(x,t)=p_-(x,t)$.
Let us first check that $W$ in \eqref{WS} satisfies the LE. Inserting \eqref{WS} in \eqref{LE2} 
we see that if $p_\pm(x,t)$ satisfy the BE \eqref{burgers} then the
LE equation is indeed obeyed. 

To recover the standard free fermion hydrodynamics one defines a local 
velocity field $v(x,t)$ through 
\be
\tilde \rho(x,t) v(x,t)= \int dp ~ p ~ W(x,p,t), \label{pv}
\ee
and using \eqref{WS} then leads to the identification $v(x,t)= \frac{1}{2} (p_+(x,t) + p_-(x,t))$.
One can also define the local kinetic energy density
\be
\epsilon(x,t) = \int dp ~ \frac{p^2}{2} W(x,p,t) = \frac{1}{2} \tilde \rho(x,t) v(x,t)^2 + \frac{\hbar^2}{6} \pi^2 \tilde \rho(x,t)^3
\ee
where in last equality  arises  from using \eqref{WS}
and \eqref{pv}. The term cubic in the fermion density is often called the pressure term resulting from the Pauli exclusion principle. 

We can now add and substract the two independent Burgers equations 
\eqref{burgers} for $p_+(x,t)$ and $p_-(x,t)$ and we obtain
\bea
&& \partial_t \tilde \rho(x,t) + \partial_x [\tilde \rho(x,t)  v(x,t)] = 0 \label{cont} \\
&& \partial_t  v(x,t) + v(x,t) \partial_x v(x,t)  + V'(x) = - \frac{1}{\tilde \rho(x,t)} \partial_x (\frac{\hbar^2 \pi^2}{3} \tilde \rho(x,t)^3 )
= - \hbar^2 \pi^2 \tilde \rho(x,t) \partial_x \tilde \rho(x,t) 
 \label{Burg}
\eea 
The first is the continuity equation (conservation of particle number) and
the second the Newton force equation. These are the standard hydrodynamical equations for fermions
\cite{Abanov,Matytsin,JPB-SM,Lama1}. They can be obtained from the variation of the action 
\bea
S_f[\tilde \rho,v] = \int dx dt \, [ \frac{1}{2} \tilde \rho({\bf x},t) v^2 - \frac{\hbar^2 \pi^2}{6} \tilde \rho^3({\bf x},t) - V({\bf x}) \tilde \rho({\bf x},t) ]
\eea 
under the constraint \eqref{cont}. Here they are simply obtained as a consequence
of the Liouville equation. 
%

The equilibrium is recovered for $V(x)=V_0(x)$, in which case
\bea \label{init} 
p_\pm(x,t)= \pm p_\mu^0(x) \quad , \quad p_\mu^0(x) = \sqrt{2 (\mu- V_0(x))} 
\eea 
and $W(x,p,t)=W(x,p,0)$ becomes time independent. This provides a time-independent solution
of the above hydrodynamic equations. 

In the case of the quantum quench, $V(x) \neq V_0(x)$, Eq. \eqref{WS} gives the solution
for the Wigner function to the LE, where $p_\pm(x,t)$ are the solution of the BE \eqref{burgers} with initial conditions
at $t=0$ given by $p_\pm(x,t=0)= \pm p_\mu^0(x)$. 

As discussed in \cite{KonikDubail2017} the more complicated cases where the parameterization using $\Omega_t(x)$
involves more than one interval is called the $2k$- hydrodynamics. Similar equations can be derived to
describe it with hydrodynamic variables. Eventually, the difficulty amounts to find a convenient
parameterization to describe
the (relatively simple) classical dynamics of the Fermi volume.

\subsection{Quantum corrections: width of the Fermi surf}

As discussed in the text, to probe the quantum corrections to the Liouville equation, 
we look for a solution of the full Wigner equation (WE), Eq. (11) in the text, near the
Fermi surf, for $p \approx p_+(x,t)$, of the form 
\bea \label{Wscal} 
W(x,p,t) \simeq F(\tilde a) \quad , \quad \tilde a = \frac{p - p_+(x,t)}{D(x,t)} 
\eea
with a time-dependent thickness $D(x,t)$ that we will determine below. Far from the Fermi surf, the approximation
by theta functions, Eq. \eqref{WS}, should hold, which leads to the boundary conditions 
$F(+\infty)=0$ and $F(-\infty)=1$. A priori the function $F$ can also depend on time, so we will
include that term below, but will see that it can be neglected. We first consider the 
WE truncated to cubic order, Eq. (25) in the text, and discuss higher order corrections below.
Inserting \eqref{Wscal} into this cubic WE reads
\bea
(\partial_t + p \partial_x - V'(x) \partial_p + \frac{\hbar^2}{24} V'''(x) \partial_p^3 ) F( \frac{p - p_+(x,t)}{D(x,t)} )=0
\eea 
Performing the derivatives and substituting $p = p_+(x,t) + \tilde a D(x,t)$, where $\tilde a$ is the scaling variable
of order $O(1)$ we obtain by dividing by $F'(\tilde a)/D(x,t)$
\bea \label{eqD} 
&& D(x,t) \frac{\partial_t F}{F'(\tilde a)} = V'(x) + \partial_t p_+(x,t) + p_+(x,t) \partial_x p_+(x,t) \\
&& + \tilde a \left( \partial_t D(x,t) + p_+(x,t) \partial_x D(x,t)  + D(x,t) \partial_x p_+(x,t) - \frac{F'''(\tilde a)}{\tilde a F'(\tilde a)} 
\frac{\hbar^2}{24 D(x,t)^2} V'''(x)  \right)
+ \tilde a^2 D(x,t) \partial_x D(x,t), 
\eea 
which should be obeyed for any $\tilde a$ of order $O(1)$. 
The r.h.s. in the first line vanishes from the Burgers equation \eqref{burgers}. 
The last term can be argued to be small (see below) and we neglect it. Now if one chooses
\bea
F'''(\tilde a) = 4 \tilde a F'(\tilde a) 
\eea 
which is obeyed by $F(\tilde a) = {\cal W}(\tilde a)={\rm Ai}_1(2^{2/3} \tilde a)$, then we obtain an evolution equation
for $D(x,t)$. Amazingly there is a simple solution to this equation
\bea \label{soluD} 
D(x,t) = - (\frac{\hbar^2}{2} \partial_{xx} p_+(x,t) )^{1/3}
\eea 
This can be checked explicitly by plugging \eqref{soluD} into the equation \eqref{eqD} and 
by using the second derivative of the Burgers equation \eqref{burgers}
\bea
3 \partial_x p_+(x,t)   \partial_x^2 p_+(x,t)  + \partial_t \partial_x^2 p_+(x,t)  + p_+(x,t)  \partial_x^3 p_+(x,t)  + V'''(x) = 0
\eea 
Now we can check that this is the (unique) correct solution, since it
satisfies the correct initial condition. Indeed, as discussed above, the initial condition for the
Burgers equation is $p_\pm(x,t=0)= \pm p_\mu^0(x)$ where $p_\mu^0(x) = \sqrt{2 (\mu- V_0(x))}$
is its "equilibrium" solution if one substitute $V(x) \to V_0(x)$. Thus we have
\be \label{D0} 
D(x,t=0) = - (\frac{\hbar^2}{2} \partial^2_{x} p_\mu^0(x))^{1/3} =
\frac{\hbar^{2/3}}{2^{1/3} p} (V'_0(x)^2 + p^2 V_0''(x))^{1/3}|_{p=p_\mu^0(x)}
= \frac{e(x,p)}{p}|_{p=p_\mu^0(x)} 
\ee
The last expression is precisely equal to $\frac{e({\sf x}_e,{\sf p}_e)}{{\sf p}_e}$ if $({\sf x}_e,{\sf p}_e)$ denotes
a generic point on the Fermi surf with here the parameterization ${\sf x}_e=x$, ${\sf p}_e=p_\mu^0(x)$
for the half top part of the Fermi surf (note that ${\sf x}_e$ should be distinguished from $x_e$ defined in
this work,
which is the upper edge of the semi circle). The factor of $p$ in the numerator of \eqref{D0} comes from
the matching of the two scaling variables in the region near the Fermi surf with $a, \tilde a = O(1)$ 
\be
a = \frac{\frac{p^2}{2} + V_0(x)-\mu}{e(x,p)} \simeq p_\mu^0(x) \frac{p- p_\mu^0(x)}{e(x,p_\mu^0(x))} = 
\frac{p- p_\mu^0(x)}{D(x,t=0)}
= \tilde a |_{t=0},
\ee 
where $a$ is the scaling variable introduced for the equilibrium result (5) of the text.

The solution corresponding to $\theta(p-p_-)$ is obtained similarly with $\tilde a\to -\tilde a$. 
Hence the complete "quantum" solution corresponding to \eqref{WS} is
\bea \label{Wscal22} 
W(x,p,t) \simeq {\cal W}(\frac{p - p_+(x,t)}{D_+(x,t)}) 
{\cal W}(\frac{p_-(x,t)-p}{D_-(x,t)}) 
\eea
where here we denote $D_+(x,t)=D(x,t)$ and 
$D_-(x,t)= - (\frac{\hbar^2}{2} \partial_{xx} p_-(x,t) )^{1/3}$. Note that \eqref{Wscal22} 
is valid only when the two edges in phase space are well separated 
$2 \pi \hbar \tilde \rho(x,t)= p_+(x,t) - p_-(x,t) \gg D_+(x,t) + D_-(x,t)$, i.e. far from the edges in real space where the density vanishes.

We can now justify that higher order terms in the exact WE, Eq. (11) in the text,
are subdominant (as compared to the cubic approximation of the WE involving only $V'''(x)$)
to predict the scaling form near the Fermi surf in the limit
of large $\mu$. We use the same counting as in the text below Eq. (25). 
The main assumption is that counting in $\mu$ and $x_e$ of $V(x)$ 
and its derivative is similar to $V_0(x)$, i.e. the two potentials are not too dissimilar. 
In the text we found the higher order terms in the WE are powers of
the two small parameters which measure the deviation
from the classical limit and the LE
\be
\frac{\hbar}{x_e p_e} \ll 1 \quad , \quad \frac{\hbar^{2/3}}{\mu^{1/3} x_e^{2/3}} \ll1 
\ee
Using the normalization $\int dx dp W(x,p,t) = N$ we see that the first parameter is simply $\sim 1/N$. The second one
is larger but is related to the quantum fluctuations near the edge. 
We first argue that the term neglected  in \eqref{eqD} is small compared to the others
proportional to $a$, i.e. 
\be
a^2 D(x,t) \partial_x D(x,t) \ll a p_+(x,t) \partial_x D(x,t)
\ee 
Recalling that $e_N \sim (\hbar \mu/x_e)^{2/3}$ (see text), we have 
$D(x,t) \sim e_N/p_e \sim \hbar^{2/3} \mu^{1/6}/x_e^{2/3}$ and
we find that $D(x,t)/p_+(x,t) \sim \hbar^{2/3}/(\mu^{1/3} x_e^{2/3}) \ll 1$ for large $\mu$. 

Consider the next order quantum corrections, e.g. a term $- g \hbar^4 V^{(5)} \partial_p^5 W$ in the
$W$ equation. The terms proportional to $a$ in \eqref{eqD} and containing the potential now read
\be
- \frac{F'''(\tilde a)}{\tilde a F'(\tilde a)} 
\frac{\hbar^2}{24 D(x,t)^2} V'''(x)   - g \hbar^4 \frac{F^{(5)}(\tilde a)}{\tilde a F'(\tilde a)} V^{(5)}(x) \frac{1}{D(x,t)^4}
\ee 
Since $\tilde a=O(1)$ the ratio of these two terms is of order 
\be
\frac{V^{(5)}(x)}{V'''(x) D(x,t)^2} \sim \frac{\hbar^{2/3}}{\mu^{1/3} x_e^{2/3}} \ll 1
\ee 

Finally, we note that if we add the term proportional to $\partial_t F$ in the l.h.s. of \eqref{eqD} 
to the previous analysis, we obtain the constraint that $\partial_t F(\tilde a) \sim \tilde a F'(\tilde a)$,
hence it corresponds simply to a redefinition of $D(x,t)$. Therefore we can assume simply that
this term is absent at this order. Note that one can in principle perform an expansion
in the small parameter $\frac{\hbar^{2/3}}{\mu^{1/3} x_e^{2/3}}$ to higher orders, 
where these neglected terms would play a role.

We note that the above derivation may be used to derive (or at least guess) the scaling form for equilibrium.
Curiously, the role of $V'''(x)$ being non zero seems crucial in the argument, and it is unclear to us
why that should be so (for equilibrium). 

In particular we note that for the harmonic oscillator the above equations do not
determine the scaling function $F(\tilde a)$, since the term containing $V'''(x)$
is absent. Nevertheless we can check independently that our above result
holds, in a non-trivial way, for the harmonic oscillator. In that case we can
use the general solution of the LE, $W(x,p,t)= W(x_0(x,p,t), p_0(x,p,t),0)$,
and the scaling form Eqs. (4) and (5) of the text, for the initial Wigner function.
The calculation for general $V_0(x)$ is quite non-trivial. Hence we choose
for simplicity, $V_0(x)= \frac{1}{2} \omega_0^2 x^2$. In this case 
\be
W(x,p,t) \simeq {\cal W}( \frac{\frac{p_0^2}{2} + V_0(x_0) - \mu}{(\hbar \omega_0)^{2/3} \mu^{1/3}}) \quad , \quad x_0(x,p,t) = 
x \cos(\omega t)- \frac{p}{\omega} \sin(\omega t) \quad , \quad 
p_0(x,p,t) = 
p \cos(\omega t) + \omega x \sin(\omega t) .
\ee 
We thus obtain
\bea \label{pxt} 
p_+(x,t) = \frac{\omega  \left(x \left(\omega _0^2-\omega ^2\right)
   \sin (2  \omega t )+2 \omega  \omega _0
   \sqrt{x_e(t)^2-x^2}\right)}{2 \left(\omega ^2 \cos
   ^2( \omega t)+\omega _0^2 \sin ^2(\omega t)\right)} \quad , \quad 
   x_e(t)= \frac{\sqrt{\mu}}{\omega_0 \omega} 
\sqrt{\omega^2+ \omega_0^2 + (\omega^2-\omega_0^2) \cos(2 \omega t)}.
\eea 
We can now expand the scaling variable for $p \approx p_+(x,t)$, which simplifies to give
\bea
a = \frac{\frac{p_0^2}{2} + V_0(x_0) - \mu}{(\hbar \omega_0)^{2/3} \mu^{1/3}})
\simeq \omega_0^{1/3} \hbar^{-2/3}  \mu^{-1/3} \sqrt{x_e(t)^2-x^2}  (p - p_+(x,t)), 
\eea 
which implies that
\be \label{Dplus} 
D_+(x,t)= \hbar^{2/3} \left(\frac{\mu}{\omega_0}\right)^{1/3} \frac{1}{\sqrt{x_e(t)^2 -x^2} } 
\ee 
Remarkably, the above prediction for $D_+(x,t)$, Eq. \eqref{soluD}, using the
formula \eqref{pxt} for $p_+(x,t)$ gives exactly the same result.

\bigskip

{\bf Semi-classical saddle point representation}. 
We can sketch here an argument which reproduces the above result
using semi-classical wave functions. It generalizes the study
 of \cite{Berry1977,Berry1979,Filippas2003,Filippas2006} for a single particle to $N$ fermions 
 and extends the calculation of \cite{Balatz1,Balatz2} at equilibrium to
 the nonequilibrium problem. It allows one  to obtain other expressions
 at non-generic points e.g. when $\partial^2_{x} p(x,t)=0$ 
 \cite{Filippas2003,UsMulticritical,UsWigner}. To study semi-classical wave functions one approximates them as 
$\psi(x,t) \simeq A(x,t) e^{\frac{i}{\hbar} S(x,t)}$. In the limit $\hbar \to 0$ the Schrodinger equation 
then leads to the eikonal equation
\be \label{eikonal}
\partial_t S + \frac{1}{2} (\partial_x S)^2 + V(x) = 0 \quad , \quad 
2 \partial_t A +  2 \partial_x A \partial_x S + A \partial_x^2 S = 0
\ee 
Consider first the initial condition at $t=0$. The kernel can be written as
\bea
&& K_\mu(x,x',0) = \int d\epsilon_0 \theta(\mu-\epsilon_0) \psi^*_{\epsilon_0}(x) 
\psi_{\epsilon_0}(x')  \quad , \quad \psi_{\epsilon_0}(x) = A_{\epsilon_0}(x,0) 
e^{\frac{i}{\hbar} S_{\epsilon_0}(x,0)} \\
&& S_{\epsilon_0}(x,0)=S_{\epsilon_0}(x) = \int^x dx' p_{\epsilon_0}(x') \quad , \quad 
A(x,0) = (2 \pi \hbar p_{\epsilon_0}(x))^{-1/2} \quad , \quad 
p_{\epsilon_0}(x) = \sqrt{2 (\epsilon_0 - V_0(x))} 
\eea
where $S_{\epsilon_0}(x,t)=- \epsilon_0 t + S_{\epsilon_0}(x)$ and $A(x,t)=A_{\epsilon_0}(x,0)$ 
are a stationary solution of \eqref{eikonal}
for $V(x)=V_0(x)$. There is a second one with $S_{\epsilon_0}(x,t)=- \epsilon_0 t - S_{\epsilon_0}(x)$,
i.e. $\partial_x S_{\epsilon_0}(x,t)= - \sqrt{2 (\epsilon_0 - V_0(x))}$ which is located at
a symmetric point along the Fermi surf, not considered here.
This choice satisfies the normalization 
\be
\int dx \, \psi_{\epsilon_0}(x) \psi^*_{\epsilon'_0}(x) =
\frac{1}{2 \pi \hbar} \int \frac{dx}{[p_{\epsilon_0}(x) p_{\epsilon'_0}(x)]^{1/2}} 
e^{\frac{i}{\hbar} (S_{\epsilon_0}(x)  - S_{\epsilon'_0}(x)) } 
\simeq 
\frac{1}{2 \pi \hbar} \int \frac{dx}{p_{\epsilon_0}(x)} 
e^{\frac{i}{\hbar} (\epsilon_0-\epsilon'_0) \partial_x S_{\epsilon_0}(x) } 
= \delta(\epsilon_0 - \epsilon'_0) 
\ee 
which ensures the reproducibility of the kernel. We used that 
$\partial_x S_{\epsilon_0}(x)=p_{\epsilon_0}(x)$ and 
$\partial_x \partial_{\epsilon_0} S_{\epsilon_0}(x)=1/p_{\epsilon_0}(x)$. We can now write the time dependent kernel as
\be
K_\mu(x,x',t) = \int d\epsilon_0 \theta(\mu-\epsilon_0) \psi^*_{\epsilon_0}(x,t) 
\psi_{\epsilon_0}(x',t)  \quad , \quad \psi_{\epsilon_0}(x,t) = A_{\epsilon_0}(x,t) 
e^{\frac{i}{\hbar} S_{\epsilon_0}(x,t)}
\ee
where $S_{\epsilon_0}(x,t)$ and $A_{\epsilon_0}(x,t)$ are the solutions of
\eqref{eikonal} with $V(x)$ (and not $V_0(x)$) and the above initial conditions. 
We note that $\partial_x S_{\epsilon_0}(x,t)=p_{\epsilon_0}(x,t) = p_+(x,t)$ satisfies the Burgers equation
\eqref{burgers} with initial condition $p_{\epsilon_0}(x,t)= p_{\epsilon_0}(x)$,
hence the eikonal equation is associated to the Liouville equation. 
It is useful to verify that 
\be \label{soluA} 
A_{\epsilon_0}(x,t) = (2 \pi \hbar)^{-1/2} [ \partial_{\epsilon_0} p_{\epsilon_0}(x,t) ]^{1/2}
\ee 
is the time dependent solution of \eqref{eikonal} with the correct initial condition given above.
Indeed taking $\partial_{\epsilon_0}$ of the Burgers equation \eqref{burgers} yieds
\be \label{burgers3} 
\partial_t \partial_{\epsilon_0} p_{\epsilon_0}(x,t)  + \partial_{\epsilon_0} p_{\epsilon_0}(x,t)  \partial_x p_{\epsilon_0}(x,t) 
+  p_{\epsilon_0}(x,t)  \partial_x \partial_{\epsilon_0} p_{\epsilon_0}(x,t)  =0
\ee
which is the same equation as obtained by inserting \eqref{soluA} into  \eqref{eikonal}.

Let us now write the associated Wigner function as
\bea
W(x,p,t) = \frac{1}{2 \pi \hbar}  \int d\epsilon_0 \theta(\mu-\epsilon_0) 
 \int dy  A^*_{\epsilon_0}(x+\frac{y}{2},t) A_{\epsilon_0}(x-\frac{y}{2},t) 
  e^{\frac{i}{\hbar} [p y - S_{\epsilon_0}(x+ \frac{y}{2},t) + S_{\epsilon_0}(x- \frac{y}{2},t)]}
\eea 
Let us expand the term in the exponential
\be
\frac{i}{\hbar} [ p y + S_{\epsilon_0}(x+ \frac{y}{2},t) - S_{\epsilon_0}(x- \frac{y}{2},t) ]
= \frac{i}{\hbar}  (p + \partial_x S_{\epsilon_0}(x,t)) y - \frac{i}{\hbar}  \frac{1}{24} y^3 
\partial_x^3 S_{\epsilon_0}(x,t) + \dots 
\ee 
In the integral over $y$ we can neglect the terms of order $O(y^5)$ and higher order terms.
Performing the integral we obtain
\bea
W(x,p,t) \simeq \frac{1}{2 \pi \hbar}  \int_{-\infty}^\mu d\epsilon_0 
\partial_{\epsilon_0} p_{\epsilon_0}(x,t) 
\frac{2}{(- \hbar^2 \partial_x^3 S_{\epsilon_0}(x,t))^{1/3} }
 {\rm Ai}\left(2 \frac{p - \partial_x S_{\epsilon_0}(x,t)}{(- \hbar^2 \partial_x^3 S_{\epsilon_0}(x,t))^{1/3} }\right)
\eea
Noting that the integral over $\epsilon_0$ is dominated by the region around $\epsilon_0 \approx \mu$, performing the expansion around $\epsilon_0=\mu$ in the argument of the Airy function,
$p - \partial_x S_{\epsilon_0}(x,t) = p - p_{\mu}(x,t) - \partial_{\mu} p_{\mu}(x,t) (\epsilon_0-\mu) + \dots$ and
approximating the term $\partial_{\epsilon_0} p_{\epsilon_0}(x,t)  \simeq 
\partial_{\mu} p_{\mu}(x,t)$ in the preexponential factor, we see that we can perform
the integral over the positive variable $\mu-\epsilon_0$ and recover exactly 
\eqref{Wscal} with $F(\tilde a)={\cal W}(\tilde a)={\rm Ai}_1(2^{2/3} \tilde a)$.

The above derivation is valid at a generic point on the Fermi surf. It must modified 
if one deals with non generic points, e.g. edges in the real space density where 
$p_+(x,t)$ and $p_-(x,t)$ coincide and annihilate.

\section{Evolution and scaling form of the finite temperature Wigner function} 

The finite temperature Wigner function $W_{\tilde \mu}(x,p,t=0)$ associated with the hamiltonian $H_0$ in the grand canonical (GC) ensemble at chemical
potential $\tilde \mu$ was defined in \cite{UsWigner}. Its definition at arbitrary time is given by \eqref{defWGC}. It is related to the time dependent GC kernel given in the text in Eq. (32)
by  Fourier transformation \eqref{Wrel}. As a consequence, it does
satisfy the WE Eq. (11) in the text, albeit with a $T,\tilde \mu$ dependent initial condition. 

It was shown in \cite{UsWigner} that at equilibrium, i.e. here at $t=0$, 
the finite temperature Wigner function $W_{\tilde \mu}(x,p,t=0)$  can be obtained from the 
Wigner function $W(x,p,t=0)|_\mu$ at $T=0$ expressed as a function of the Fermi energy $\mu$.
The relation reads 
\be
W_{\tilde \mu}(x,p,t=0)=\int d\mu' \frac{\partial_{\mu'} W(x,p,t=0)|_{\mu=\mu'}}{1+ e^{\beta(\mu'-\tilde \mu)}} 
\ee 
Since the WE is linear, it is then clear that the same relation will hold for any $t$, leading
to Eq. (34) in the text. For large $\tilde \mu$, i.e. large $\bar N$ [see Eq. (\ref{Nbar})], the integral
is controlled by large $\mu'$. In the bulk we can thus insert the form (26) of the text
for $W(x,p,t)$ evaluated at $\mu=\mu'$, and we obtain the formula given above (33) in the text for 
the time dependent Wigner function at finite temperature. We recall that
$x_0(x,p,t)$ and $p_0(x,p,t)$ are the starting points (at $t=0$) of the classical trajectory
(which goes from $(x_0,p_0)$ to $(x,p)$ in phase space) expressed as
functions of the endpoints (at time $t$).

Let us now study the edge, i.e. the vicinity of the transported Fermi surf $S_t$. 
The regime of temperature where the non trivial scaling arises at equilibrium, i.e. for $t=0$, is
$T \sim e_N/b \ll \mu$, where $b$ is the important dimensionless parameter. We can first assume the same temperature regime for $t>0$,
but we will be more precise below. 
We insert the scaling form for $p$ near $p_+(x,t,\mu)$ given by Eq. (30) and (31)
of the text (valid at $T=0$) inside Eq. (35) in the text. We obtain
\bea \label{rel4} 
W_{\tilde \mu}(x,p,t)=\int d\mu' \frac{1}{1+ e^{\beta(\mu'-\tilde \mu)}}
\partial_{\mu'} {\cal W}(\frac{p- p_+(x,t,\mu')}{- (\frac{\hbar^2}{2} \partial_x^2 p_+(x,t,\mu'))^{1/3}}) 
\eea 
with $F(a)={\cal W}(a)={\rm Ai}_1(2^{2/3} a)$. Note that here we have indicated explicitly the
dependence in $\mu$, so $p_+(x,t,\mu)$ denotes the solution of Burgers equation (28) in the text,
with initial condition 
$p(x,t=0,\mu)=p^0_{\mu}(x) = \sqrt{2 (\mu - V_0(x))}$. In the low temperature regime
of interest one has $\tilde \mu \simeq \mu$ (see \cite{UsWigner,fermions_review}).
The analysis is quite similar to the one performed at equilibrium in \cite{UsWigner} (see end of Section III there).
The integral in \eqref{rel4} is then dominated by $\mu'$ near $\mu$ 
such that $\beta(\mu'-\mu)=-b u$ and $u=O(1)$. Let us denote $\tilde a_\mu = \frac{p- p_+(x,t,\mu)}{D(x,t,\mu)}$ the scaling variable in the function ${\cal W}$ in \eqref{rel4}. One has
\be
 \partial_\mu \tilde a_\mu = - \frac{1}{D(x,t,\mu)} \partial_\mu p_+(x,t,\mu) - a_\mu \partial_\mu \log D(x,t,\mu)
\ee
As for equilibrium \cite{UsWigner}, the second term on the r.h.s. is subdominant at large $\mu$,
hence we can write 
$\tilde a_{\mu'} \simeq \tilde a_\mu + \frac{b}{\beta} u \frac{\partial_\mu p_+(x,t,\mu)}{D(x,t,\mu)}$.
Taking into account all factors we obtain
\be
W_{\mu}(x,p,t)
  = \int_{-\infty}^{+\infty} \frac{2^{2/3} du}{1+ e^{-b u}} {\rm Ai}(2^{2/3} (\tilde a_\mu + 
 \frac{b u}{\beta} \frac{\partial_\mu p_+(x,t,\mu)}{- (\frac{\hbar^2}{2} \partial_x^2 p(x,t,\mu))^{1/3}} )) = {\cal W}_{b(x,t)}(\tilde a_\mu) ~,~ 
 b(x,t) = \beta \frac{- (\frac{\hbar^2}{2} \partial_x^2 p(x,t,\mu))^{1/3}}{\partial_\mu p_+(x,t,\mu)}
\ee  
which recovers Eq. (36) of the text. As noted there it matches the equilibrium result for $t=0$ since one has
$- (\frac{\hbar^2}{2} \partial^2_{x} p_\mu^0(x))^{1/3} =\frac{e_N(x,p)}{p}|_{p=p_\mu^0(x)}$ on the Fermi surf
and $p_\mu^0(x) \partial_\mu p_\mu^0(x)=1$, hence $b(x,t=0)= \beta e_N= b$. In the dynamics, the same scaling function as in equilibrium thus describes the quantum and thermal rounding of the transported Fermi surf $S_t$.
It depends on the parameter $b$ which now becomes time dependent. It is also dependent of the position
along the Fermi surf, i.e. on $x$. The regime of thermal crossover is thus now $b(x,t) = O(1)$.

Finally, as mentionned in the text below Eq. (35), for the harmonic oscillator one has $b(x,t)=b$, i.e. it
is constant. This can be verified using that $b(x,t)=\beta D_+(x,t)/\partial_\mu p_+(x,t,\mu)=b$
together with Eq. \eqref{Dplus} and 
\be
\partial_\mu p_+(x,t,\mu) = \frac{1}{\omega_0 \sqrt{x_e(t)^2 -x^2}}
\ee
which can be checked from \eqref{pxt}. 
%
%

\section{Multi-point correlations}

\subsection{Determinantal structure of correlations and Eynard-Mehta theorem}


Here we address the multi-point, multi-time correlations. We show their extended determinantal structure and
obtain Eq. (36) in the text for the two time kernel. We follow closely the presentation in \cite{UsPeriodic}
and extend it to the non equilibrium case. We want to calculate, with the notation $X^{(a)}=(x_1^a,\dots, x_N^a)$ the quantum correlations
\bea
&& C_{t_1, \dots, t_m}(X^{(1)},\dots, X^{(m)}) = \\
&& \langle E_0 | e^{\frac{i}{\hbar} t_m {\cal H}} | X^{(m)} \rangle \langle  X^{(m)} |
e^{- \frac{i}{\hbar} (t_m-t_{m-1}) {\cal H}}  | X^{(m-1)} \rangle \langle  X^{(m-1)}  | \dots
| X^{(2)} \rangle \langle  X^{(2)}  |
e^{- \frac{i}{\hbar} (t_2-t_1) {\cal H}} | X^{(1)} \rangle \langle  X^{(1)}  | e^{- \frac{i}{\hbar} t_1 {\cal H}} | E_0 \rangle 
\nonumber 
\eea
where here $|E_0 \rangle$ is the ground state of ${\cal H}_0$. 
It is easy to check that each $C_{t_1, \dots, t_m}(X^{(1)},\dots X^{(m)})$ is normalized to unity. 
Nevertheless, for $m \geq 2$, they do not have the interpretation of a probability. Indeed they
are not necessarily real, since complex conjugation corresponds to time reversal $
C^*_{t_1, \dots, t_m}(X^{(1)},\dots, X^{(m)}) = C_{t_m, \dots, t_1}(X^{(m)},\dots, X^{(1)})$. For $m=1$ it is the equal time
quantum joint probability $C_t(X) = |\Psi(X,t)|^2$, with $\Psi(X,t)=\langle  X  | e^{- \frac{i}{\hbar} t {\cal H}} | E_0 \rangle$.

For non-interacting fermions, using the determinantal form of the real time many body propagator
in terms of the single particle one defined in \eqref{defG} and the Slater determinant form of $\langle X|E_0\rangle$
\be
\langle  X | e^{- \frac{i}{\hbar} t {\cal H}}  | Y \rangle = \frac{1}{N!} \det_{1 \leq j,k \leq N} G(x_j,y_k; i t )  \quad , \quad 
\langle X |E_0 \rangle = \frac{1}{\sqrt{N!}} \det_{1 \leq j,k \leq N} \phi_j^0(x_k) 
\ee
we can rewrite
\bea
&& C_{t_1,\dots, t_m}(X^{(1)},\dots, X^{(m)}) =  
\int dX^{(0)} dX^{(m+1)} 
C_{t_0,t_1, \dots, t_m,t_{m+1}}(X^{(0)},X^{(1)},\dots, X^{(m)},X^{(m+1)}) |_{t_0 \to 0, t_{m+1} \to 0} 
\eea
where we have inserted two complete sets of states $|X^{(0)} \rangle$ and $|X^{(m+1)} \rangle$
at (fictitious) times $t_0$ and $t_{m+1}$ (later set to zero) so that (in simplified notation) 
\bea \label{d1} 
&& C_{t_0,t_1, \dots, t_m,t_{m+1}}(X^{(0)},X^{(1)},\dots, X^{(m)},X^{(m+1)})  
= \frac{1}{N!^{m+2}} \det \phi_j^0(x^0_k) 
\det G(x^0_j,x^1_k;i (t_1-t_0)) 
\det G(x^1_j,x^2_k;i (t_2-t_1)) \nonumber \\
&& \times
\dots \times
\det G(x^{m-1}_j,x^m_k;i (t_m-t_{m-1})) 
\det G(x^m_j,x^{m+1}_k; i(t_{m+1}- t_m) )
\det \phi_j^0(x^{m+1}_k) 
\eea 
where the notation $\det$ stands for $\det_{1 \leq j,k \le N}$ and where we have used
that the eigenfunctions $\phi_j^0$ are real for a $1d$ confining potential
(which is not a restriction for the present considerations). Note that
consequently $G(x,x',\tau)$ is symmetric w.r.t. $x,x'$.
We have reversed the order of the terms for convenience. It is
now exactly the form required to apply the Eynard-Mehta theorem,
see the formula (4.1) in Ref. \cite{borodin_determinantal} (see also 
Proposition 19 in
Ref. \cite{FerrariBegRohu}). 
We can thus apply the Theorem (4.2) in \cite{borodin_determinantal}.
The Gram matrix ${\cal G}_{jk}$ defined there becomes trivial in the limit of interest here, $t_0, t_{m+1} \to 0$, i.e. ${\cal G}_{jk}=\delta_{jk}$. Indeed, using the convolution property of the propagator
$\int dy \, G(x,y,\tau) G(y,x',\tau')=G(x,x',\tau+\tau')$, as well as $G(x,x',0)=\delta(x-x')$, one has
\bea
&& {\cal G}_{jk} = \int \prod_{a=0}^{m+1} dx^a 
\phi^0_j(x^0) \prod_{a=0}^m G(x^a,x^{a+1};i(t_{a+1}-t_a))  \phi_k^0(x^{m+1}) |_{t_0 \to 0, t_{m+1} \to 0} = \int dx^0 
\phi^0_j(x^0)  \phi_k^0(x^0)
= \delta_{jk}
\eea
We thus obtain that $C_{t_0,t_1,..t_m,t_{m+1}}(X^{(0)},X^{(1)},\dots, X^{(m)},X^{(m+1)})$ is 
a determinantal complex valued measure with the kernel in Eq. (4.2) in \cite{borodin_determinantal}, 
which becomes, using again the
properties of the propagator,
for $0 \leq r \leq m+1$, $0 \leq s \leq m+1$ and setting $t_0=t_{m+1}=0$
\be
K_{\mu}(x ,t_r ; y , t_s) = - 
G(x,y;i (t_s-t_r)) \theta_{0 \leq  r<s \leq m+1} 
+ \sum_{j=1}^N
\int dz_1 dz_2  \, G(x,z_1,- i t_r)  \phi^0_j(z_1) \phi^0_j(z_2) \, G(z_2,y;i t_{s} )
\ee
Upon integration over the extra coordinates $X^0$ and $X^{m+1}$ (which is
immediate, since using the determinantal property it simply amounts to removing them
and multiply by $N!^2$),
we finally obtain the following result. One defines (as in \cite{UsPeriodic}) general 
multi space-time correlation functions involving an arbitrary number 
$\ell_j$ of fermions in each time slice $t_j$, with $j=1,\dots,m$,
and $t_1<t_2<\dots < t_m$ as (see Fig. 3 of \cite{UsPeriodic})
\cite{footnote10} 
\be
R_{\ell_1,\dots,\ell_m}(\{x_1^1,t_1;\dots;x^1_{\ell_1},t_1\},\dots,
\{x_1^m,t_1;\dots;x_{\ell_m}^m,t_m\})
= \prod_{j=1}^q \frac{N!}{(N-{\ell_j})!} \int \prod_{j=1}^q \prod_{i=\ell_j+1}^N dx_i^{(j)} 
C_{t_1, \dots, t_m}(X^{(1)},\dots, X^{(m)})
\ee 
The EM theorem states that any such correlation equals the determinant
\bea \label{detprop} 
R_{\ell_1,\dots,\ell_m}(\{x_1^1,t_1;\dots;x^1_{\ell_1},t_1\},\dots,
\{x_1^m,t_1;\dots;x_{\ell_m}^m,t_m\})
= \det_{1 \leq r,s \leq m, 1 \leq i \leq \ell_r , 1\leq j \leq \ell_s} K_\mu(x_i^r, t_r ; x_j^s, t_s) 
\eea 
with the following extended real time kernel, for $1 \leq r,s \leq m$
%
%
\bea \label{kernelr} 
 K_\mu(x ,t_r ; y , t_s) &=& 
\int dz_1 dz_2  G(x,z_1,- i t_r) \left( \sum_k (n^0_k   -  \theta_{+}(t_s-t_r) \phi_k^0(z_1) 
\phi_k^0(z_2) \right)
G(z_2,y;i t_{s}) \\
& =&  \sum_k (n^0_k   -   \theta_{+}(t_s-t_r) ) \psi^*_k(x,t_r) 
\psi(y,t_s) 
\eea
where $\theta_+(x)=1$ if $x>0$ and $\theta_+(x)=0$ if $x \leq 0$. Here we 
have introduced the occupation numbers of the initial state. It is clear that
this derivation can be performed for any eigenstate of ${\cal H}_0$ with arbitrary set of
occupation numbers $n_k \in \{0,1\}$. Performing the 
GC average at the end, amounts to replacing the $n^0_k$ by their GC averages $\bar n^0_k$. It leads to the corresponding determinantal property \eqref{detprop} in the GC
ensemble at any temperature with the GC extended kernel given in Eq. (36) in the text. 

In the case $H=H_0$, we have $\psi_k(x,t) = e^{- i \omega (k+ \frac{1}{2})  t} \phi_k^0(x)$
and Eq. \eqref{kernelr} identifies with the formula (153) of Ref. \cite{UsPeriodic} (arXiv version)
for the kernel at equilibrium.

The simplest application of the above formula is to calculate the correlation functions of the density operator 
at different times and $T=0$. We define the density operator
\be
\langle X | \hat \rho(z) | X' \rangle = \sum_{i=1}^N \delta(x_i-z) \delta(X-X') \quad , \quad 
\hat \rho(z,t) = e^{- \frac{i}{\hbar} {\cal H} t} \hat \rho(z) e^{\frac{i}{\hbar} {\cal H} t} .
\ee 
Applying the formula \eqref{detprop} with one fermion per time slice ($\ell_1=\ell_m=1$) we 
obtain 
\be
\langle \Psi_0 | \hat \rho(x_1,t_1) \dots \hat \rho(x_m,t_m) | \Psi_0 \rangle 
= \det K_\mu(x_i, t_i ; x_j, t_j) 
\ee 
with $t_1<t_2<\dots <t_m$ where $| \Psi_0 \rangle $ denotes the ground-state of $H_0$.


\subsection{Bulk and edge two time correlations and extended kernels}

Consider the case  where $H(t)$ is the  harmonic oscillator with time dependent frequency $\omega(t)$
and $H_0$ is the harmonic oscillator with fixed frequency $\omega_0$. Using
the rescaling method, i.e. formula (15) in the text, and
the representation of the kernel, Eq (36) in the text, we can
relate the nonequilibrium two time
kernel to the equilibrium one
\be \label{relationkernel} 
K_\mu(x,t;x't') = \frac{e^{- \frac{i}{\hbar} ( \frac{L'(t)}{2 L(t)} x^2 
- \frac{L'(t')}{2 L(t')} {x'}^2 )}}{\sqrt{L(t) L(t')}} K_\mu^{\rm eq}(\frac{x}{L(t)}, \tau(t); \frac{x'}{L(t')}, \tau(t')).
\ee
where $L(t)$ is the solution of the Ermakov equation (\ref{erm}) (setting $\omega_1 = \omega_0$) with $L(0)=1,L'(0)=0$ and
$\tau'(t)=1/L(t)^2$. Note that the right hand side only depends on $\tau(t)-\tau(t')$. 

Let us now recall 
the results for the equilibrium two time kernel, $K_{\tilde \mu}^{\rm eq}$,
for the $\omega_0$ harmonic oscillator obtained in \cite{UsPeriodic} in the large $N$ limit.

{\it Bulk regime}. In the bulk it takes the scaling form 
\be \label{scalb} 
K^{\rm eq}_{\mu}(x,t;x',t') = e^{\frac{i}{2} \omega_0  (t-t')}
 \frac{1}{\xi_x} {\cal K}^{\rm bulk,r}( \frac{|x-x'|}{\xi_x} 
, \hbar \frac{t-t'}{\xi_x^2} ) 
\ee
where the scaling function has the simple form at $T=0$
\bea
&& {\cal K}^{\rm bulk,r}(z,t)=\frac{1}{\pi} \int_0^2 e^{i v^2 t/2} \cos(v z) dv \quad , \quad t \geq 0 \\
&& {\cal K}^{\rm bulk,r}(z,t)= - \frac{1}{\pi} \int_2^{+\infty} e^{i v^2 t/2} \cos(v z) dv \quad , \quad t < 0
\eea
where $\xi_x = 2/(\pi N \rho_N(x))=\frac{2 \hbar}{\omega_0 \sqrt{x_e^2 - x^2}}$ where 
$x_e= \sqrt{2 \mu}/\omega_0 = \sqrt{2 N \hbar/\omega_0}$. 
At finite $T$ it can be read off from formula (90-92) in \cite{UsPeriodic}. 
\\

We note that $\xi_x \sim 1/\sqrt{N}$ at large $N$, hence the equilibrium kernel
decays very fast in $t-t'$. One can thus expand for $t-t'$ small, $t-t' \sim N^{-1}$,
and $x-x' \sim N^{-1/2}$.
Inserting the scaling form \eqref{scalb} into Eq. \eqref{relationkernel} 
and using the expansion
\be
\tau(t)-\tau(t') \simeq (t-t') \tau'(t) = \frac{t-t'}{L(t)^2} 
\quad , \quad \frac{x}{L(t)} - \frac{x'}{L(t')} \simeq \frac{x-x'}{L(t)} - x (t-t') \frac{L'(t)}{L(t)^2} 
\ee
we obtain the formula (38) of the text (where we have discarded the
global phase factor) with 
\be
\xi_{x,t}= L(t) \xi_{x/L(t)} = \frac{2 \hbar L(t)^2}{\omega_0 \sqrt{x_e(t)^2 - x^2}}
\ee
where $x_e(t)$ is the time dependent edge, $x_e(t)=L(t) x_e$. One can
thus define a velocity of propagation of correlations $v_{x,t}= x L'(t)/L(t)$,
at the edge $x=x_e(t)$ is equal to the velocity of the edge
$\frac{d}{dt} x_e(t)= x_e L'(t)$ (since $x_e(t) = x_e\,L(t)$). Note that this velocity is
of order $N^{1/2}$ hence the two terms in the argument of the scaling function in the Eq. (38) of the text, i.e.  
$x-x'$ and $v_{xt} (t-t')$, are of the same order, i.e. $N^{-1/2}$. 

\begin{figure}
\includegraphics[width = 0.7\linewidth]{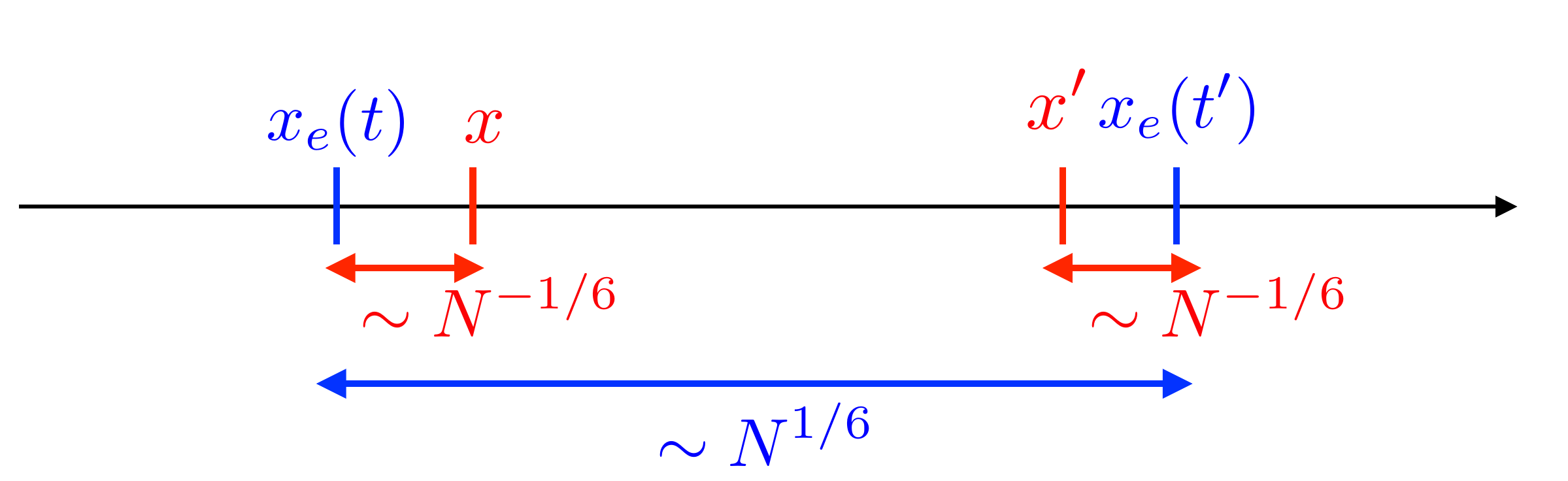}
\caption{Illustration of the two different scales in the nonequilibrium dynamics at the edge [see Eq. (39)
for the two time correlation in the text]: the two regions (in red) $x-x_e(t) \sim w_N(t) \sim N^{-1/6}$ and $x' - x_e(t') \sim w_N(t) \sim N^{-1/6}$ 
(i.e. the scale over which the density vanishes)
are separated by a much larger length scale (in blue) which is due to the motion of the edge $x_e(t') - x_e(t) \sim N^{1/6}$.}\label{Fig_scale_edge}
\end{figure}

{\it Edge regime}. We recall that near the edge, i.e. for $x,x' \approx x_e$,
the equilibrium kernel (at finite temperature) takes the scaling form 
\bea \label{scaledge}
&& K^{\rm eq}_{\tilde \mu}(x,t;x',t') = \frac{1}{w_N} {\cal K}_b^{\rm edge,r}(\frac{x-x_e}{w_N},\frac{x'-x_e}{w_N},
(t-t') \omega_0 N^{1/3})
\eea 
where $w_N = \sqrt{\frac{\hbar}{2 \omega_0} } N^{-1/6} \sim \mu^{-1/6}$
and $b=\frac{\hbar \omega_0}{T} N^{1/3}$. The scaling functions are \cite{UsPeriodic}
\bea
&& {\cal K}_b^{\rm edge,r}(s,s',u)=\int_{-\infty}^{+\infty} dv \frac{e^{- i v u}}{e^{-b v} +1} 
{\rm Ai}(s+v) {\rm Ai}(s'+v) \quad , \quad \;\;\; u \geq 0 \\
&&  {\cal K}_b^{\rm edge,r}(s,s',u)=- \int_{-\infty}^{+\infty} dv \frac{e^{- i v u}}{e^{-b v} +1}  
{\rm Ai}(s+v) {\rm Ai}(s'+v) 
\quad , \quad u <0
\eea

We now insert the scaling form \eqref{scaledge} into Eq. \eqref{relationkernel} 
and use the expansion, for small time differences $t'-t \sim N^{-1/3}$
\bea
&& (\tau(t)-\tau(t')) \, \omega_0 \, N^{1/3}  \simeq \frac{t-t'}{L(t)^2}  \omega_0 N^{1/3} \\
&& \frac{x'/L(t') -x_e}{w_N}  =  \frac{1}{w_N L(t)}  (x'-x_e(t) - x' (t'-t) \frac{L'(t)}{L(t)} )
\simeq \frac{1}{w_N L(t)}  (x'-x_e(t) - x_e(t) (t'-t) \frac{L'(t)}{L(t)} ).
\eea 
This leads to Eq. (39) in the text (up to a global phase factor which was discarded)
where $w_N(t)=w_N L(t)$ and $v_e(t) = x_e L'(t)$
is the velocity of the edge. Note that this velocity is
of order $N^{1/2}$ hence now the term 
$v_e(t-t')$ is of order $N^{1/6}$, while $x-x_e(t) \sim N^{-1/6}$.
Hence the two regions $x-x_e(t) \sim N^{-1/6}$ and $x'-x_e(t') \sim N^{-1/6}$ 
are separated in space by a much larger distance $N^{1/6}$ (see Figure \ref{Fig_scale_edge}).


\section{Large time limit}

As discussed in the text, it was found in \cite{Kulkarni2018} that in the classical
limit $\hbar = 1/N$ and large $N$, the Wigner function converges at large time 
to a limit which was calculated. This was claimed to hold for ``generic'' potentials $V(x)$,but not for the harmonic oscillator. 

In the text we have studied the quantum case, and obtained the formula for
the diagonal approximation (DA)
\be \label{DA}
K^{\rm di}_{\tilde \mu}(x,x') = \sum_{\ell=1}^{\infty} \nu_\ell \phi_\ell^*(x) \phi_\ell(x') 
\ee 
where we recall that the $\phi_\ell(x)$ are the eigenstates of $H$ and
\be
\nu_\ell = \sum_k \bar n^0_k ~ |\langle \phi_\ell | \phi_k^0 \rangle|^2  = 
\langle \phi_\ell  | \frac{1}{1+ e^{\beta(H_0-\tilde \mu)}} | \phi_\ell \rangle
\ee
using that $\bar n^0_k = (1+ e^{\beta(\tilde \mu - \epsilon^0_k)})^{-1}$, implying
$\sum_k \bar n^0_k |\phi_0^k \rangle \langle \phi^0_k| = [1+ e^{\beta(H_0 - \tilde \mu)}]^{-1}$.
As discussed in the text, the DA gives the exact {\it time average} of the kernel at large time,
$K^{\rm di}_{\tilde \mu}(x,x') = \lim_{\tau \to +\infty} \frac{1}{\tau} \int_0^{\tau} dt K_{\tilde \mu}(x,x',t)$.
This is an exact statement, a priori valid for any $N$ and $\hbar$, 
since the eigenstates of $H$ are non degenerate. If the large time 
limit of the kernel exists, i.e. $K_{\tilde \mu}^\infty(x,x')=\lim_{t \to +\infty} K_{\tilde \mu}(x,x',t)$, then
it is given by the DA, i.e. $K_{\tilde \mu}^\infty(x,x')=K^{\rm di}_{\tilde \mu}$.
However, showing that this limit exists, e.g. for large $N$ and ``generic'' potentials, and quantify the decay towards it,
is much more delicate and a priori requires further knowledge of
the spectrum of $H$. There are potentials, such as the harmonic oscillator, and, as we
have shown here, the harmonic oscillator plus a $1/x^2$ potential (which is a fully quantum case which does not obey the Liouville equation), 
where for any $N$ the system is time-periodic and the limit thus does not exist. 
This program was achieved in the classical
limit $\hbar = 1/N$ and large $N$ in \cite{Kulkarni2018}, where a power law
decay of the Wigner function to the limit was obtained. This implies the
same for the kernel since they are related by simply Fourier transforming.
For the quantum case, to our knowledge, the question of large time convergence is open.


Here we derive Eq. (41) in the text for the Wigner function in the DA. Then we take the
classical limit and 
verify that it reproduces the result of \cite{Kulkarni2018}. As shown in the text,
the DA is also equivalent to the result from the simplest Generalized Gibbs Ensemble (GGE) prediction. This settles the
question for correlation embodied in the kernel, e.g. the one point density.
Next we derive the formula (45) in the text for multipoint correlations
in the DA, which deviates from the predictions of the simplest GGE. 

\subsection{Derivation of Eq. (41) of the text} 

To calculate the Wigner function associated with the DA kernel \eqref{DA} we define
the kernel of the ground state of ${\cal H}$
\bea
K_\mu^H(x,x') =  \sum_{\ell} \theta(\mu-\epsilon_\ell)  | \phi_\ell \rangle \langle \phi_\ell| . 
\eea 
Now, let us define a function $\nu_{\tilde \mu}(\epsilon)$ such that 
\be
\nu_{\tilde \mu}(\epsilon_\ell) = \nu_\ell
\ee 
This function will be evaluated only at these points hence its value elsewhere is unimportant. 
Using $\partial_\mu K_\mu^H(x,x') =  \sum_{\ell} \delta(\mu-\epsilon_\ell)  | \phi_\ell \rangle \langle \phi_\ell|$
we see that we can rewrite 
\be 
K^{\rm di}_{\tilde \mu}(x,x') = \sum_{\ell=1}^{\infty} \nu_{\tilde \mu}(\epsilon_\ell) \phi_\ell^*(x) \phi_\ell(x') 
= \int d\mu' \nu_{\tilde \mu}(\mu') K^H_{\mu'}(x,x') 
\ee 
If we define the Wigner function $W^H_\mu(x,x')$ 
associated to the ground state of ${\cal H}$ from $K^H_{\mu'}(x,x')$
using the Eq. (9) of the text, by linearity of the Fourier transform we obtain
Eq. (41) of the text. 

\subsection{Classical limit}

To deal with the classical limit it is convenient to introduce the density of occupation numbers
and the density of states of $H$, together with their relation
\be
\hat \nu_{\tilde \mu}(\epsilon) = \sum_\ell \nu_\ell \delta(\epsilon - \epsilon_\ell) \quad , \quad 
\rho(\epsilon) = \sum_\ell \delta(\epsilon - \epsilon_\ell) \quad , \quad \hat \nu_{\tilde \mu}(\epsilon)  =  \nu_{\tilde \mu}(\epsilon)  \rho(\epsilon)
\ee 
In the classical limit $\hbar \sim 1/N$, and large $N$, we can use the result that the Wigner function
$W_\mu^H(x,p)$ becomes $W_\mu^H(x,p) \simeq \frac{1}{2 \pi \hbar} \theta(\mu- H(x,p))$. Inserting this form in
Eq. (41) of the text we obtain Eq. (43) of the text, i.e.
\be \label{Wsc} 
W^{\rm di}_{\tilde \mu}(x,p) \simeq \frac{1}{2 \pi \hbar} \nu_{\tilde \mu}(H(x,p)) \quad , \quad H(x,p)= \frac{p^2}{2} + V(x)
\ee 

We now restrict ourselves to the case  $T=0$, and consider the probability that a measure of the energy operator $H$ of one fermion 
in the initial state gives a value $\epsilon$, $P(\epsilon)$. In the classical limit we know that 
this can be evaluated as an average over the Wigner function of the initial state of $H(x,p)$
\be
P^{\rm sc}(\epsilon) = \frac{1}{N} \int dx_0 dp_0 W(x_0,p_0,0) \delta(\epsilon - H(x_0,p_0)) 
\ee 
which is normalized to unity. On the other hand the same observable in the quantum case
can be calculated as follows
\be
P(\epsilon) = \frac{1}{N}  \sum_i \langle \Psi(t=0) | \delta(\epsilon - H_i) |\Psi(t=0) \rangle \quad , \quad 
\langle x_1,\dots, x_n |\Psi(t=0) \rangle  = \frac{1}{\sqrt{N!}} \det_{1 \leq i,k \leq N} 
\phi^0_k(x_i)
\ee 
where $H_i$ is the single particle Hamiltonian of the $i$-th particle, leading to
\be
P(\epsilon) = \frac{1}{N}  \sum_{k=1}^N  \langle \phi_k^0 | \delta(\epsilon - H) | \phi_k^0 \rangle
=
\frac{1}{N}  \sum_k \theta(\mu - \epsilon^0_k) \langle \phi_k^0 | \delta(\epsilon - H) | \phi_k^0 \rangle
\ee 
as can be seen e.g. expanding the Slater determinants in sums over permutations and integrating over
all coordinates except one. One can check that it is also normalized to unity. 
This formula can be rewritten as
\be
P(\epsilon) = \frac{1}{N} {\rm Tr}[ \theta(\mu - H_0) \delta(\epsilon - H) ]
=  \frac{1}{N}  \sum_\ell \langle \phi_\ell | \theta(\mu - \epsilon_\ell) | \phi_\ell \rangle
\delta(\epsilon - \epsilon_\ell) = \frac{1}{N}  \hat \nu_{\mu}(\epsilon) 
\ee
where in the second equality we expressed the trace in the eigenbasis of $H$
and used that $\nu_\ell = \langle \phi_\ell | \theta(\mu - \epsilon_\ell) | \phi_\ell\rangle$.

It is thus clear that in the classical limit at $T=0$ we have 
\be \label{Psc} 
P(\epsilon) =  \frac{1}{N}  \hat \nu_{\mu}(\epsilon)  \to 
P^{\rm sc}(\epsilon) = \frac{1}{2 \pi \hbar N} \int dx_0 dp_0 \, \delta(\epsilon - \frac{p_0^2}{2} - V(x_0))  
\theta(\mu - \frac{p_0^2}{2} - V_0(x_0)) 
\ee 

We now use our result \eqref{Wsc} for the Wigner function in the classical limit
\bea \label{resultW} 
W^{\rm di}_{\tilde \mu}(x,p) &\simeq& \frac{1}{2 \pi \hbar} \nu_{\tilde \mu}(H(x,p)) 
\to \frac{1}{2 \pi \hbar}  \frac{\hat \nu^{\rm sc}_{\mu}(\epsilon)}{\rho^{\rm sc}(\epsilon)} |_{\epsilon=H(x,p)} 
= \frac{N}{2 \pi \hbar}  \frac{P^{\rm sc}(\epsilon)}{\rho^{\rm sc}(\epsilon)} |_{\epsilon=H(x,p)} \\
&
=& \frac{1}{\rho^{\rm sc}(H(x,p))} \int \frac{dx_0 dp_0}{2 \pi \hbar} ~\delta(H(x,p)-H(x_0,p_0)) 
W(x_0,p_0,0) 
\eea
where $\rho^{\rm sc}(\epsilon)$ is the semi-classical density of states of $H$. This is
our main result for the classical limit of the DA Wigner function, which we recall
is exact for the time averaged Wigner function in all cases (i.e. solution of the 
Liouville solution) and equal to the
large time limit when this limit exists. Although derived here from the quantum formula
at $T=0$ it is actually easy to see that it extends to a finite temperature quench, and in fact
to any initial condition $W(x_0,p_0,0)$ (see below). 

We can now obtain a more explicit form for the numerator and the denominator of the first line of
\eqref{resultW}, which will allow us to compare with
the result of \cite{Kulkarni2018}, obtained by a completely different
method. Let us consider,  again, only the case $T=0$. If $V(x)$ 
is a confining potential, there are only periodic classical trajectories. The semi-classical density
of states $\rho^{\rm sc}(\epsilon)$ is related to the period $T(\epsilon)$ of the classical orbits at 
energy $\epsilon$. Let us
assume for simplicity that there is only one periodic orbit at energy $\epsilon$, with
turning points $x_{\pm}(\epsilon)$. In the limit of small $\hbar$, the semi-classical quantization condition for level $k$ at energy $\epsilon=\epsilon_k$ reads
$2 \int_{x_-(\epsilon)}^{x_+(\epsilon)} dx \sqrt{2 (\epsilon-V(x))}|_{\epsilon=\epsilon_k} = 2 \pi \hbar k$.
From this one obtains 
\bea \label{rhoT} 
\rho^{sc}(\epsilon) \simeq \frac{dk}{d\epsilon_k}|_{\epsilon_k=\epsilon} = 
\frac{1}{\hbar \pi} \int_{x_-(\epsilon)}^{x_+(\epsilon)}\frac{dx}{\sqrt{2 (\epsilon-V(x))} }
= \frac{T(\epsilon)}{2 \pi \hbar} 
\eea 
In case of multiple classical orbits at the same energy, with periods $T_i(\epsilon)$, as is the case e.g. for a double well potential, a more general formula is
$\rho^{sc}(\epsilon) \simeq  
\frac{1}{\hbar \pi} \int_{-\infty}^{+\infty}\frac{dx}{\sqrt{2 (\epsilon-V(x))_+} }= \sum_i T_i(\epsilon)/(2 \pi \hbar)$
where $\frac{1}{\sqrt{(x)_+}}=\frac{1}{\sqrt{x}} \theta(x)$.

Next one can integrate over $p_0$ in \eqref{Psc} and obtain
\bea
P^{\rm sc}(\epsilon) &=& \frac{1}{\pi \hbar N} \int dx_0 \frac{1}{\sqrt{2 (\epsilon - V(x_0))_+} } 
\theta(\mu- \epsilon + V(x_0) - V_0(x_0)) 
\eea
{Note that there are two roots for $p_0$ leading to an overall factor of 2.}
Putting all together, we find our final formula for the Wigner function in the DA and the classical
limit, valid in all cases
\be \label{WscUS} 
W^{\rm di}_{\tilde \mu}(x,p) \simeq 
\frac{1}{2 \pi \hbar}  
\frac{\int_{-\infty}^{+\infty} dx_0 \frac{1}{\sqrt{2 (\epsilon - V(x_0))_+} } \theta(\mu- \epsilon + V(x_0) - V_0(x_0))
 }{
\int_{-\infty}^{+\infty}\frac{dx_0}{\sqrt{2 (\epsilon-V(x_0))_+} }} ~\bigg|_{\epsilon=H(x,p)} 
\ee

In Ref. \cite{Kulkarni2018} the simplest case was studied, i.e. when there are at most two
roots to the equation
\be \label{eq1} 
\mu - H_0(x_0,p_0) = \epsilon - H(x_0,p_0)  \quad \Leftrightarrow \quad \mu- \epsilon + V(x_0) - V_0(x_0) = 0 
\ee 
denoted there $x_0^\pm(\epsilon)$, and a single periodic orbit per energy. In that case our formula \eqref{WscUS} can be rewritten as
\bea
&& W^{\rm di}_{\tilde \mu}(x,p) \simeq     \frac{1}{\pi \hbar T(\epsilon)}
 \int_{x_0^{-}(\epsilon)}^{x_0^{+}(\epsilon)} dx_0 \frac{1}{\sqrt{2 (\epsilon - V(x_0))} } |_{\epsilon=H(x,p)} 
\eea
which can be compared with Eq. (56) in \cite{Kulkarni2018} for the different
case of open trajectories in a spatially periodic system which takes a
similar form. 

%

In general however there can be more than two roots to the equation \eqref{eq1}. Indeed
for a given $\epsilon$ this equation describes the intersection of the contour line of $H$ defined by the curve 
$H(x_0,p_0)=\epsilon$, with the initial Fermi surf $H_0(x,p) < \mu$, which may have 
an arbitrary even number of roots $x_0^{\pm,j}(\epsilon)$. The difference
between these situations is illustrated in Figs. \ref{Fig_scale_edge}
and \ref{Fig_scale_edge2}. In that case the formula
\eqref{WscUS} obtained here is likely to be more convenient.

\begin{figure}[h]
\includegraphics[width = 0.6\linewidth]{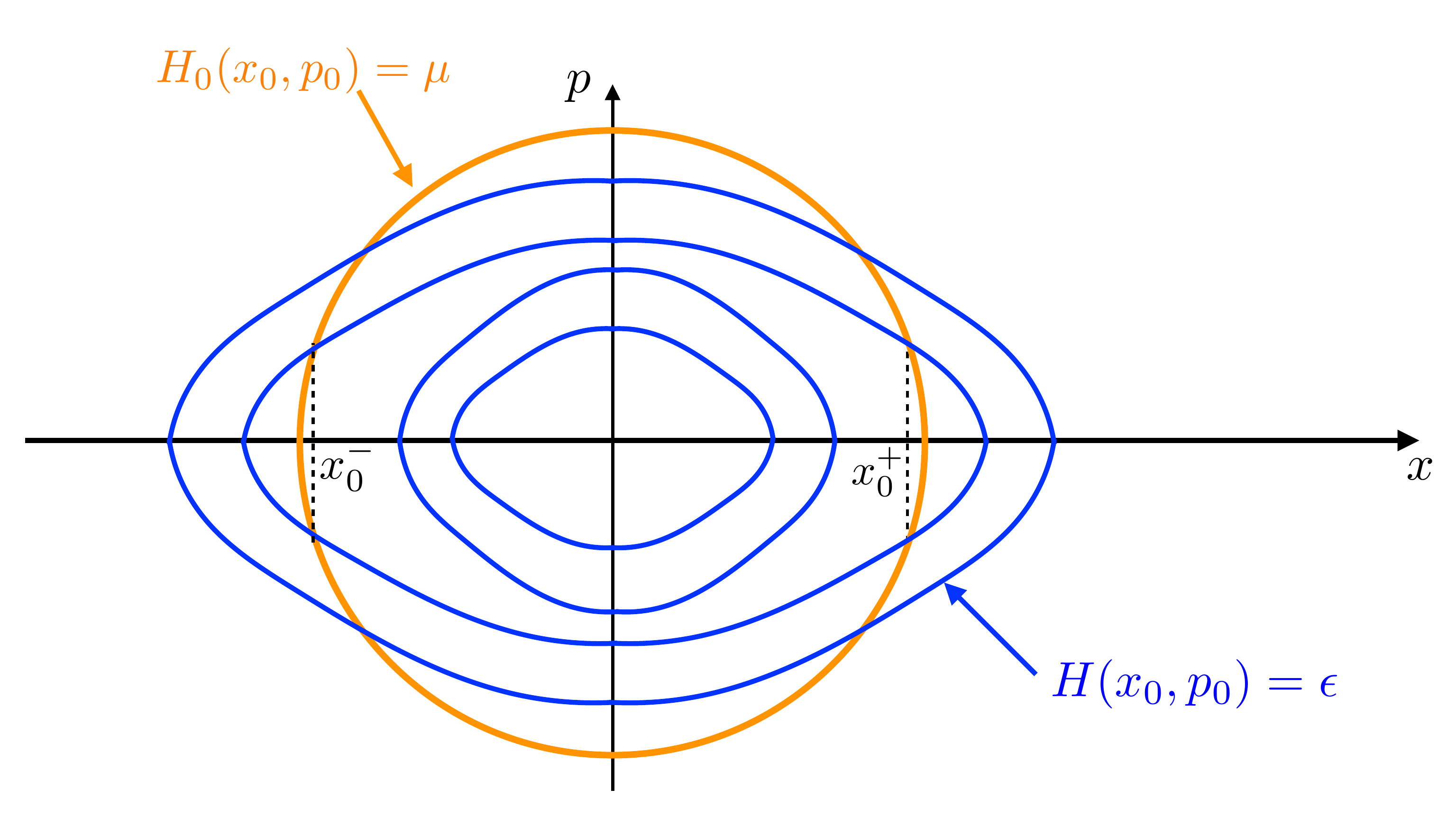}
\caption{{Case where there are at most two roots to the equation \eqref{eq1}.
In yellow the initial Fermi surf $H_0(x,p)=\mu$ in the $(x,p)$ phase space. The intersection
with the surfaces of fixed energy $H(x,p)=\epsilon$ determine the two roots
$x_0^{\pm}(\epsilon)$. Note that there are in fact 4 intersections of these
two surfaces because of the symmetry $p \to -p$.}}\label{Fig_scale_edge}
\end{figure}

\begin{figure}[h]
\includegraphics[width = 0.6\linewidth]{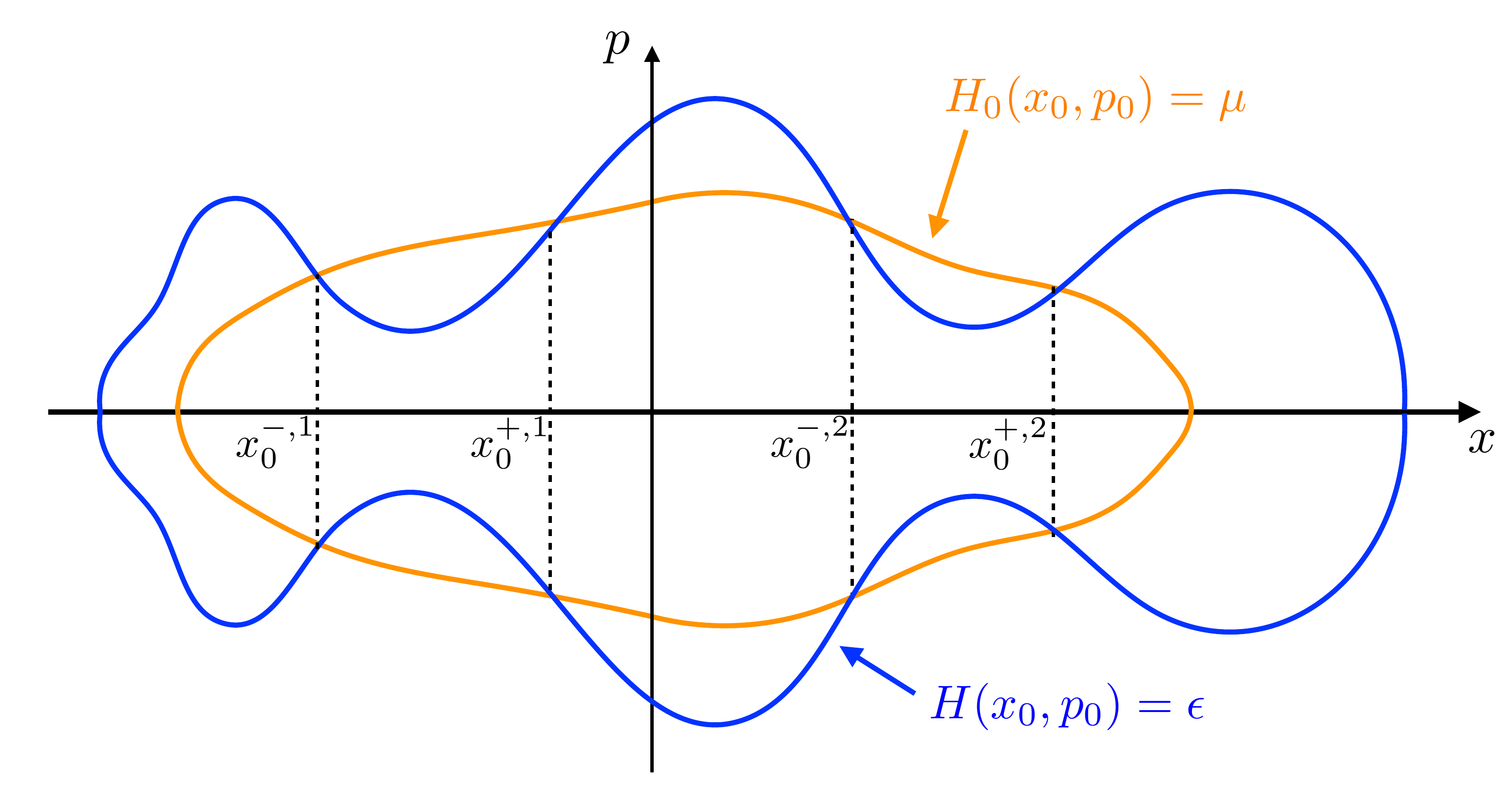}
\caption{{Same as Fig. \ref{Fig_scale_edge} except now there can be more
than two roots to \eqref{eq1}, hence there can be more than 4 intersections of the two surfaces.}}\label{Fig_scale_edge2}
\end{figure}


Let us mention an alternative approach to the above derivation. One defines the 
Wigner function for a single particle in the eigenstate $\phi_\ell$ of $H$
\be
W_\ell(x,p)= \frac{1}{2 \pi \hbar} \int dp e^{i p y/\hbar} \phi_\ell^*(x+\frac{y}{2}) 
\phi_\ell(x-\frac{y}{2}) \quad , \quad \sum_\ell 
\theta(\mu- \epsilon_\ell) W_\ell(x,p) = W^H(x,x') 
\ee
One can then show
that $\nu_\ell = 2 \pi \hbar \int dy dp W(y,p,0) W_\ell(y,p)$, which
implies that
\be \label{Dean} 
W^{\rm di}_{\tilde \mu}(x,p)= \sum_\ell \nu_\ell W_\ell(x,p) =
2 \pi \hbar \sum_\ell \bigg[ \int dy dp' W(y,p',0) W_\ell(y,p') \bigg]
W_\ell(x,p)
\ee 
This shows that the $N$ fermion Wigner function in the DA is a sum of
single particle Wigner function, weighted by the overlap with the initial Wigner function.
This overlap is, heuristically, proportional to the volume of phase space of each
single particle state which lies inside the initial Fermi volume. 
In the semi-classical limit we thus have
\be \label{Well} 
W_\ell(x,p) \simeq \frac{1}{T(\epsilon_\ell)} \delta(\epsilon_\ell - \frac{p^2}{2} - V(x)) 
\ee 
which is clearly normalized to unity $\int dp dx W_\ell(x,p)=1$. Inserting 
\eqref{Well} into \eqref{Dean} we see that it exactly agrees with the final formula in
\eqref{resultW} using that 
\be
\frac{1}{\rho^{sc}(\epsilon)} = (2 \pi \hbar)^2 \sum_\ell \frac{1}{T(\epsilon_\ell)^2} \delta(\epsilon_\ell-\epsilon)
\ee
which is equivalent to \eqref{rhoT}.

{Finally let us note that in the classical limit $2 \pi \hbar W(x,p,t)$ is either $0$ or $1$ pointwise, as discussed
above. However we note that the infinite time limit obtained here and in \cite{Kulkarni2018}, 
$2 \pi \hbar  W(x,p,t=\infty)$ is a smooth function with values in $[0,1]$. These two facts can
be reconciled as follows. As can be seen, e.g.
in the leftmost plot in Fig. 2 in \cite{Kulkarni2018}, it is expected that this limit is indeed distributional
and not pointwise. This means that its value is proportional the local density of points where $2 \pi \hbar  W$ takes
the value 1. It would be interesting to understand how this rather unusual feature is modified in
presence of the quantum effects, in the light of what we have shown here i.e. that at
finite time the quantum corrections (i.e. a finite $\hbar$)
induce a quantum ``width'' to the Fermi surf.}

\subsection{Diagonal ensemble and $m$-point correlations} 

We now derive the equation (45) in the text. We consider the evolution
from an eigenstate $| {\bf n}^0 \rangle$ 
of ${\cal H}_0$ with a given set of occupation numbers 
${\bf n}_0= \{ n_k^0 \}_{k \geq 1}$ with $n^0_k \in \{0,1\}$, 
with $\sum_{k=1}^\infty n_k^0=N$. As discussed in
Section \ref{sec:temperature} the correlations in this state 
are determinantal and given by 
\be
R_{m,{\bf n}_0}(x_1,\dots,x_m,t) = \det_{1 \leq i,j \leq m} K(x_i,x_j,t;{\bf n}_0)  \quad , \quad 
K(x,x',t;{\bf n}_0) = \sum_{k=1}^{+\infty} n^0_k \psi_k^*(x,t) \psi_k(x',t)
= \langle x | K[{\bf n}^0,t] | x' \rangle
\ee
where we define the operator form of  the kernel as
\be
K[{\bf n}^0,t] = \sum_{k} n^0_{k} |\psi_k(t) \rangle \langle \psi_k(t)|
\quad , \quad |\psi_k(t) \rangle = e^{- i H t/\hbar} |\phi^0_k \rangle 
\ee 
This can also be written as 
\bea
R_{m,{\bf n}_0}(x_1,\dots,x_m,t) =  \langle A:x_1,\dots, x_m | K[{\bf n}^0,t] \otimes \dots  \otimes K[{\bf n}^0,t]| A:x_1,\dots, x_m \rangle
\eea 
where $\otimes$ is the tensorial product and 
$| A:x_1,\dots, x_m \rangle = \frac{1}{\sqrt{m!}} \sum_{\sigma \in S_m} (-1)^\sigma |x_{\sigma(1)}, \dots, x_{\sigma(m)}\rangle$.
The tensor product evolves as
\be
K[{\bf n}^0,t] \otimes \dots  \otimes K[{\bf n}^0,t] = e^{i {\cal H}_m t} 
K[{\bf n}^0,0] \otimes \dots  \otimes K[{\bf n}^0,0] e^{- i {\cal H}_m t} 
\ee
where ${\cal H}_m$ 
is the $m$ fermion Hamiltonian, since we retain only antisymmetric states. We can parameterize
its orthonormal eigenbasis by $|{\bf n} \rangle$ where $n_\ell=0,1$ are occupation numbers with 
$\sum_\ell n_\ell=m$. These eigenstates are Slater determinants $\langle x_1,\dots, x_m |{\bf n} \rangle =
\frac{1}{\sqrt{m!}} \det_{1 \leq i,j \leq m} \phi_{\ell_i}(x_j)$ (such that 
$n_\ell= \sum_{j=1}^m \delta_{\ell,\ell_j}$) and thus antisymmetric, hence
\bea
&&    R_{m,{\bf n}_0}(x_1,\dots,x_m,t)  \\
&& = m! \sum_{{\bf n},{\bf n}',\sum_\ell n_\ell=m,\sum_\ell n'_\ell=m} 
 \langle x_1,\dots, x_m |{\bf n} \rangle 
\langle {\bf n}' | x_1,\dots, x_m \rangle
e^{- i t \sum_\ell \epsilon_\ell (n'_\ell - n_\ell))} 
\langle {\bf n}|  K[{\bf n}^0,0] \otimes K[{\bf n}^0,0] \otimes \dots  \otimes K[{\bf n}^0,0]  |{\bf n}' \rangle \nonumber 
\eea
The diagonal ensemble is defined such that we keep only the terms ${\bf n}={\bf n}'$ hence
\bea
&&  R^{\rm di}_{m,{\bf n}_0}(x_1,\dots,x_m)  = { m!} \sum_{{\bf n},\sum_\ell n_\ell=m} 
 |\langle x_1,\dots, x_m |{\bf n} \rangle|^2 
\langle {\bf n}|  K[{\bf n}^0,0] \otimes  \dots  \otimes K[{\bf n}^0,0]  |{\bf n}\rangle
\eea
On the other hand it is easy to see that 
\bea
\langle {\bf n}|  K[{\bf n}^0,0] \otimes \dots  \otimes K[{\bf n}^0,0]  |{\bf n}\rangle
= \det_{1 \leq i,j  \leq m} \langle \phi_{\ell_i} | K[{\bf n}^0,0]  | \phi_{\ell_j}  \rangle.
\eea 
We can now perform the average over the occupation numbers 
${\bf n}^0$ with the grand canonical weight and we obtain a first expression
for the $m$-point GC correlation in the DA 
\bea
R^{\rm di}_{m}(x_1,\dots,x_m)  = { m!}
\sum_{{\bf n},\sum_\ell n_\ell=m}  |\langle x_1,\dots, x_p |{\bf n} \rangle|^2
\det_{1 \leq i,j  \leq m} \langle \phi_{\ell_i} | \frac{1}{1 + e^{\beta(H_0-\mu)}}  | \phi_{\ell_j}  \rangle
\eea 
Defining as in the text $\nu_{\ell,\ell'} = \langle \phi_{\ell} | \frac{1}{1 + e^{\beta(H_0-\mu)}}  | \phi_{\ell'}  \rangle
= \langle c_\ell^\dagger c_{\ell'} \rangle_0$ and using the Slater determinant form of the
states $|{\bf n} \rangle$ given above we obtain the formula (45) in the text. 
We give here several other equivalent forms
\bea
&& R^{\rm di}_{m}(x_1,\dots,x_m)  = 
\sum_{\ell_1<\dots<\ell_m} \det_{1 \leq i,j  \leq m}[ \sum_{k=1}^m \phi_{\ell_k}(x_i) \phi_{\ell_k}(x_j) ]
\det_{1 \leq i,j  \leq m} \nu_{\ell_i,\ell_j} \\
&& = 
\sum_{{\bf n},\sum_\ell n_\ell=m} \det_{1 \leq r,s \leq m} \bigg[ \sum_{i,j=1}^m \phi_{\ell_i}(x_r) 
\langle \phi_{\ell_i} | \frac{1}{1 + e^{\beta(H_0-\mu)}}  | \phi_{\ell_j}  \rangle
\phi_{\ell_j}(x_s)  \bigg] =  
\sum_{{\bf n},\sum_\ell n_\ell=m} \det_{1 \leq i,j  \leq m}    \langle x_i | K_{\bf n} \frac{1}{1 + e^{\beta(H_0-\mu)}}  K_{\bf n} | x_j \rangle \nonumber 
\eea

%
%
%

This DA for the-$m$ point correlation is the exact limit of the time averaged kernel
if there are no degeneracies in the $m$-fermion Hamiltonian ${\cal H}_m$. Such degeneracies
would happen if $\sum_\ell \epsilon_\ell (n'_\ell - n_\ell)) =0$ for some pairs of 
distincts ${\bf n} \neq {\bf n}'$.

As mentioned in the text the result for the DA is different from the prediction of the simplest GGE.
This prediction amounts making the  replacement $\det_{1 \leq i,j \leq m} \nu_{\ell_i,\ell_j}  \to \nu_{\ell_1} \dots \nu_{\ell_m}$,
leading to 
\be
R^{\rm di}_{m}(x_1,\dots,x_m)  \to
\frac{1}{m!} \sum_{\ell_1,\dots,\ell_m} \det_{1 \leq i,j  \leq m} [\nu_{\ell_i} \phi_{\ell_i}(x_j)]
\det \phi_{\ell_i}(x_j) = \det_{1 \leq i,j  \leq m} [\sum_\ell \nu_\ell \phi_\ell(x_i) \phi_\ell(x_j) ]
= R^{\rm GGE}_{m}(x_1,\dots,x_m)
\ee 
where we have used the Cauchy-Binet identity.
This shows that unless one can neglect the off-diagonal elements,
the time averaged correlation (i.e. as obtained from the DA) does not coincide with the
GGE prediction for $m \geq 2$.